\documentclass[12pt,oneside,english]{book}
\usepackage[T1]{fontenc}
\usepackage[latin1]{inputenc}
\usepackage{a4wide}
\usepackage{graphicx}
\usepackage{amssymb}

\makeatletter

\providecommand{\LyX}{L\kern-.1667em\lower.25em\hbox{Y}\kern-.125emX\@}

\usepackage{graphicx}

\usepackage{babel}
\makeatother
\begin{document}

\newcommand{\ob}[1]{\left(#1\right)}

\newcommand{\rb}[1]{\left[#1\right]}

\newcommand{\cb}[1]{\left\{ #1\right\} }

\newcommand{\tb}[1]{\left\langle #1\right\rangle }

\thispagestyle{empty}

\begin{center}\textbf{\huge A Transverse Lattice QCD Model for Mesons}\end{center}{\huge \par}

{\huge \par{}}{\huge \par}

{\huge \par{}

\par{}}{\huge \par}

{\huge \par{}}{\huge \par}

{\huge \par{}

\par{}}{\huge \par}

\bigskip{}

\begin{center}A thesis submitted for the degree of\\
 Doctor of Philosophy in the Faculty of Science\end{center}

\vspace*{\fill}

\begin{center}\textbf{\large Raghunath Ratabole}\end{center}{\large \par}

{\large \par{}}{\large \par}

{\large \par{}

\par{}}{\large \par}

\vspace*{\fill}

\noindent \begin{center}\includegraphics{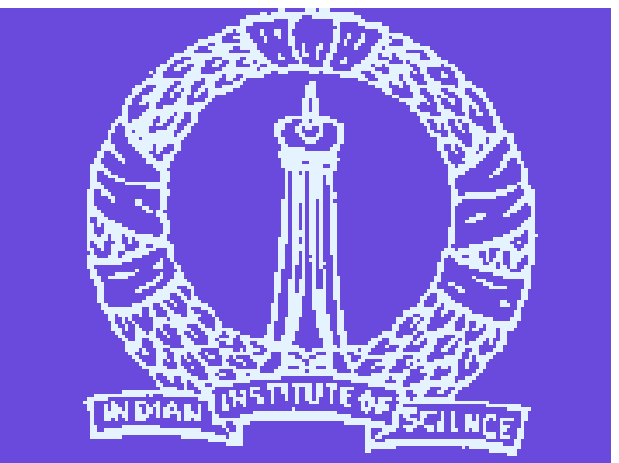}\end{center}

\begin{center}Centre for High Energy Physics\\
 Indian Institute of Science\\
 Bangalore - 560012, India \end{center}

\noindent \bigskip{}

\begin{center}Submitted: August 2003\\
              Defended: July 2004\end{center}

\newpage

\thispagestyle{empty}

\vspace*{\fill}

\begin{center}\textbf{\huge Declaration}\end{center}{\huge \par}

{\huge \par{}}{\huge \par}

{\huge \par{}

\par{}}{\huge \par}

{\huge \par{}

\par{}}{\huge \par}

{\huge \bigskip{}}{\huge \par}

{\huge \par{}

\par{}}{\huge \par}

This thesis describes work done by me during my tenure as a Ph. D.
student at the Centre for High Energy Physics, Indian Institute of
Science, Bangalore. This thesis has not formed the basis for the award
of any degree, diploma, membership, associateship or similar title
of any university or institution.\\

\begin{flushleft}Raghunath Ratabole\\
 August 2003\end{flushleft}

\begin{flushleft}Centre for High Energy Physics\\
 Indian Institute of Science\\
 Bangalore - 560012\\
 India\end{flushleft}

\vfill{}

\newpage

\chapter*{Acknowledgements}

This thesis wouldn't have been possible without the help of

\begin{itemize}
\item Apoorva Patel: My thesis advisor. Working with him has been a great
learning experience. 
\item Avaroth Harindranath: I developed an interest in light-front field
theory through his lectures at the SERC school, Santiniketan. I eventually
collaborated with him and Asmita Mukherjee in the area of light-front QCD. 
\item Asmita Mukherjee: I was lucky to have her as a collaborator who would
willingly and patiently trace out little errors in my calculations. 
\item The \LyX{} Team: Their creation, \LyX{}, a frontend for \LaTeX{},
made thesis writing an easy task. 
\item CSIR: I thank them for providing research fellowship. 
\item DST: I acknowledge support from the project \char`\"{}Some aspects
of low energy hadronic physics, SP/S2/K-28/99\char`\"{}. 
\item Anand Ratabole: My father. He has been most kind and patient in
dealing with me. 
\item Sanjeevani Jain: My college physics teacher. If it weren't for her,
I would probably have majored in chemistry. 
\item Arvind Kumar: I learnt Quantum Mechanics from his weekly pedagogical
lectures at the Homi Bhabha Centre for Science education. 
\item I have made many friends during my stay on the IISc campus. Amongst
them are those who get voyeuristic pleasure out of reading thesis
acknowlegdements. I dedicate this thesis to them. 
\end{itemize}
\pagenumbering{roman}\tableofcontents{}

\chapter{Introduction}

\pagenumbering{arabic}

\setcounter{page}{ 1}

\section{Reasons for gauge theory of the strong interactions}

Quantum Chromodynamics (QCD) is an $SU_{c}(3)$ colour gauge theory,
describing interactions of quarks in the fundamental representation.
We begin by summarizing the standard arguments in favour of QCD as
the theory for strong interaction physics (see for instance \cite{pokorski}).

\begin{enumerate}
\item At an empirical level, light hadrons (mesons and baryons) can be classified
according to representations of $SU_{f}(3)$ (the subscript $f$ indicates
flavour). It is observed that precisely those representations of $SU_{f}(3)$
are physically realized that may be obtained by decomposing the direct
products $\underline{3}\otimes \underline{3}^{\star }$ (mesons) or
$\underline{3}\otimes \underline{3}\otimes \underline{3}$ (baryons)
as direct sums of irreducible representations of $SU_{f}(3)$. \begin{eqnarray*}
\underline{3}\otimes \underline{3}^{\star } & = & \underline{8}\oplus \underline{1}\\
\underline{3}\otimes \underline{3}\otimes \underline{3} & = & \underline{10}\oplus \underline{8}\oplus \underline{8}\oplus \underline{1}
\end{eqnarray*}
 It is an important fact that the fundamental representations of $SU_{f}(3)$,
$\underline{3}$ and $\underline{3}^{\star }$, are not among the
physically realized representations. 
\item This leads to the hypothesis of quarks and antiquarks as the fundamental
constituents of hadrons (Quark Model of hadrons). One finds that there
are three {}``flavours'' of light quarks (antiquarks), which transform
according to the fundamental representation $\underline{3}$ ($\underline{3}^{\star }$)
of $SU_{f}(3)$. These three flavours are referred to as {}``up''($u$),
{}``down''($d$) and {}``strange''($s$), and they have electric
charges $(\frac{2}{3},-\frac{1}{3},-\frac{1}{3})$ respectively. Consideration
of $SU_{f}(3)$ as an approximate symmetry of the strong interactions
leads to the above mentioned classification of hadrons. The approximate
nature of $SU_{f}(3)$ is due to the fact that the three quark flavours
have different masses and electric charges. In the Non-relativistic
Quark Model, the intrinsic angular momenta associated with hadrons
are obtained by combining $SU_{\mathrm{spin}}(2)$ with $SU_{f}(3)$.
This leads to a reasonably successful $SU(6)$ classification scheme.
The orbital excitations are then included by classifying hadrons under
$SU(6)\otimes O_{\mathrm{space}}(3)$. 
\item Absence of quarks among the observed hadrons requires them to exist
only as constituents of bound states. The quarks are assigned colour
charge, and the strong interactions bind them into colour neutral
hadrons. This is the phenomenon of confinement. In the Non-relativistic
Quark Model, a meson is a bound state of a quark and an antiquark,
while a baryon is a bound state of three quarks. 
\item Mesons are bosons, while baryons are fermions. This requires quarks
to be fermions, and baryon wavefunctions to be fully antisymmetric
when all their degrees of freedom are taken into account. In the Non-relativistic
Quark Model, the lowest lying baryons are in zero relative angular
momentum states, and their wavefunctions are symmetric in spin-flavour
degrees of freedom. For example, the wavefunction of the $\Delta ^{++}\ob{\frac{3}{2}^{+}}$
is $|u_{\uparrow }u_{\uparrow }u_{\uparrow }\rangle $, where the
up-arrow ($\uparrow $) denotes the spin $S_{z}=\frac{1}{2}$ for
each quark. These wavefunctions are reconciled with Fermi statistics
by postulating a new internal quantum number for quarks, i.e. colour.
If each quark has three, otherwise indistinguishable, colour states,
which transform according to the fundamental representation of $SU_{c}(3)$
(subscript $c$ stands for colour), then Fermi statistics is saved
by using a totally antisymmetric colour wavefunction for baryons. 
\item This assignment makes strong interactions invariant under global $SU_{c}(3)$
transformations. The states may then be classified according to various
irreducible representations of $SU_{c}(3)$. Using the fact that physical
hadrons are colour neutral, i.e. singlets under $SU_{c}(3)$, we can
understand why only $q\overline{q}$ and $qqq$ states (and not $q$,
$qq$, $qqqq$ etc.) exist in nature---the singlet representation
appears only in the $\underline{3}\otimes \underline{3}^{\star }$
and $\underline{3}\otimes \underline{3}\otimes \underline{3}$ products. 
\item Data from deep inelastic $e-p$ scattering (which involves a large
momentum transfer to the proton via the exchange of a virtual photon)
suggest that the proton can be understood as a bound state of constituents
called {}``partons'', which interact very weakly with each other
at high energies. At low energies, the interactions among the partons
becomes strong and they are strongly bound to each other. These ideas
form the basis of Feynman's Parton Model. Non-abelian gauge theories
can provide a field-theoretic understanding for Parton Model, because
it has been shown that only non-abelian gauge theories are asymptotically
free, i.e. their interactions vanish as energy $E\rightarrow \infty $
but become strong at a scale $E\sim \Lambda _{\mathrm{QCD}}$ ($\Lambda _{\mathrm{QCD}}$
denotes the $O(\mathrm{GeV})$ confinement scale). 
\item Experimentally, there is no evidence for any flavour dependence of
strong interactions; it is found that flavour-dependent effects can
be explained by quark mass and electric charge differences. We have
already argued that strong interactions should have an exact global
$SU_{c}(3)$ symmetry. Converting this to a local symmetry, $SU_{c}(3)$
becomes the non-abelian gauge theory of strong interactions. 
\item The concept of three colour states for quarks is supported by at least
two other powerful experimental results.

\begin{enumerate}
\item The parton model cross-section calculation for the process $e^{+}+e^{-}\rightarrow \mathrm{hadrons}$:
At large energies ($E\gg \Lambda _{\mathrm{QCD}}$), the ratio \[
R=\frac{\sigma \ob{e^{+}e^{-}\rightarrow \mathrm{hadrons}}}{\sigma \ob{e^{+}e^{-}\rightarrow \mu ^{+}\mu ^{-}}}\]
 is predicted to be \begin{eqnarray*}
R=3\sum _{q}Q_{q}^{2} & = & 3\times \ob{\frac{4}{9}+\frac{1}{9}+\frac{1}{9}+\frac{4}{9}+\frac{1}{9}+\cdots }\\
 & = & \frac{11}{3}\; (\mathrm{including}\; \mathrm{quarks}\; \mathrm{up}\; \mathrm{to}\; b)
\end{eqnarray*}
 The experimental value of $R$ is in good agreement with this prediction
and in poor agreement with the colourless prediction $\frac{11}{9}$. 
\item The decay rate for $\pi ^{0}\rightarrow 2\gamma $: This is again
proportional to the number of colours, and the observed decay rate
agrees with three colour states for the quarks. 
\end{enumerate}
\end{enumerate}
All this analysis suggests that the theory of strong interactions
is a non-abelian gauge theory based on the gauge group $SU_{c}(3)$
with quarks in the fundamental representation. This theory is precisely
defined with a single gauge coupling and quark masses as its parameters.
For its confirmation, it must quantitatively explain quark confinement,
spontaneous breakdown of chiral symmetry (i.e. light pions), the hadron
spectrum, partial decay widths, cross-sections, form-factors, structure
functions, and so on.

\section{QCD}

QCD is a non-abelian gauge theory based on the gauge group $SU_{c}(3)$.
We assume that there are $N_{f}$ quark flavours (experimentally $N_{f}=6$).
Each quark of a specific flavour $n$ comes in $3$ colours, and transforms
according to the fundamental representation of $SU_{c}(3)$. The QCD
Lagrangian for $N_{f}$ flavours of quarks is given by \[
\mathcal{L}=-\frac{1}{4}F_{a}^{\mu \nu }(x)F_{\mu \nu }^{a}(x)+\sum _{n=1}^{N_{f}}\overline{\Psi }_{n}(x)\rb{i\gamma ^{\mu }D_{\mu }-m_{n}}\Psi (x).\]
 Various symmetries of the QCD follow from $\mathcal{L}$.

\begin{itemize}
\item Invariance under the $(3+1)$-dim Poincar\'{e} group, which consists
of time translations, space translations, spatial rotations, Lorentz
boosts, and their combinations. Conservation of energy, linear momentum
and angular momentum follow as a result of these space-time symmetries. 
\item Charge conjugation, Parity and Time-reversal invariance. 
\item $SU_{c}(3)$ colour gauge invariance. 
\item Invariance under $\Psi _{n}\rightarrow \exp \rb{i\theta _{n}}\Psi _{n}$
and $\overline{\Psi }_{n}\rightarrow \overline{\Psi }_{n}\exp \rb{-i\theta _{n}}$.
This corresponds to \[
\underbrace{U(1)\otimes U(1)\otimes \cdots \otimes U(1)}_{N_{f}\; \mathrm{times}}\equiv \rb{U(1)}^{N_{f}}\]
 global invariance, which implies conservation of the number of quarks
minus the number of antiquarks for every flavour. 
\item When all $m_{n}=0$, invariance under \[
\left.\begin{array}{ccc}
 \Psi _{n}\rightarrow \ob{\exp \rb{i\theta _{k}t_{k}+i\tilde{\theta }_{k}t_{k}\gamma _{5}}}_{nm}\Psi _{m} & , & \overline{\Psi }_{n}\rightarrow \overline{\Psi }_{m}\ob{\exp \rb{-i\theta _{k}t_{k}+i\tilde{\theta }_{k}t_{k}\gamma _{5}}}_{mn}\\
 \Psi _{n}\rightarrow \exp \rb{i\theta +i\tilde{\theta }\gamma _{5}}\Psi _{n} & , & \overline{\Psi }_{n}\rightarrow \overline{\Psi }_{n}\exp \rb{-i\theta +i\tilde{\theta }\gamma _{5}}\end{array}\right.,\]
 where $t_{k}$ are the $N_{f}^{2}-1$ matrices generating the Lie
algebra of $SU\ob{N_{f}}$ group, and $\theta _{k}$ and $\tilde{\theta }_{k}$
are arbitrary real parameters (see Appendix~A). Thus the global $\rb{U(1)}^{N_{f}}$
gets extended to $SU_{L}\ob{N_{f}}\otimes SU_{R}\ob{N_{f}}\otimes U_{V}(1)\otimes U_{A}(1)$
which is known as the the chiral group of symmetries. The $U_{V}(1)$
symmetry is exact and corresponds to the conservation of Baryon number.
The $U_{A}(1)$ symmetry is anomalous, i.e. although it is a symmetry
of the Lagrangian, it is not a symmetry of the quantum dynamics. Since
parity doubling is absent in the observed hadron spectrum, one expects
the $SU_{L}\ob{N_{f}}\otimes SU_{R}\ob{N_{f}}$ symmetry to be spontaneously
broken to its diagonal subgroup $SU_{V}\ob{N_{f}}$. A signature of
this spontaneous symmetry breaking is provided by the non-zero expectation
value of the chiral condensate, $\tb{\overline{\Psi }\Psi }\neq 0$,
and $N_{f}^{2}-1$ massless Goldstone bosons. (For $N_{f}=2$, the
triplet of pions are identified as the pseudo-Goldstone bosons). 
\item The axial ($A$) part of the chiral symmetry is broken explicitly
by the non-zero bare quark masses, while the vector ($V$) part is
explicitly broken due to the fact that the bare quark masses are unequal.
Thus the mass terms account for explicit flavour symmetry breaking
effects in QCD. 
\end{itemize}
An important aspect of $4$-dim QCD which arises as a result of the
non-abelian nature of the gauge group $SU_{c}(3)$ is asymptotic freedom.
The coupling constant goes to zero as the energy scale at which it
is defined is increased. Thus the behaviour of the Green's functions
when all the momenta are scaled up by a common factor is governed
by a theory where $g\rightarrow 0$. For processes which involve large
momentum transfers, perturbation theory in powers of $g$ can be reliably
used, and it is in this domain that QCD has been extensively tested.
Factorization of long distance physics (which is process independent)
from the short distance distance physics (which is process specific)
gives perturbative QCD a clear predictive power. Perturbation theory
also indicates that $g\rightarrow \infty $ at a low energy scale,
typically labeled $\Lambda _{\mathrm{QCD}}$. The strong coupling
region ($g$ very large) has been studied for QCD formulated on a
space-time lattice using high temperature expansions. Various non-perturbative
features of hadronic physics (e.g. confinement, chiral symmetry breaking
and quantum numbers of light hadrons) emerge naturally when $g$ is
large. To connect the two regions, QCD has to be analysed at intermediate
values of $g$. It has not been possible to do this analytically so
far. Numerical simulations have extensively explored this intervening
region, however, and they show that the region of large gauge coupling
is analytically connected to the region of small gauge coupling, without
any phase transition in between. This implies that the symmetries
of the theory are the same in both the strong coupling and the weak
coupling regions.

\section{Summary of the thesis}

There is little doubt that QCD is the theory behind hadronic physics.
Our main difficulty lies in performing quantitative calculations using
it. Such calculations are important because they test the success
of QCD in low energy domain. For example, the cross-section for a
process involving hadrons, in general, has a sizable non-perturbative
component related to hadronic substructure. If such processes are
to be used to identify physics beyond the Standard Model, then it
is important to have a good theoretical control over the non-perturbative
contribution.

The study of bound states in QCD is made difficult due to \cite{wilson2}

\begin{itemize}
\item the unlimited growth of the running coupling $g$ in the infrared
region, which invalidates use of perturbation theory; 
\item confinement, which requires potentials that diverge at long distances
as opposed to the Coulomb/Yukawa potentials used in perturbation theory; 
\item spontaneous chiral symmetry breaking, which cannot be demonstrated
using perturbation theory; and 
\item the non-perturbative structure of the QCD vacuum. 
\end{itemize}
Before one can quantitatively understand QCD, it is necessary to establish
a suitable framework for studying these non-perturbative effects.
A lattice formulation is a natural way to define QCD so that all the
non-perturbative physics is accessible. Much progress has been made
since QCD was formulated on a space-time lattice by Wilson \cite{wilson1}
in 1974. In this thesis, we study QCD formulated on a transverse lattice,
which combines techniques from light-front field theory and lattice
gauge theory. An approximation scheme based on $1/N_{c}$ expansion
and transverse strong gauge coupling expansion allows us to systematically
study this transverse lattice QCD.

The use of light-front framework in studying bound states has several
advantages over the equal-time framework. In this framework, the theory
is evolved along a light-like direction ($x^{+}$). The light-front
wavefunctions, which encode the bound state structure, are the eigenstates
of the light-front Hamiltonian $P^{-}$. Since the longitudinal and
transverse boosts are purely kinematical, it becomes easy to separate
the centre-of-mass motion from the internal motion. Furthermore one
can choose to work with internal variables which are manifestly boost
invariant. The computation of the light-front wavefunctions leads
to estimates of a wide range of observables. Implementing a cut-off
that removes degrees of freedom with exactly zero longitudinal momentum,
one can force the vacuum to be trivial. The bound state wavefunctions
can then be expanded in the light-front Fock space of its constituents.
In the large-$N_{c}$ limit, only the lowest Fock space amplitudes
survive. The non-dynamical longitudinal gauge field provides a linear
confining potential in the $x^{-}$-direction (see Appendix~B).

The lattice cut-off in the transverse directions provides a non-perturbative
gauge-invariant regularization, albeit at the cost of rotational invariance.
In the strong transverse gauge coupling limit, there is a linear confining
potential in the transverse directions. We expect QCD formulated with
two continuum (light-front) and two lattice (transverse) dimensions
to be closer to real continuum QCD, than the extensively studied version
with all the four dimensions on lattice. In the limit of large-$N_{c}$
and strong transverse gauge coupling, the dynamics of QCD along the
light-front and the transverse directions essentially factorizes,
and the theory can be solved in a closed form. This approach to studying
QCD was outlined by Patel \cite{patel}. Here we present the explicit
solution, and that is the original contribution of this thesis.

In Chapter 2, we formulate QCD in the large-$N_{c}$ and strong transverse
coupling limits, and then exactly integrate out all the gauge degrees
of freedom to obtain the generating functional for quark-antiquark
bilinears. We study the chiral properties of our limiting theory in
Chapter 3, for naive as well as Wilson fermions, and obtain a recursive
relation for the chiral condensate. In Chapter 4, we obtain the homogeneous
integral equation satisfied by the meson states of our theory. Comparison
of this equation with the corresponding one for the 't~Hooft model
allows us to infer many physical meson properties. Our results are
consistent with phenomenological expectations, and this is the first
time that such results have been obtained from $(3+1)$-dim QCD, with
only quark and gluon degrees of freedom and no ad hoc model assumptions.
Extraction of precise values requires numerical solution of the integral
equation; we have not yet carried that out and we describe our outlook
for further investigations at the end of Chapter 4.

Three appendices supplement our analysis. Our notation and conventions
are listed in Appendix A. The known results for mesons in the 't~Hooft
model and in strong coupling lattice QCD are rederived, in Appendix
B and in Appendix C respectively, using the same methodology as followed
in the thesis. That allows convenient comparison, as well as demonstrates
the advantage of our approach over these two well-studied approximations
to QCD.

\chapter{QCD at large $N_{c}$ and strong transverse gauge coupling}

\section{Large-$N_{c}$ limit}

QCD is a non-abelian gauge theory based on the gauge group $SU(3)$.
One would like to see all the qualitative and quantitative features
of hadronic physics (confinement, chiral symmetry breaking, the hadron
spectrum, form factors, partial decay widths, cross-sections, etc.)
emerge from QCD. For most of these properties, the low momentum structure
of Green's functions is important. But QCD is an asymptotically free
theory, with the running gauge coupling increasing as the scale at
which it is defined decreases. The low momentum behaviour of Green's
functions, therefore, is governed by a theory with a large coupling,
making ordinary perturbation theory a completely ineffective tool
of computation.

In such a situation, one has to look for alternative approximation
schemes. A promising approach is the $1/N_{c}$ expansion of a gauge
theory based on the $SU\ob{N_{c}}$ group. The subject of the large-$N_{c}$
limit of $SU\ob{N_{c}}$ gauge theories with quarks in the fundamental
representation was initiated by 't~Hooft \cite{thooft1}, wherein
he showed that $1/N_{c}$ can be treated as a small expansion parameter
in these theories. The usual perturbative expansion reveals very little
information about the spectrum of the theory, while the large-$N_{c}$
limit retains many important non-perturbative aspects of the theory.

Witten \cite{witten} showed that many qualitative features of hadronic
physics can be understood in the framework of the $1/N_{c}$ expansion.
In particular, if one assumes confinement, the large-$N_{c}$ limit
produces a weakly interacting theory of an infinite number of mesons
with coupling of the order of $1/N_{c}$. Baryons are solitons of
the theory, with mass proportional to $N_{c}$, and with size and
shape $N_{c}$-independent. Let us now have a quick look at some aspects
of the large-$N_{c}$ limit of QCD. This section is largely based
on reviews by Coleman \cite{coleman}, Witten \cite{witten}, Manohar
\cite{manohar} and Makeenko \cite{makeenko}, and follows the power-counting
conventions of Witten.

A sensible large-$N_{c}$ limit of any quantity in QCD is obtained
as follows:

\begin{itemize}
\item Generalize the colour gauge group from $SU(3)$ to $SU\ob{N_{c}}$, 
\item holding $g$ fixed, take large-$N_{c}$ limit of each Feynman diagram
contributing to that quantity, and sum the leading contributions to
all orders in $g$. 
\end{itemize}
The central idea of the large-$N_{c}$ limit of QCD is that gluons
have many more colour states than quarks ($N_{c}^{2}-1$ compared
to $N_{c}$), and so quark radiative corrections are suppressed relative
to gluonic radiative corrections when one holds $g$ fixed. Also planar
type of gluonic radiative corrections are more important (larger powers
of $N_{c}$) than non-planar type. Thus the gluons effectively provide
a mean field environment through which the quarks propagate.

In more concrete terms, the Lagrangian density for $SU\ob{N_{c}}$
gauge field theory with quarks in the fundamental representation is
given by \[
\mathcal{L}=-\frac{1}{4}F_{a}^{\mu \nu }F_{\mu \nu }^{a}+\overline{\Psi }_{i}\rb{i\gamma ^{\mu }D_{\mu }-m}_{ij}\Psi _{j}\; ,\]
 with \[
\ob{D^{\mu }\Psi }_{i}=\partial ^{\mu }\Psi _{i}-i\frac{g}{\sqrt{N_{c}}}\ob{t_{a}}_{ij}A_{a}^{\mu }\Psi _{j},\; \]
 \[
F_{a}^{\mu \nu }=\partial ^{\mu }A_{a}^{\nu }-\partial ^{\nu }A_{a}^{\mu }+\frac{g}{\sqrt{N_{c}}}f_{abc}A_{b}^{\mu }A_{c}^{\nu }\; ,\]
 and \[
\rb{t_{a},t_{b}}_{-}=if_{abc}t_{c}\; ,\; \mathrm{tr}\ob{t_{a}t_{b}}=\delta _{ab}\; .\]
 Repeated colour indices are summed here; see Appendix~A for notations
and conventions. We have left out gauge-fixing terms and renormalization
counter-terms. Inclusion of these terms is trivial and does not alter
the nature of qualitative results obtained in their absence. Holding
$g$ fixed and taking $N_{c}\rightarrow \infty $, one can show the
following:

\begin{itemize}
\item The leading diagrams in the large-$N_{c}$ limit are planar diagrams
with quarks (if any) on boundaries.

\begin{itemize}
\item If $G_{i}$ represent local gauge invariant operators (which cannot
be split further into smaller gauge invariant parts) made up exclusively
of gauge fields, then the $n$-point Green's function, the connected
$n$-point Green's function and the $n$-point 1PI vertex function
are of order $N_{c}^{2n}$, $N_{c}^{2}$ and $N_{c}^{2(1-n)}$ respectively,
\begin{eqnarray*}
\tb{G_{1}G_{2}\ldots G_{n}} & \propto  & N_{c}^{2n}\; ,\\
\tb{G_{1}G_{2}\ldots G_{n}}_{\mathrm{conn}} & \propto  & N_{c}^{2}\; ,\\
\Gamma _{\cb{G}}^{(n)}\ob{1,2,\ldots ,n} & \propto  & N_{c}^{2(1-n)}\; .
\end{eqnarray*}

\item If $B_{i}$ represent gauge invariant (local or non-local) quark bilinear
operators (which cannot be split further into smaller gauge invariant
parts), then the $m$-point Green's function, the connected $m$-point
Green's function and the $m$-point 1PI vertex function are of order
$N_{c}^{m}$, $N_{c}$ and $N_{c}^{1-m}$ respectively, \begin{eqnarray*}
\tb{B_{1}B_{2}\ldots B_{m}} & \propto  & N_{c}^{m}\; ,\\
\tb{B_{1}B_{2}\ldots B_{m}}_{\mathrm{conn}} & \propto  & N_{c}\; ,\\
\Gamma _{\cb{B}}^{(m)}\ob{1,2,\ldots ,m} & \propto  & N_{c}^{1-m}\; .
\end{eqnarray*}

\item In general, with $n$-gluonic and $m$-quark bilinear operators ($m>0$),
we have for the $(n+m)$-point Green's function, the connected $(n+m)$-point
Green's function and the $(n+m)$-point 1PI vertex function, \begin{eqnarray*}
\tb{G_{1}\ldots G_{n}B_{1}\ldots B_{m}} & \propto  & N_{c}^{2n+m}\; ,\\
\tb{G_{1}\ldots G_{n}B_{1}\ldots B_{m}}_{\mathrm{conn}} & \propto  & N_{c}\; ,\\
\Gamma _{\cb{G;B}}^{(n;m)}\ob{1,\ldots ,n;1,\ldots ,m} & \propto  & N_{c}^{1-2n-m}\; .
\end{eqnarray*}

\end{itemize}
\item The diagrams with minimum number of quark loops dominate; each internal
quark loop is suppressed by a factor of $1/N_{c}$. 
\item Expansion in powers of $1/N_{c}$ becomes a topological expansion
for the diagrams. Consider a general diagram contributing to the connected
correlation function with $n$-gluonic and $m$-quark bilinear operators.
Any such diagram can be associated with a $2$-dim surface by associating
colour index loops with faces of polygons, gluon propagators with
edges where polygons touch each other and vertices with points where
corners of different polygons fuse. The contribution of such a diagram
is $O\ob{N_{c}^{\chi }}$, where $\chi $ is the topologically invariant
Euler characteristic of the associated $2$-dim surface. If the $2$-dim
surface has $H$ handles and $L$ boundaries, then $\chi =2-2H-L$.
($L$ is the sum of the number of internal virtual quark loops and
the number of external boundaries in the diagram.) For a connected
$(n+m)$-point correlation function, a maximum of $m$ external boundaries
are allowed. The leading contribution arises from {}``planar'' diagrams
which have genus $H=0$, and contribution of diagrams with larger
genus $H$ or more boundaries $L$ is suppressed. 
\item Distinction between gauge groups $SU\ob{N_{c}}$ and $U\ob{N_{c}}$
is immaterial at the leading order; they differ from each other by
$O\ob{1/N_{c}^{2}}$ terms. 
\end{itemize}
It follows that in the large-$N_{c}$ limit, all gauge invariant correlation
factorize; the fluctuations are suppressed by powers of $1/N_{c}$
and vanish when $N_{c}=\infty $. This essentially means that the
functional integral for the theory gets saturated by contribution
from a single gauge orbit, known as the master orbit and represented
by a set of gauge equivalent vector potentials. If one knows the master
gauge field, then, in principle, one can compute the leading $N_{c}$
contribution to any quantity. In this sense, the large-$N_{c}$ limit
is like a classical limit of the theory.

Assuming that large-$N_{c}$ QCD is a confining theory, the following
qualitative picture of glueballs and mesons emerges \cite{witten}:

\begin{itemize}
\item At $N_{c}=\infty $, glueballs (mesons) are free, stable and non-interacting.
Glueball (meson) masses have smooth finite limits as $N_{c}\rightarrow \infty $,
and the number of glueball (meson) states is infinite. Specifically,
if $G$ and $B$ are gauge invariant field operators for a meson and
a glueball respectively, with appropriate quantum numbers, then \begin{eqnarray*}
\tb{\; 0\mid G\mid \mathrm{glueball}\; } & \propto  & N_{c}\; ,\\
\tb{\; 0\mid B\mid \mathrm{meson}\; } & \propto  & \sqrt{N_{c}}\; .
\end{eqnarray*}

\item With $\Gamma _{\cb{G}}^{(n)}\ob{1,2,\ldots n}\propto N_{c}^{2(1-n)}$
and $\tb{0\mid G\mid \mathrm{glueball}}\propto N_{c}$, it follows
that the amplitude for a process involving $n$ glueballs is proportional
to $N_{c}^{2-n}$. In particular, the amplitude for a glueball decaying
in to two other glueballs is $O\ob{1/N_{c}}$, and glueball-glueball
scattering amplitude is $O\ob{1/N_{c}^{2}}$. 
\item With $\Gamma _{\cb{B}}^{(m)}\ob{1,2,\ldots m}\propto N_{c}^{1-m}$
and $\tb{0\mid B\mid \mathrm{meson}}\propto \sqrt{N_{c}}$, it follows
that the amplitude for a process involving $m$ mesons is proportional
to $N_{c}^{1-m/2}$. In particular, the amplitude for a meson decaying
in to two mesons is $O\ob{1/\sqrt{N_{c}}}$, and meson-meson scattering
amplitude is $O\ob{1/N_{c}}$. 
\item More generally, the amplitude for a process involving $n$ glueballs
and $m$ mesons ($m>0$) is of order $O\ob{N_{c}^{1-n-m/2}}$. In
particular, the amplitude for a glueball to mix with a meson is $O\ob{1/\sqrt{N_{c}}}$,
the amplitude for a glueball to decay in to two mesons is $O\ob{1/N_{c}}$,
the amplitude for a meson to decay in to two glueballs is $O\ob{1/N_{c}^{3/2}}$,
and the glueball-meson scattering amplitude is $O\ob{1/N_{c}^{2}}$. 
\item Arbitrary $n$-point Green's functions and $n$-point scattering amplitudes
are given by sums of tree diagrams with effective local vertices and
propagators corresponding to exchange of physical hadrons (glueballs
and mesons). 
\end{itemize}
This qualitative picture agrees very well with observed phenomenology:

\begin{enumerate}
\item Suppression of the quark sea in hadronic physics and the fact that
mesons are approximately pure $\overline{q}q$ states: In large-$N_{c}$
limit quark loops are suppressed by factors of $1/N_{c}$, which is
a consequence of the fact that there are $O\ob{N_{c}^{2}}$ gluon
degrees of freedom, but only $O\ob{N_{c}}$ quark degrees of freedom.
Thus, the quark sea is absent at $N_{c}=\infty $. 
\item The absence, or at least suppression, of $\overline{q}\overline{q}qq$
exotics: Mesons are non-interacting at $N_{c}=\infty $, which means
that two mesons would not be bound together in to an exotic state.
Presence of exotics would show up as simple poles in the connected
Green's function $\tb{E(x)E(y)}_{\mathrm{conn}}$, where $E(x)=\overline{q_{i}}q_{i}\overline{q_{j}}q_{j}(x)$.
But to leading order in $1/N_{c}$, $\tb{E(x)E(y)}_{\mathrm{conn}}$
factorizes to $\tb{\overline{q}_{i}q_{i}(x)\overline{q}_{k}q_{k}(y)}_{\mathrm{conn}}^{2}$,
and we have freely propagating non-interacting mesons instead of exotics. 
\item Resonant two-particle final state domination (whenever kinematically
allowed) of multi-particle decays of unstable mesons: It is observed
that meson decays proceed mainly through resonant two-particle states.
For example, $B$($1237$ MeV) decays to $4\pi $, but mainly via
$B\rightarrow \omega \pi $, with subsequent $\omega \rightarrow 3\pi $.
(It is generally believed that the tendency for decays to be two-particle
dominated persists, even when the larger phase space of multi-particle
decays is accounted for.) The explanation for this feature from the
point of view of large-$N_{c}$ limit is that the decay $B\rightarrow \omega \pi $
is $O\ob{1/\sqrt{N_{c}}}$, while the direct, non-resonant decay $B\rightarrow \pi \pi \pi \pi $
is $O\ob{1/N_{c}^{3/2}}$. 
\item Zweig's rule: Constituent quark-antiquark production is suppressed
compared to rearrangement of existing constituents inside hadrons.
For example, the decay $\phi (\overline{s}s)\rightarrow K^{+}(\overline{s}u)+K^{-}(\overline{u}s)$
has considerably less phase space compared to $\phi (\overline{s}s)\rightarrow \pi ^{+}(\overline{d}u)+\pi ^{0}(\overline{u}u-\overline{d}d)+\pi ^{-}(\overline{u}d)$.
Nevertheless, $\phi \rightarrow K^{+}K^{-}$ is the dominant decay
mode. In large-$N_{c}$ limit, creation/annihilation of every quark-antiquark
pair gives a suppression factor of $1/N_{c}$. In $\phi $-decay,
there is an extra $1/\sqrt{N_{c}}$ suppression factor due to the
fact that $\phi \rightarrow \pi ^{+}\pi ^{0}\pi ^{-}$ is a three-particle
decay while $\phi \rightarrow K^{+}K^{-}$ is a two-particle decay.
Overall, the $\pi ^{+}\pi ^{0}\pi ^{-}$ decay mode is suppressed
relative to the $K^{+}K^{-}$ decay mode by $1/N_{c}^{3/2}$. 
\item Meson multiplet structure as nonets of flavour $SU(3)$: If $u$,
$d$ and $s$ quark masses were equal, then there would be singlet-octet
degeneracy for mesons to leading order in $1/N_{c}$. This is because
the diagrams that split singlets from octets involve $\overline{q}q$
annihilation, and are of order $1/N_{c}$. 
\end{enumerate}
If one restricts QCD to two dimensions (see Appendix~B), it is possible
to study various dynamical features of the theory analytically (many
of these features continue to hold good in four dimensions). But it
is only the $N_{c}=\infty $ limit which has been solved in a closed
form. 't~Hooft \cite{thooft2} obtained an integral equation for
the meson wavefunction in 2-dim $N_{c}=\infty $ QCD, with eigenstates
corresponding to an infinite number of bound states. Subsequently,
Callan, Coote and Gross \cite{ccg}, Einhorn \cite{einhorn1}, Brower
et al. \cite{brower}, Cardy \cite{cardy}, Einhorn et al. \cite{einhorn2},
and Shei and Tsao \cite{shei} further explored questions relating
to quark confinement, hadron states, scattering amplitudes and high
energy behaviour. The key features of this solution are:

\begin{itemize}
\item Choosing the light-front gauge $A^{+}=A^{0}+A^{1}=0$ eliminates the
non-linear self-coupling of the gauge fields (gluon trilinear and
quartic terms). This gauge is also free of ghosts. 
\item The dynamics of gauge fields (terms involving $A^{-}$) can be replaced
by an instantaneous linearly confining potential between quark bilinears. 
\item The confining potential is defined using a principal value prescription,
which keeps its Fourier transform infrared finite. 
\item To study mesons in the light-front gauge, one needs to sum only the
planar {}``rainbow'' and {}``ladder'' diagrams. 
\item The choice of light-front coordinates together with the light-front
gauge simplifies the Lorentz index structure of the theory, and eliminates
the non-dynamical components in terms of the dynamical ones. 
\item The mass term in the quark propagator gets additively renormalized,
while the gluon propagator and the quark-gluon vertex receive no corrections. 
\item The quark-antiquark scattering amplitude shows that there are no continuum
states in the spectrum; only discrete meson bound states exist. 
\item The meson spectrum (masses and wavefunctions) is determined by the
eigenvalues of an integral equation that characterizes the singularities
of the quark-antiquark scattering amplitude. It lies on an approximately
linear trajectory in the $M^{2}-n$ plane, where $n$ is the radial
excitation quantum number. 
\item Spin does not exist in two dimensions, but parity is a good symmetry.
The lightest meson has negative parity, and the subsequent meson states
alternate in parity as a function of $n$. 
\item In the large-$N_{c}$ limit, there is a non-zero chiral condensate,
$\tb{\overline{\Psi }\Psi }\neq 0$, even in two dimensions. 
\end{itemize}
Many of these features are explicitly described in Appendix~B. Thus
$(1+1)$-dim large-$N_{c}$ QCD displays confinement and chiral symmetry
breaking as anticipated in $(3+1)$-dim QCD, even though the origins
of these features are quite different in the two theories. The ability
to solve $(1+1)$-dim large-$N_{c}$ QCD proves extremely useful,
when attempting to solve $(3+1)$-dim QCD on a transverse lattice.

\section{Strong coupling limit}

QCD is an asymptotically free theory, so its running gauge coupling
grows as its cut-off is lowered. Perturbative renormalization group
evolution shows that the gauge coupling depends logarithmically on
the cut-off, and diverges at a scale $\Lambda _{\mathrm{QCD}}=O(\mathrm{GeV})$.
Strong coupling expansion, i.e. an expansion in inverse powers of
the gauge coupling in the region where the cut-off is $O(\Lambda _{\mathrm{QCD}})$,
is yet another approximation scheme which elucidates some of the observed
features of hadronic physics. Strong coupling methods are usually
employed in the context of gauge theories formulated on a lattice.
In the strong coupling region, one can show that quarks are confined,
chiral symmetry is spontaneously broken, the lightest hadrons have
the correct quantum numbers, and so on. These features are expected
to survive in the weak coupling region (i.e. as one proceeds toward
the continuum limit by increasing the cut-off), since there is no
phase transition separating the two regions.

Let us illustrate the idea behind the strong coupling expansion in
the context of pure gauge theory on a Euclidean $4$-dim hypercubic
lattice with spacing $a$ (see for instance \cite{montvay-munster}).
The fundamental variables of this theory are the parallel transporters
$U\ob{x,\mu }$, associated with the oriented link from site $x$
to site $x+a\mu $ (see Appendix~A for details of the notation).
The Wilson action for this theory is defined in terms of the oriented
plaquette variables $U_{x;\mu \nu }$ as \[
S_{W}=-\frac{N_{c}}{2g^{2}}\sum _{x;\mu \nu }\mathrm{Re}\; \mathrm{tr}\rb{U_{x;\mu \nu }-1},\]
 where the plaquette variables are defined in terms of the parallel
transporters as \[
U_{x;\mu \nu }=U^{\dagger }(x,\mu )U^{\dagger }(x+a\mu ,\nu )U(x+a\nu ,\mu )U\ob{x,\nu }\]
 and $g\equiv g(a)$ is the lattice gauge coupling constant. The action
$S_{W}$ is often abbreviated as \[
S_{W}=-\frac{N_{c}}{2g^{2}}\sum _{p}\mathrm{Re}\; \mathrm{tr}\rb{U_{p}-1},\]
 where the sum includes every plaquette with both orientations. This
is the simplest gauge invariant lattice action which reduces to the
Yang-Mills action in the continuum limit $a\rightarrow 0$.

To demonstrate confinement, one needs to show that it requires infinite
energy to separate a quark-antiquark pair by an infinite distance.
The expectation value of the Wilson loop, $W\ob{\mathcal{C}}=\tb{\mathrm{tr}U\ob{\mathcal{C}}}$,
probes the interaction potential between a static quark-antiquark
pair. Consider $\mathcal{C}$ to be a large flat rectangular loop
with sides $Ra$ and $Ta$. In a linearly confining theory, $W\ob{\mathcal{C}}$
obeys the area law \[
W\ob{\mathcal{C}}\simeq \exp \ob{-\sigma RTa^{2}},\]
 with a non-zero string tension $\sigma $. If the theory does not
confine, then the Wilson loop would follow a perimeter law \[
W\ob{\mathcal{C}}\simeq \exp \rb{-\mu \ob{R+T}a}.\]

At the leading order in the strong coupling expansion, one can easily
show that Wilson loops obey area law as follows. The expectation value
of the Wilson loop is \[
W\ob{\mathcal{C}}=\frac{1}{Z'}\int \rb{\prod _{x,\mu }dU(x,\mu )}\mathrm{tr}\ob{\mathrm{P}\prod _{l\in \mathcal{C}}U(l)}\exp \rb{\frac{N_{c}}{2g^{2}}\sum _{p}\mathrm{tr}U_{p}},\]
 where $\mathrm{P}$ denotes the orientation of loop $C$ and \[
Z'=\int \rb{\prod _{x,\mu }dU(x,\mu )}\exp \rb{\frac{N_{c}}{2g^{2}}\sum _{p}\mathrm{tr}U_{p}}.\]
 Expanding the exponential in a Taylor series, we get \[
\exp \rb{\frac{N_{c}}{2g^{2}}\sum _{p}\mathrm{tr}U_{p}}=\sum _{m_{1},m_{2},\ldots }\frac{\ob{N_{c}/2g^{2}}^{m_{1}+m_{2}+\cdots }}{m_{1}!m_{2}!\cdots }\ob{\mathrm{tr}U_{p_{1}}}^{m_{1}}\ob{\mathrm{tr}U_{p_{2}}}^{m_{2}}\cdots .\]
 The group integration over $\rb{dU(x,\mu )}$ produces a non-zero
result only when all the factors of $U\ob{x,\mu }$ for that link
combine into a gauge singlet. The simplest instance is when one factor
of $U$ is combined with another factor of $U^{\dagger }$, \[
\int dU=1\; ,\qquad \int dU\; U_{ij}U_{kl}^{\dagger }=\frac{1}{N_{c}}\delta _{jk}\delta _{il}\; .\]
 The leading contribution to $W\ob{\mathcal{C}}$ thus arises when
the smallest number of plaquettes (with matching orientations) tile
the loop $\mathcal{C}$. On explicitly carrying out the integration
over the link variables, we get \[
W\ob{\mathcal{C}}=N_{c}\ob{\frac{1}{2g^{2}}}^{RT}.\]
 This shows that at strong coupling Wilson loops obey area law with
string tension \[
\sigma =\frac{\ln \ob{2g^{2}}}{a^{2}}\; ,\qquad \mathrm{for}\; N_{c}\geq 3\; .\]
 For $N_{c}=2$, the orientation of the plaquette does not matter
in group integration, and so each plaquette is associated with a factor
of $1/g^{2}$. This yields the string tension \[
\sigma =\frac{\ln \ob{g^{2}}}{a^{2}}\; ,\qquad \mathrm{for}\; N_{c}=2\; .\]
 Subleading corrections to this strong coupling limit are conveniently
obtained by expanding the exponential of the gauge action in terms
of the characters of the gauge group. Such character expansions effectively
sum a subset of graphs, and considerably simplify the computation.

In contrast to the weak coupling expansions which are asymptotic in
nature, strong coupling expansions have a non-zero radius of convergence.
Within its domain of convergence, one can rigorously show:

\begin{itemize}
\item Static quark confinement: Wilson loops in the fundamental representation
obey area law at strong coupling. 
\item Existence of a mass gap: The plaquette-plaquette correlation function
decays exponentially at strong coupling. 
\item Possible phase transitions: Power series expansions (in inverse powers
of the lattice gauge coupling) for quantities like internal energy,
mass gap, string tension etc. can be used to locate phase transitions. 
\end{itemize}
When dynamical quarks are added to the gauge theory, calculations
get technically more involved. Firstly, it becomes difficult to carry
out strong coupling expansions to high orders. Secondly, phenomena
associated with chiral symmetry breaking can only be seen if the quark
part of the theory is taken into account exactly. This is due to the
fact that chiral symmetry breaking produces pions as pseudo-Goldstone
bosons, which are very light for small bare quark masses and hence
give rise to long distance correlations.

The results for the meson spectrum, at the leading order in strong
coupling expansion, are described in Appendix~C. They correspond
to propagation of quark-antiquark pairs without any separation at
all, and show that there is only one meson state for every spin-parity
quantum number (i.e. all radial excitations are pushed away to infinite
energy).

\section{Combined large-$N_{c}$ and strong transverse \protect \protect \\
 gauge coupling limits}

We argued in Section 2.1 that large-$N_{c}$ QCD in a good phenomenological
approximation to many aspects of hadronic physics. This was deduced
by power counting, combining the assumption of colour confinement
for large-$N_{c}$ QCD with the fact that at large~$N_{c}$ only
planar diagrams with quarks on boundaries survive. In particular,
to leading order in $1/N_{c}$, meson scattering amplitudes are given
by sums of tree diagrams with exchange of physical mesons. The sum
is always an infinite sum, because there are always an infinite number
of mesons that could be exchanged in any given channel. This feature
has a strong resemblance to the successful {}``Regge phenomenology''
description of strong interactions. In Regge phenomenology, strong
interactions are interpreted as an infinite sum of hadron-exchange
tree diagrams, with the infinite number of hadrons lying on approximately
linear trajectories in the $M^{2}-J$ plane. Thus it seems likely
that the large-$N_{c}$ limit of QCD should be an important cornerstone
in the eventual derivation of Regge phenomenology \cite{witten}.
It is also believed that the straightness of Regge trajectories is
related, by analyticity, to the narrowness of hadron resonances. If
there is any approximation to QCD in which the trajectories are linear,
this must be an approximation in which the resonances are narrow.
Again it is an attractive feature of large-$N_{c}$ expansion that
the hadron resonances are narrow, with widths $O\ob{1/N_{c}}$. This
offers hope that at $N_{c}=\infty $ the Regge trajectories are linear.

With the above in mind, the problem that needs to be addressed is
how to sum the planar diagrams of large-$N_{c}$ QCD. In order to
tackle a quantum field theory, with its continuously infinite number
of degrees of freedom, one has to impose kinematical cut-offs (i.e.
a suitable regularization procedure). The dynamics of the degrees
of freedom eliminated by the cut-off is accounted for by

\begin{itemize}
\item renormalization of the existing couplings in the theory, and 
\item introduction of new counter-terms with cut-off dependent couplings,
that are consistent with the symmetries of the theory left unbroken
by the selected cut-off. 
\end{itemize}
The renormalized couplings and the induced counter-terms remove the
cut-off dependence from physical observables and help restore continuum
physics. The number and symmetries of the counter-terms depend on
the symmetry properties of the cut-off. For example, in the weak coupling
perturbative treatment of gauge theories, if one adopts a cut-off
which maintains manifest residual gauge invariance (some gauge choice
is essential to make the perturbative scheme well defined; usually
a Lorentz covariant gauge is chosen) and Lorentz invariance at all
intermediate stages of the calculation, then the counter-terms also
respect the very same symmetries.

In a confining theory such as QCD, gauge fields have large variations,
and a non-perturbative cut-off should be such that it is manifestly
gauge invariant. This criterion in the choice of cut-off is far more
important than the criterion of manifest Lorentz invariance, because
such a starting point is closer to the real world of strong interaction
physics. (Of course, quantitative description of strong interaction
physics will be attained only after taking in to account all the cut-off
dependent counter-terms.) Such a cut-off is provided by formulating
the gauge theory on a lattice. The discussion in Section 2.2 shows
that for very coarse lattice spacing, the gauge theory produces linear
confinement, i.e. large Wilson loops obey the area law. Numerical
simulations for QCD, at different values of the lattice spacing, show
that this behaviour persists all the way to very fine lattices with
no phase transitions in the intervening region.

An extremely coarse $4$-dim lattice cut-off, however, has the disadvantage
of drastically breaking Poincar\'{e} invariance (actually $4$-dim
Euclidean invariance) all the way down to $4$-dim hypercubic group.
Though confinement is manifest, glueballs become infinitely massive
and are pushed out of the spectrum. In the hadronic sector, only the
lowest lying states survive in each quantum number sector. This is
explicitly demonstrated in Appendix~C, which analyses $SU\ob{N_{c}\rightarrow \infty }$
lattice gauge theory at strong coupling with naive fermions in the
fundamental representation. It has not been possible to perform analytical
calculations on less coarse lattices; only numerical results have
been obtained using powerful computers. Can we do something better
without spoiling the analytical simplicity of working with a coarse
lattice, and yet retain the higher energy excitations? Appendix~B
demonstrates that the 't~Hooft model shares two key phenomenological
features of strong interaction physics, namely, confinement and chiral
symmetry breaking. Is there any way of incorporating these features
in a systematic framework when studying $(3+1)$-dim QCD? The answer
to these two questions leads us to formulating QCD on a transverse
lattice.

To explicitly incorporate the 't~Hooft model in the solution to large-$N_{c}$
$(3+1)$-dim QCD, one must keep the temporal and one spatial dimension
continuous. Crucial to the explicit solution of the 't~Hooft model
is the choice of the light-front $x^{+}=\ob{x^{0}+x^{1}}/\sqrt{2}=0$
as the quantization surface, and $P^{-}=\ob{P^{0}-P^{1}}/\sqrt{2}$
as the `time' ($x^{+}$) evolution operator. Since $P^{0}\geq \left|P^{1}\right|$,
it follows that $P^{\pm }\geq 0$. On the light-front, the `energy'
dispersion relation for a free particle of mass $m$ is given by \[
p^{-}=\frac{p_{\bot }^{2}+m^{2}}{2p^{+}}\; ,\]
 where $p^{+}$ is the longitudinal momentum and $p_{\bot }$ is the
transverse momentum. Two important observations follow from this dispersion
relation:

\begin{itemize}
\item $p^{+}=0$ states have infinite energy unless $p_{\bot }=0$ and $m=0$.
These `infrared' divergences associated with longitudinal zero modes
are special to field theories analysed on the light-front. These are
separate from and in addition to the infrared problems encountered
in a covariant formulation. The infrared divergences of the latter
type show up in the light-front formulation as divergences associated
with small $p_{\bot }$ states. 
\item There are ultraviolet divergences associated with large $p_{\bot }$
states. 
\end{itemize}
One needs to regulate these divergences before any concrete calculation
is attempted. The large $p_{\bot }$ divergences can be regulated
by imposing a lattice cut-off $a_{\bot }$ in the transverse dimensions.
We will regulate the longitudinal infrared divergences by explicitly
eliminating the longitudinal zero mode using the principal value prescription.

There are distinct advantages associated with this choice of cut-offs.

\begin{itemize}
\item The total longitudinal momentum $P^{+}$ of a state is just the sum
of the longitudinal momenta of its constituents, $P^{+}=\sum _{i}k_{i}^{+}$.
It follows that the vacuum of the cut-off theory (for which $P^{+}=0$)
contains no constituents, i.e. the vacuum is trivial. 
\item In the $A^{+}=0$ gauge, $A^{-}$ is non-dynamical and plays the role
of providing a linear confining interaction between quark-antiquark
pairs in the longitudinal ($x^{-}$) direction. The infrared singularity
of this confining potential is regularized by the principal value
prescription. 
\item After fixing the $A^{+}=0$ gauge, one is still left with residual
gauge degrees of freedom corresponding to $x^{-}$-independent gauge
transformations. These can be manifestly maintained on a transverse
lattice. 
\item Working with a very coarse transverse lattice makes confinement in
the transverse directions automatic. Wilson loops in the transverse
plane then obey area law. 
\end{itemize}
In our study of transverse lattice QCD, we restrict ourselves to the
large-$N_{c}$ and strong transverse gauge coupling limits \cite{patel}.
This choice is dictated by the fact that only in these limits we can
analytically solve the theory in a closed form. Consequently, our
approach has following limitations:

\begin{itemize}
\item Although linear confinement is built in to the theory, the dynamical
mechanism behind it is different in longitudinal and transverse directions.
Confinement arises in the longitudinal direction because the Coulomb
potential in $(1+1)$-dim is linear for any value of the gauge coupling,
while it arises in the transverse direction because of the strong
gauge coupling (i.e. completely disordered gluon fields). Thus perfect
rotational symmetry is not expected in our results. 
\item Gluons have no transverse propagating degrees of freedom, so they
can be completely integrated out of the theory. The glueballs are
infinitely massive, just as in $4$-dim strong coupling lattice QCD. 
\item The spectrum consists exclusively of towers of meson states, and the
large-$N_{c}$ limit brings out the constituent picture of mesons
in an explicit way. But the large-$N_{c}$ limit also makes the baryons
infinitely heavy. Although the baryons can be brought back in to the
theory as semi-classical soliton states, that requires another set
of field theoretical techniques. The case of baryons is not dealt
with in this thesis. 
\item Renormalization group scaling does not hold in the strong coupling
lattice theory, and hence our results for dimensionless quantities
are not expected to be the same as in the continuum theory. The scaling
behaviour can be improved by calculating higher order terms in the
strong coupling expansion or by adding counter-terms to the simplest
lattice action. We have not done that, and so our results can only
be viewed as a model of real QCD. We hope that by having two dimensions
already in the continuum, our results would not be too far off the
real world, and our model would have practical use. 
\item With the principal value prescription, the infrared cut-off introduced
to eliminate the longitudinal momentum zero mode disappears from the
final results. On the other hand, the transverse lattice ultraviolet
cut-off that eliminates the transverse momentum modes with $\left|p_{\bot }\right|>\pi /a_{\bot }$
explicitly appears in the final results. So our results can be interpreted
only in the sense of an effective theory, i.e. valid at a scale below
the cut-off scale where the contribution of the neglected counter-terms
is suppressed. 
\item Dynamical fermions on the lattice have the well-known conflict between
chiral symmetry and species doubling. For computational convenience,
we have studied fermion actions with only the nearest-neighbour hopping
term. The two popular choices in this case are:

\begin{itemize}
\item Naive fermions: These retain exact chiral symmetry, but lead to multiple
species of fermions---one for every corner of the Brillouin zone.
Exact spin-diagonalization converts naive fermions to staggered fermions
and reduces the number of fermion species, but that cannot eliminate
the doublers completely. In the large-$N_{c}$ limit, however, fermion
doubling is not a serious problem because the fermion determinant
does not contribute. 
\item Wilson fermions: These avoid fermion doubling, but explicitly break
chiral symmetry. The {}``chiral limit'' is defined as the location
where pions become massless, but even at this location pions do not
have all the properties of Goldstone bosons. 
\end{itemize}
\end{itemize}
Despite these limitations, our approach is tailor-made for the study
of deep inelastic hadronic physics. Experimental results for deep
inelastic structure functions are expressed in terms of light-front
variables, and their scaling behaviour implies that transverse dynamics
contributes only at the subleading order. So even if the transverse
directions are severely distorted using a coarse lattice, the leading
scaling behaviour can remain largely undisturbed. Note that for any
$2$-point or $3$-point Green's function (e.g. propagators and form
factors), it is always possible to choose a reference frame where
the transverse momenta of the external states vanish. We make this
choice in our analysis to express our results in the continuum language
in a straightforward manner.

\section{$SU\left(N_{c}\right)$ transverse lattice gauge theory}

Elements from both continuum gauge theory (Yang and Mills) and lattice
gauge theory (Wilson) are needed in constructing transverse lattice
gauge theory. Such a theory was first formulated by Bardeen and Pearson
\cite{bardeen-pearson}.

\subsection{Basic gauge invariant quantities}

We start by introducing pure $SU\ob{N_{c}}$ lattice gauge theory
on a transverse lattice in four dimensions. The underlying space-time
is the direct product of two-dimensional Minkowski space and planar
square lattice $\mathbb{Z}^{2}$ (with lattice spacing $a_{\bot }$).

Let $C_{y,y_{\bot };x,x_{\bot }}$ be a particular directed curve
from $\ob{x,x_{\bot }}$ to $\ob{y,y_{\bot }}$, and let the parallel
transporter for the gauge field along the curve $C$ be $U\ob{C_{y,y_{\bot };x,x_{\bot }}}\in SU\ob{N_{c}}$.
Under a gauge transformation, \begin{equation}
U\ob{C_{y,y_{\bot };x,x_{\bot }}}\rightarrow U'\ob{C_{y,y_{\bot };x,x_{\bot }}}=V\ob{y,y_{\bot }}U\ob{C_{y,y_{\bot };x,x_{\bot }}}V^{-1}\ob{x,x_{\bot }},\label{eq:24.1}\end{equation}
 where $V\ob{x,x_{\bot }},V\ob{y,y_{\bot }}$ are arbitrary $SU\ob{N_{c}}$
matrices. For an infinitesimal straight segment from $\ob{x,x_{\bot }}$
to $\ob{x+dx,x_{\bot }}$, \begin{equation}
U\ob{C_{x+dx,x_{\bot };x,x_{\bot }}}=1-A_{\mu }\ob{x,x_{\bot }}dx^{\mu }+\cdots ,\label{eq:24.2}\end{equation}
 where $A_{\mu }\ob{x,x_{\bot }}$ is the continuum gauge field, i.e.
an element of the Lie algebra of $SU\ob{N_{c}}$. The gauge transformation
rule for $A_{\mu }\ob{x,x_{\bot }}$ follows from the gauge transformation
rule for parallel transporters, eq.(\ref{eq:24.1}), \begin{eqnarray}
A\ob{x,x_{\bot }}\rightarrow A'\ob{x,x_{\bot }} & = & V\ob{x,x_{\bot }}A_{\mu }\ob{x,x_{\bot }}V^{-1}\ob{x,x_{\bot }}\nonumber \\
 & + & V\ob{x,x_{\bot }}\cdot \partial _{\mu }V^{-1}\ob{x,x_{\bot }}\label{eq:24.3}
\end{eqnarray}
 Given the gauge field $A_{\mu }\ob{x,x_{\bot }}$, the parallel transporters
can be constructed as, \begin{eqnarray}
U\ob{C} & = & \mathcal{P}\exp \cb{-\int _{0}^{1}A_{\mu }\ob{c(t),x_{\bot }}\frac{dc^{\mu }}{dt}dt}\nonumber \\
 & = & \mathcal{P}\exp \cb{-\int _{C}A_{\mu }\ob{x,x_{\bot }}dx^{\mu }},\label{eq:24.4}
\end{eqnarray}
 where $c^{\mu }(t)$ parametrizes the curve $C$ with $c^{\mu }(0)=x^{\mu }$
and $c^{\mu }(1)=y^{\mu }$, and the symbol $\mathcal{P}$ denotes
path ordering with respect to the variable $t$.

The simplest gauge invariant quantity one can construct using the
parallel transporters is the trace of the parallel transporter around
a closed curve, also known as the Wilson loop variable. For a closed
curve $C_{o}$ based at point $\ob{x,x_{\bot }}$, \begin{equation}
\mathrm{tr}\, U\ob{C_{o}}=\mathrm{tr}\left[\mathcal{P}\exp \cb{-\int _{C_{o}}A^{\mu }\ob{x,x_{\bot }}dx_{\mu }}\right].\label{eq:24.5}\end{equation}
 When $C_{o}$ is an infinitesimal rectangle based at $\ob{x,x_{\bot }}$,
with sides $dx^{\mu }$ and $dy^{\mu }$, \begin{eqnarray}
U\ob{C_{o}} & = & U\ob{C_{x,x_{\bot };x+dy.x_{\bot }}}U\ob{C_{x+dy,x_{\bot };x+dx+dy,x_{\bot }}}\nonumber \\
 &  & \cdot \; U\ob{C_{x+dx+dy,x_{\bot };x+dx,x_{\bot }}}U\ob{C_{x+dx,x_{\bot };xx_{\bot }}}\label{eq:24.6}\\
 & = & 1-\cb{\partial _{\mu }A_{\nu }\ob{x,x_{\bot }}-\partial _{\nu }A_{\mu }\ob{x,x_{\bot }}+\rb{A_{\mu }\ob{x,x_{\bot }},A_{\nu }\ob{x,x_{\bot }}}_{-}}dx^{\mu }dy^{\nu }+\cdots \nonumber 
\end{eqnarray}
 The quantity in the curly brackets in the last line of eq.(\ref{eq:24.6})
is the field strength of the continuum gauge field \begin{equation}
F_{\mu \nu }\ob{x,x_{\bot }}\equiv \partial _{\mu }A_{\nu }\ob{x,x_{\bot }}-\partial _{\nu }A_{\mu }\ob{x,x_{\bot }}+\rb{A_{\mu }\ob{x,x_{\bot }},A_{\nu }\ob{x,x_{\bot }}}_{-}.\label{eq:24.7}\end{equation}
 It follows from eq.(\ref{eq:24.1}) and eq.(\ref{eq:24.6}) that
under a gauge transformation the field strength transforms as \begin{equation}
F_{\mu \nu }\ob{x,x_{\bot }}\rightarrow F'_{\mu \nu }\ob{x,x_{\bot }}=V\ob{x,x_{\bot }}F_{\mu \nu }\ob{x,x_{\bot }}V^{-1}\ob{x,x_{\bot }}.\label{eq:24.8}\end{equation}
 So the simplest local gauge invariant quantity in the continuum is
\begin{equation}
\mathrm{tr}\rb{F^{\mu \nu }\ob{x,x_{\bot }}F^{\rho \sigma }\ob{x,x_{\bot }}}.\label{eq:24.9}\end{equation}

Let us now consider the finite straight segment from $\ob{x,x_{\bot }}$
to $\ob{x,x_{\bot }+a_{\bot }n}$, and the corresponding parallel
transporter $U\ob{C_{x,x_{\bot }+a_{\bot }n;x,x_{\bot }}}$. We abbreviate
this transverse link variable as $U\ob{x,x_{\bot },n}$. It gauge
transforms as \begin{equation}
U\ob{x,x_{\bot },n}\rightarrow U'\ob{x,x_{\bot },n}=V\ob{x,x_{\bot }+a_{\bot }n}U\ob{x,x_{\bot },n}V^{-1}\ob{x,x_{\bot }}.\label{eq:24.10}\end{equation}
 The parallel transporter around the smallest lattice loop (i.e. the
elementary plaquette), \begin{equation}
U_{mn}\ob{x,x_{\bot }}\equiv U^{\dagger }(x,x_{\bot },m)U^{\dagger }(x,x_{\bot }+a_{\bot }m,n)U(x,x_{\bot }+a_{\bot }n,m)U\ob{x,x_{\bot },n},\label{eq:24.11}\end{equation}
 gauge transforms as \begin{equation}
U_{mn}\ob{x,x_{\bot }}\rightarrow U'_{mn}\ob{x,x_{\bot }}=V\ob{x,x_{\bot }}U_{mn}\ob{x,x_{\bot }}V^{-1}\ob{x,x_{\bot }}.\label{eq:24.12}\end{equation}
 So the simplest gauge invariant quantity one can construct on a planar
square lattice is \begin{equation}
\mathrm{tr}\rb{U_{mn}\ob{x,x_{\bot }}}.\label{eq:24.13}\end{equation}

In a transverse lattice space-time, we also need to consider {}``mixed''
loops which traverse partly in the continuum and partly on the lattice.
For a mixed loop $C_{n}$ based at $\ob{x,x_{\bot }}$, with sides
$dx$ and $a_{\bot }n$, \begin{eqnarray}
U\ob{C_{m}} & = & U^{\dagger }\ob{x,x_{\bot },n}U\ob{C_{x,x_{\bot }+a_{\bot }n;x+dx,x_{\bot }+a_{\bot }n}}U\ob{x+dx,x_{\bot },n}U\ob{C_{x+dx,x_{\bot };x,x_{\bot }}}\nonumber \\
 & = & 1+U^{\dagger }\ob{x;x_{\bot },n}\left\{ \partial _{\rho }U\ob{x;x_{\bot },n}+A_{\rho }\ob{x,x_{\bot }+a_{\bot }n}U\ob{x,x_{\bot },n}\right.\label{eq:24.14}\\
 &  & \qquad \qquad \qquad \qquad \qquad \qquad \qquad \left.-U\ob{x,x_{\bot },n}A_{\rho }\ob{x,x_{\bot }}\right\} dx^{\rho }+\cdots \nonumber 
\end{eqnarray}
 The quantity in the curly brackets in the last line of eq.(\ref{eq:24.14})
is the mixed gauge covariant derivative of the transverse link variable
\begin{equation}
D_{\mu }U\ob{x,x_{\bot },n}=\partial _{\mu }U\ob{x,x_{\bot },n}+A_{\mu }\ob{x,x_{\bot }+a_{\bot }n}U\ob{x,x_{\bot },n}-U\ob{x;x_{\bot },n}A_{\mu }\ob{x,x_{\bot }}.\label{eq:24.15}\end{equation}
 From the gauge transformation properties of parallel transporters,
eq.(\ref{eq:24.3}) and eq.(\ref{eq:24.10}), one finds that \begin{equation}
D^{\rho }U\ob{x,x_{\bot },n}\rightarrow V\ob{x,x_{\bot }+a_{\bot }n}D^{\rho }U\ob{x,x_{\bot },n}V^{-1}\ob{x,x_{\bot }},\label{eq:24.16}\end{equation}
 and hence the simplest two-derivative gauge invariant quantity of
{}``mixed'' type is \begin{equation}
\mathrm{tr}\ob{\rb{D^{\mu }U\ob{x;x_{\bot },n}}^{\dagger }D^{\nu }U\ob{x;x_{\bot },n}}.\label{eq:24.17}\end{equation}

Now let us introduce fermions, in the fundamental representation of
$SU\ob{N_{c}}$, in to the theory. Under a gauge transformation $V\ob{x,x_{\bot }}$,
the fermion fields $\Psi $ and $\overline{\Psi }$ transform as \begin{eqnarray}
\Psi \ob{x,x_{\bot }} & \rightarrow  & \Psi '\ob{x,x_{\bot }}=V\ob{x,x_{\bot }}\Psi \ob{x,x_{\bot }}\nonumber \\
\overline{\Psi }\ob{x,x_{\bot }} & \rightarrow  & \overline{\Psi }'\ob{x,x_{\bot }}=\overline{\Psi }\ob{x,x_{\bot }}V^{-1}\ob{x,x_{\bot }}.\label{eq:24.18}
\end{eqnarray}
 The simplest gauge invariant fermion bilinear terms are therefore
of the form \begin{equation}
\overline{\Psi }\ob{y,y_{\bot }}U\ob{C_{y,y_{\bot };x,x_{\bot }}}\Psi \ob{x,x_{\bot }}.\label{eq:24.19}\end{equation}
 In a transverse lattice space-time, there are two distinct choices
for the curve $C$ in this term:

\begin{enumerate}
\item $C$ is an infinitesimal segment from $\ob{x-dx,x_{\bot }}$ to $\ob{x,x_{\bot }}$.
Expanding the fermion gauge invariant term, we get \begin{eqnarray}
 &  & \overline{\Psi }\ob{x,x_{\bot }}U\ob{C_{x,x_{\bot };x-dx,x_{\bot }}}\Psi \ob{x-dx,x_{\bot }}\nonumber \\
 & = & \overline{\Psi }\ob{x,x_{\bot }}\cb{1-A_{\mu }\ob{x,x_{\bot }}dx^{\mu }}\cb{\Psi \ob{x,x_{\bot }}-\partial _{\mu }\Psi \ob{x,x_{\bot }}dx^{\mu }}+\cdots \label{eq:24.20}\\
 & = & \overline{\Psi }\ob{x,x_{\bot }}\Psi \ob{x,x_{\bot }}-\overline{\Psi }\ob{x,x_{\bot }}\rb{\partial _{\mu }+A_{\mu }\ob{x,x_{\bot }}}\Psi \ob{x,x_{\bot }}dx^{\mu }+\cdots .\nonumber 
\end{eqnarray}
 The first term in last line of the above equation is the fermion
mass term. In the second term, the combination $D_{\mu }=\partial _{\mu }+A_{\mu }$
is the gauge covariant derivative operator for fermions. Gauge covariant
derivatives of the fermion field transform in a manner identical to
the fermion field, \begin{equation}
D_{\mu }\Psi \ob{x,x_{\bot }}\rightarrow D'_{\mu }\Psi '\ob{x,x_{\bot }}=V\ob{x,x_{\bot }}D_{\mu }\Psi \ob{x,x_{\bot }}.\label{eq:24.21}\end{equation}

\item $C$ is the finite straight segment from $\ob{x,x_{\bot }}$ to $\ob{x,x_{\bot }+a_{\bot }n}$.
The corresponding gauge invariant fermion bilinear term is the transverse
hopping term, \begin{equation}
\overline{\Psi }\ob{x,x_{\bot }+a_{\bot }n}U\ob{x,x_{\bot },n}\Psi \ob{x,x_{\bot }}.\label{eq:24.22}\end{equation}

\end{enumerate}
In summary, the maximally localized gauge invariant terms involving
the gauge fields $A_{\mu }\ob{x,x_{\bot }}$, $U\ob{x,x_{\bot },n}$
and the fermion fields $\Psi \ob{x,x_{\bot }}$, $\overline{\Psi }\ob{x,x_{\bot }}$
are:

\begin{itemize}
\item $\mathrm{tr}\ob{F^{\mu \nu }(x,x_{\bot })F^{\rho \sigma }(x,x_{\bot })}$
\item $\mathrm{tr}\ob{\rb{D_{\mu }U(x,x_{\bot },n)}^{\dagger }D_{\nu }U(x,x_{\bot },n)}$
\item $\mathrm{tr}\ob{U_{mn}(x,x_{\bot })}$
\item $\overline{\Psi }(x,x_{\bot })\Psi (x,x_{\bot })$
\item $\overline{\Psi }(x,x_{\bot })\rb{D^{\mu }\Psi (x,x_{\bot })}$
\item $\overline{\Psi }\ob{x,x_{\bot }+a_{\bot }n}U\ob{x,x_{\bot },n}\Psi \ob{x,x_{\bot }}$
\end{itemize}

\subsection{Transverse lattice QCD action}

Using the above terms, and demanding two-dimensional Poincare invariance
as well as planar square lattice symmetry, we can write down a general
action for the $SU\ob{N_{c}}$ transverse lattice gauge theory: \begin{eqnarray}
S & = & \frac{N_{c}a_{\bot }^{2}}{4g^{2}}\int d^{2}x\sum _{x_{\bot }}\mathrm{tr}\ob{F^{\mu \nu }\ob{x,x_{\bot }}F_{\mu \nu }\ob{x,x_{\bot }}}\nonumber \\
 & + & \frac{N_{c}}{4g^{2}}\lambda _{cl}\int d^{2}x\sum _{x_{\bot },n}\mathrm{tr}\ob{\rb{D_{\mu }U\ob{x,x_{\bot },n}}^{\dagger }D^{\mu }U\ob{x,x_{\bot },n}}\nonumber \\
 & + & \frac{N_{c}}{2g^{2}a_{\bot }^{2}}\lambda _{ll}\int d^{2}x\sum _{x_{\bot },m,n}\mathrm{Re}\; \mathrm{tr}\rb{U_{mn}\ob{x,x_{\bot }}-1}\label{eq:24.23}\\
 & + & a_{\bot }^{2}\int d^{2}x\sum _{x_{\bot }}\overline{\Psi }\ob{x,x_{\bot }}\rb{i\gamma ^{\mu }D_{\mu }-m}\Psi \ob{x,x_{\bot }}\nonumber \\
 & + & \frac{\kappa a_{\bot }}{2}\int d^{2}x\sum _{x_{\bot },n}\left[\overline{\Psi }\ob{x,x_{\bot }+a_{\bot }n}\ob{r-i\gamma ^{n}}U\ob{x,x_{\bot },n}\Psi \ob{x,x_{\bot }}\right.\nonumber \\
 &  & \qquad \qquad \qquad \left.+\overline{\Psi }\ob{x,x_{\bot }}\ob{r+i\gamma ^{n}}U^{\dagger }\ob{x,x_{\bot },n}\Psi \ob{x,x_{\bot }+a_{\bot }n}\right].\nonumber 
\end{eqnarray}
 Here we have adopted the following conventions:

\begin{itemize}
\item The normalizations for the gauge and the fermion fields are chosen
such that each term in the action is proportional to $N_{c}$. With
this choice, the large-$N_{c}$ limit amounts to finding the stationary
point of the action, while holding $g$ fixed \cite{coleman}. 
\item Anisotropy parameters $\lambda _{cl},\lambda _{ll},\kappa $ have
been introduced to take care of explicit breaking of rotational symmetry. 
\item The Wilson parameter $r$ has been introduced in the fermion transverse
hopping term. As $a_{\bot }\rightarrow 0$, a non-zero value for $r$
adds an irrelevant operator to the fermion Lagrangian, $a_{\bot }(D_{n}\overline{\Psi })(D^{n}\Psi )$.
That removes the fermion doublers without changing the continuum limit,
but breaks the chiral symmetry explicitly. As mentioned earlier, we
study two special cases:

\begin{itemize}
\item $r=0$: Transverse lattice QCD with naive fermions. 
\item $r=1$: Transverse lattice QCD with Wilson fermions. 
\end{itemize}
\end{itemize}
In order to reveal the connection of the above action with continuum
$4$-dim $SU\ob{N_{c}}$ gauge theory, we take the naive continuum
limit $a_{\bot }\rightarrow 0$. The transverse link variable can
be parametrized as \begin{equation}
U\ob{x,x_{\bot },n}=\exp \rb{-a_{\bot }A_{n}\ob{x,x_{\bot }}}=1-a_{\bot }A_{n}\ob{x,x_{\bot }}+\frac{a_{\bot }^{2}}{2}A_{n}^{2}\ob{x,x_{\bot }}+\cdots \label{eq:24.24}\end{equation}
 In the pure gauge field transverse lattice terms, using the forward
lattice derivative \begin{equation}
\Delta _{n}^{f}A_{m}\ob{x,x_{\bot }}\equiv \frac{A_{m}\ob{x,x_{\bot }+a_{\bot }n}-A_{m}\ob{x,x_{\bot }}}{a_{\bot }},\label{eq:24.25}\end{equation}
 and the Baker-Campbell-Hausdorff formula \begin{equation}
\exp (x)\exp (y)=\exp \ob{x+y+\frac{1}{2}[x,y]_{-}+\cdots },\label{eq:24.26}\end{equation}
 one finds that \begin{equation}
\begin{array}{c}
 D_{\mu }U\ob{x,x_{\bot },n}=-a_{\bot }\cb{F_{\mu n}\ob{x,x_{\bot }}}+O(a_{\bot }^{2})\\
 U_{mn}\ob{x,x_{\bot }}=\exp \cb{-a_{\bot }^{2}F_{mn}\ob{x,x_{\bot }}+O(a_{\bot }^{3})}\end{array},\label{eq:24.27}\end{equation}
 with\begin{equation}
\begin{array}{c}
 F_{\mu n}\ob{x,x_{\bot }}=\partial _{\mu }A_{n}\ob{x,x_{\bot }}-\Delta _{n}^{f}A_{\mu }\ob{x,x_{\bot }}+\left[A_{\mu }\ob{x,x_{\bot }},A_{n}\ob{x,x_{\bot }}\right]_{-}\\
 F_{mn}\ob{x,x_{\bot }}=\Delta _{m}^{f}A_{n}\ob{x,x_{\bot }}-\Delta _{n}^{f}A_{m}\ob{x,x_{\bot }}+\rb{A_{m}\ob{x,x_{\bot }},A_{n}\ob{x,x_{\bot }}}_{-}\end{array}.\label{eq:24.28}\end{equation}
 Thus the leading terms, as $a_{\bot }\rightarrow 0$, coincide with
the continuum Yang-Mills action, \begin{eqnarray}
 &  & \int d^{2}x\sum _{x_{\bot },n}\mathrm{tr}\ob{\rb{D_{\mu }U\ob{x,x_{\bot },n}}^{\dagger }D^{\mu }U\ob{x,,x_{\bot },n}}\nonumber \\
 &  & =\int d^{2}x\sum _{x_{\bot },n}\rb{a_{\bot }^{2}\mathrm{tr}\ob{F_{\mu n}\ob{x,x_{\bot }}F^{\mu n}\ob{x,x_{\bot }}}+O(a_{\bot }^{3})},\label{eq:24.29}\\
 &  & \int d^{2}x\sum _{x_{\bot },m,n}\mathrm{Re}\; \mathrm{tr}\rb{U_{mn}\ob{x,x_{\bot }}-1}\nonumber \\
 &  & =\int d^{2}x\sum _{x_{\bot },m,n}\rb{\frac{a_{\bot }^{4}}{2}\mathrm{tr}\ob{F^{mn}\ob{x,x_{\bot }}F_{mn}\ob{x,x_{\bot }}}+O(a_{\bot }^{5})}.\label{eq:24.30}
\end{eqnarray}
 In case of the fermionic transverse lattice terms, the $a_{\bot }\rightarrow 0$
limit yields, \begin{eqnarray}
 &  & \frac{1}{2}\int d^{2}x\sum _{x_{\bot },n}a_{\bot }\left[\overline{\Psi }\ob{x,x_{\bot }+a_{\bot }n}\ob{r-i\gamma ^{n}}U\ob{x,x_{\bot },n}\Psi \ob{x,x_{\bot }}\right.\nonumber \\
 &  & \qquad \qquad \qquad \left.+\overline{\Psi }\ob{x,x_{\bot }}\ob{r+i\gamma ^{n}}U^{\dagger }\ob{x,x_{\bot },n}\Psi \ob{x,x_{\bot }+a_{\bot }n}\right]\nonumber \\
 & = & \int d^{2}x\sum _{x_{\bot },n}a_{\bot }^{2}\left[\frac{i}{2}\cb{\overline{\Psi }\ob{x,x_{\bot }}\gamma ^{n}\Delta _{n}^{f}\Psi \ob{x,x_{\bot }}-\ob{\Delta _{n}^{f}\overline{\Psi }\ob{x,x_{\bot }}}\gamma ^{n}\Psi \ob{x,x_{\bot }}}\right.\nonumber \\
 &  & \qquad \qquad \qquad \left.+i\; \overline{\Psi }\ob{x,x_{\bot }}\gamma ^{n}A_{n}\ob{x,x_{\bot }}\Psi \ob{x,x_{\bot }}\right]\label{eq:24.31}\\
 & + & \int d^{2}x\sum _{x_{\bot },n}a_{\bot }^{3}\rb{\frac{r}{a_{\bot }^{2}}\overline{\Psi }\ob{x,x_{\bot }}\Psi \ob{x,x_{\bot }}-\frac{r}{2}\ob{\Delta _{n}^{f}\overline{\Psi }\ob{x,x_{\bot }}}\ob{\Delta _{n}^{f}\Psi \ob{x,x_{\bot }}}}+O\ob{a_{\bot }^{4}}.\nonumber 
\end{eqnarray}
 Thus the terms proportional to $r$ shift the fermion mass by $2\kappa r/a_{\bot }$,
and add an irrelevant operator to the fermion action, while the other
terms agree with the continuum Dirac action.

\subsection{Determination of parameters}

The preceding analysis shows that in the continuum limit, all the
anisotropy coefficients should tend to unity, \begin{equation}
a_{\bot }\rightarrow 0:\quad \lambda _{cl},\; \lambda _{ll},\kappa \rightarrow 1.\label{eq:24.32}\end{equation}
 The transverse lattice action then reduces to the conventional continuum
action

\begin{eqnarray}
S & = & S_{G}+S_{F}\nonumber \\
 & \rightarrow  & \frac{N_{c}}{4g^{2}}\int d^{2}x\int d^{2}x_{\bot }\left[F^{\mu \nu }(x,x_{\bot })F_{\mu \nu }(x,x_{\bot })+F^{\mu n}(x,x_{\bot })F_{\mu n}(x,x_{\bot })\right.\nonumber \\
 &  & \qquad \qquad \qquad \qquad \left.+F^{mn}(x,x_{\bot })F_{mn}(x,x_{\bot })\right]\label{eq:24.33}\\
 &  & +\int d^{2}x\int d^{2}x_{\bot }\; \overline{\Psi }\ob{x,x_{\bot }}\left[i\gamma ^{\mu }\left(\partial _{\mu }+A_{\mu }\ob{x,x_{\bot }}\right)\right.\nonumber \\
 &  & \qquad \qquad \qquad \qquad \qquad \quad \left.+i\gamma ^{n}\ob{\partial _{n}+A_{n}\ob{x,x_{\bot }}}-m\right]\Psi \ob{x,x_{\bot }}.\nonumber 
\end{eqnarray}
 In the quantum theory, the gauge coupling $g$ gets converted to
the QCD scale $\Lambda _{\mathrm{QCD}}$ by dimensional transmutation,
while the quark mass $m$ is not a physical observable. They have
to be determined using experimentally observed hadronic properties
as input.

When the transverse lattice cut-off is finite, quantum renormalization
effects make the anisotropy coefficients different from unity, i.e.
they become functions of $a_{\bot }$. For small values of $a_{\bot }$,
these coefficients can be computed using perturbation theory, and
they remain finite. In general, they must be determined non-perturbatively,
by demanding that physical observables satisfy rotational symmetry
as closely as possible. Convenient choices for enforcing rotational
symmetry are isotropy of the energy-momentum dispersion relation,
and degeneracy of multiple helicity states of hadrons with non-zero
angular momentum. $\Lambda _{\mathrm{QCD}}$ and quark masses have
to be determined from hadronic properties as usual, and so more physical
conditions have to be imposed to fix all the parameters of the theory.
For Wilson fermions in particular, $m$ contains a power-law divergence
as $a_{\bot }\rightarrow 0$, and therefore it always needs to be
fixed using a non-perturbative criterion. The conventional choice
is to define the {}``chiral limit'' for Wilson fermions as the value
of $m$ (not necessarily zero) where the pseudoscalar meson mass vanishes.

The strong transverse gauge coupling limit can be viewed as the extreme
case of anisotropic renormalization group evolution, where the transverse
lattice cut-off scale is lowered as far as possible. In this limit
$a_{\bot }$ is $O(\Lambda _{\mathrm{QCD}})$, and we expect the anisotropy
coefficients for the gauge field terms ($\lambda _{cl},\lambda _{ll}$)
to vanish. We also do not expect any phase transition to take place
during the renormalization group evolution from small $a_{\bot }$
to its largest value. These expectations are based on the non-perturbative
behaviour of $4$-dim lattice QCD, and are supported by leading order
perturbative calculations \cite{burgio-etal} that show logarithmic
behaviour for $\lambda _{cl},\lambda _{ll}$. Thus the action we have
studied in this thesis is \begin{eqnarray}
S_{\bot } & = & \frac{N_{c}a_{\bot }^{2}}{4g^{2}}\int d^{2}x\sum _{x_{\bot }}\mathrm{tr}\ob{F^{\mu \nu }\ob{x,x_{\bot }}F_{\mu \nu }\ob{x,x_{\bot }}}\nonumber \\
 & + & a_{\bot }^{2}\int d^{2}x\sum _{x_{\bot }}\overline{\Psi }\ob{x,x_{\bot }}\rb{i\gamma ^{\mu }D_{\mu }-m}\Psi \ob{x,x_{\bot }}\label{eq:24.34}\\
 & + & \frac{\kappa a_{\bot }}{2}\int d^{2}x\sum _{x_{\bot },n}\left[\overline{\Psi }\ob{x,x_{\bot }+a_{\bot }n}\ob{r-i\gamma ^{n}}U\ob{x,x_{\bot },n}\Psi \ob{x,x_{\bot }}\right.\nonumber \\
 &  & \qquad \qquad \qquad \left.+\overline{\Psi }\ob{x,x_{\bot }}\ob{r+i\gamma ^{n}}U^{\dagger }\ob{x,x_{\bot },n}\Psi \ob{x,x_{\bot }+a_{\bot }n}\right].\nonumber 
\end{eqnarray}
 Compared to the continuum action, eq.(\ref{eq:24.33}), this action
has only one more parameter, i.e. $\kappa $. Since the powers of
$a_{\bot }$ that accompany factors of $g^{2}$ and $\kappa $ are
obvious, in order to simplify expressions, we choose the units such
that $a_{\bot }=1$.

\section{The generating functional}

The hadronic physics is buried in various quark-antiquark correlations
functions. As explained in previous sections, large-$N_{c}$ and strong
transverse gauge coupling expansions combine efficiently to produce
reasonable mesonic structure in the leading order. This section is
devoted to computing the generating functional of quark-antiquark
correlation functions in these limits.

Since we are primarily interested in the quark-antiquark correlation
functions, we add a suitable source term to the action, eq.(\ref{eq:24.34}),
\begin{equation}
I=S+\int d^{2}xd^{2}y\sum _{x_{\bot },y_{\bot }}J_{ij}^{\alpha \beta }\ob{x,x_{\bot };y,y_{\bot }}\overline{\Psi }_{i}^{\alpha }\ob{x,x_{\bot }}\Psi _{j}^{\beta }\ob{y,y_{\bot }}.\label{eq:25.1}\end{equation}
 The generating functional, $Z\rb{J}$, is given by \begin{equation}
Z\rb{J}=\int \rb{DA}\rb{DU}\rb{D\Psi \cdot D\overline{\Psi }}\exp \rb{iI+\ob{\mathrm{gauge}\; \mathrm{terms}}},\label{eq:25.2}\end{equation}
 where {}``gauge terms'' stand for the gauge fixing terms as well
as the corresponding ghost determinant. The time-ordered quark-antiquark
correlation functions can be obtained by taking functional derivatives
of $W\rb{J}\equiv -i\ln Z\rb{J}$ with respect to $J$. For example,
\begin{equation}
\left\langle \mathrm{T}\ob{\overline{\Psi }_{i}^{\alpha }\left(x,x_{\bot }\right)\Psi _{j}^{\beta }\left(y,y_{\bot }\right)}\right\rangle =\left.\frac{\partial W\left[J\right]}{\partial J_{ij}^{\alpha \beta }\left(x,x_{\bot };y,y_{\bot }\right)}\right|_{J=0},\label{eq:25.3}\end{equation}
\begin{eqnarray}
 &  & \left\langle \mathrm{T}\ob{\overline{\Psi }_{i}^{\alpha }\left(x,x_{\bot }\right)\Psi _{j}^{\beta }\left(y,y_{\bot }\right)\overline{\Psi }_{k}^{\gamma }\left(z,z_{\bot }\right)\Psi _{l}^{\delta }\left(w,w_{\bot }\right)}\right\rangle \nonumber \\
 &  & -\left\langle \mathrm{T}\ob{\overline{\Psi }_{i}^{\alpha }\left(x,x_{\bot }\right)\Psi _{j}^{\beta }\left(y,y_{\bot }\right)}\right\rangle \left\langle \mathrm{T}\ob{\overline{\Psi }_{k}^{\gamma }\left(z,z_{\bot }\right)\Psi _{l}^{\delta }\left(w,w_{\bot }\right)}\right\rangle \label{eq:25.4}\\
 &  & =\left.-i\frac{\partial ^{2}W\left[J\right]}{\partial J_{ij}^{\alpha \beta }\left(x,x_{\bot };y,y_{\bot }\right)\partial J_{kl}^{\gamma \delta }\left(z,z_{\bot };w,w_{\bot }\right)}\right|_{J=0}.\nonumber 
\end{eqnarray}
 So if one is interested in computing the quark propagator and the
quark-antiquark scattering amplitude, then $W\rb{J}$ should be known
to $O\ob{J^{2}}$. Before presenting the technical details of the
calculation, we outline the essential features of the computation
of $W\rb{J}$.

\begin{enumerate}
\item In the strong transverse gauge coupling limit, the dynamics of different
$x_{\bot }$-hyperplanes does not completely decouple. Colour singlet
hadron states can jump from one hyperplane to the next, and the fermion
transverse hopping term (with coupling $\kappa $) continues to contribute
in this limit. 
\item We work in the light front gauge $A^{+}=0$. There are no ghost terms
in this gauge. Also the functional integral over $A^{-}$ becomes
purely Gaussian, and can be evaluated exactly. This integration shows
that the sole outcome of the light-front longitudinal gauge field
is to mediate a linear Coulomb interaction between quarks and antiquarks,
when they are propagating in a given hyperplane. 
\item The integration over transverse gauge link variables can be performed
exactly in the large-$N_{c}$ limit. This results in non-local interactions
between local fermion bilinears. 
\item The fermionic functional integration is carried out in the large-$N_{c}$
limit by rewriting the integral in terms of non-local bosonic variables.
The large-$N_{c}$ limit of the integral is then equivalent to the
stationary point limit of the bosonized functional integral. 
\end{enumerate}
Now we demonstrate this calculation in detail. By scaling all fields
by their canonical dimensions using the lattice spacing $a_{\bot }$,
one can work exclusively with dimensionless quantities: \begin{eqnarray}
x'_{\mu }=x_{\mu }/a_{\bot } & \qquad  & x'_{\bot }=x_{\bot }/a_{\bot }\label{eq:25.5}
\end{eqnarray}
\begin{eqnarray}
\Psi '\ob{x',x'_{\bot }}=\ob{a_{\bot }}^{\frac{3}{2}}\Psi \ob{x,x_{\bot }} & \quad  & \overline{\Psi }'\ob{x',x'_{\bot }}=\ob{a_{\bot }}^{\frac{3}{2}}\overline{\Psi }\ob{x,x_{\bot }}\nonumber \\
U'\ob{x';x'_{\bot },n}=U\ob{x;x_{\bot },n} & \quad  & A'_{\mu }\ob{x',x'_{\bot }}=a_{\bot }A_{\mu }\ob{x,x_{\bot }}\label{eq:25.6}
\end{eqnarray}
 Henceforth we assume that this has been done, and \begin{eqnarray}
S_{\bot } & = & \frac{N_{c}}{4g^{2}}\int d^{2}x\sum _{x_{\bot }}\mathrm{tr}\rb{F^{\mu \nu }\ob{x,x_{\bot }}F_{\mu \nu }\ob{x,x_{\bot }}}\nonumber \\
 & + & \int d^{2}x\sum _{x_{\bot }}\overline{\Psi }\ob{x,x_{\bot }}\rb{i\gamma ^{\mu }D_{\mu }-m}\Psi \ob{x,x_{\bot }}\label{eq:25.7}\\
 & + & \frac{\kappa }{2}\int d^{2}x\sum _{x_{\bot },n}\left[\overline{\Psi }\ob{x,x_{\bot }+n}\ob{r-i\gamma ^{n}}U\ob{x,x_{\bot },n}\Psi \ob{x,x_{\bot }}\right.\nonumber \\
 &  & \qquad \qquad \quad \left.+\overline{\Psi }\ob{x,x_{\bot }}\ob{r+i\gamma ^{n}}U^{\dagger }\ob{x,x_{\bot },n}\Psi \ob{x,x_{\bot }+n}\right]\nonumber 
\end{eqnarray}
 The functional integral in the light-front gauge, $A^{+}=0$, is
\begin{eqnarray}
Z\left[J\right] & = & \int \left[DA^{-}\right]\rb{DU}\left[D\Psi \cdot D\overline{\Psi }\right]\nonumber \\
 & \times  & \exp \Bigg [i\int d^{2}x\sum _{x_{\bot }}\overline{\Psi }_{i}^{\alpha }\left(x,x_{\bot }\right)\left(i\gamma ^{\mu }\partial _{\mu }-m\right)_{ij}^{\alpha \beta }\Psi _{j}^{\beta }\left(x,x_{\bot }\right)\nonumber \\
 &  & \qquad +i\int d^{2}xd^{2}y\sum _{x_{\bot },y_{\bot }}\overline{\Psi }_{i}^{\alpha }\left(x,x_{\bot }\right)J_{ij}^{\alpha \beta }\left(x,x_{\bot };y,y_{\bot }\right)\Psi _{j}^{\beta }\left(y,y_{\bot }\right)\Bigg ]\label{eq:25.8}\\
 & \times  & \exp \Bigg [\frac{i\kappa }{2}\int d^{2}x\sum _{x_{\bot },n}\left\{ \overline{\Psi }_{i}^{\alpha }\ob{x,x_{\bot }+n_{\bot }}\ob{r-i\gamma ^{n}}^{\alpha \beta }U_{ij}\ob{x;x_{\bot },n}\Psi _{j}^{\beta }\ob{x,x_{\bot }}\right.\nonumber \\
 &  & \qquad \qquad \qquad \qquad \left.+\overline{\Psi }_{i}^{\alpha }\ob{x,x_{\bot }}\ob{r+i\gamma ^{n}}^{\alpha \beta }U_{ij}^{\dagger }\ob{x;x_{\bot },n}\Psi _{j}^{\beta }\ob{x,x_{\bot }+n_{\bot }}\right\} \Bigg ]\nonumber \\
 & \times  & \exp \Bigg [-\frac{1}{2}\int d^{2}x\sum _{x_{\bot }}A_{ij}^{-}\left(x,x_{\bot }\right)\left(-\frac{iN_{c}}{g^{2}}\delta _{il}\delta _{jk}\partial _{-}^{2}A_{kl}^{-}\left(x,x_{\bot }\right)\right)\nonumber \\
 &  & \qquad -\int d^{2}x\sum _{x_{\bot }}\left(\overline{\Psi }_{i}\left(x,x_{\bot }\right)\gamma ^{+}\Psi _{j}\left(x,x_{\bot }\right)\right)A_{ij}^{-}\left(xx_{\bot }\right)\Bigg ]\nonumber 
\end{eqnarray}
 The Gaussian functional integral over the $A^{-}$ variables produces
a non-local interaction term, which is quadratic in fermion bilinears
(we suppress the overall normalization constant that does not affect
expectation values). \begin{eqnarray}
Z\left[J\right] & = & \int \rb{DU}\left[D\Psi \cdot D\overline{\Psi }\right]\nonumber \\
 & \times  & \exp \Bigg [i\int d^{2}x\sum _{x_{\bot }}\overline{\Psi }_{i}^{\alpha }\left(x,x_{\bot }\right)\left(i\gamma ^{\mu }\partial _{\mu }-m\right)_{ij}^{\alpha \beta }\Psi _{j}^{\beta }\left(x,x_{\bot }\right)\nonumber \\
 &  & \qquad +i\int d^{2}xd^{2}y\sum _{x_{\bot },y_{\bot }}\overline{\Psi }_{i}^{\alpha }\left(x,x_{\bot }\right)J_{ij}^{\alpha \beta }\left(x,x_{\bot };y,y_{\bot }\right)\Psi _{j}^{\beta }\left(y,y_{\bot }\right)\Bigg ]\nonumber \\
 & \times  & \exp \Bigg [-\frac{ig^{2}}{2N_{c}}\int d^{2}xd^{2}y\sum _{x_{\bot },y_{\bot }}\ob{\gamma ^{+}}^{\alpha \beta }\ob{\gamma ^{+}}^{\gamma \delta }\delta _{x_{\bot }y_{\bot }}h\ob{x-y}\label{eq:25.9}\\
 &  & \qquad \cdot \; \overline{\Psi }_{i}^{\alpha }\left(x,x_{\bot }\right)\Psi _{i}^{\delta }\left(y,y_{\bot }\right)\overline{\Psi }_{j}^{\gamma }\left(y,y_{\bot }\right)\Psi _{j}^{\beta }\left(x,x_{\bot }\right)\Bigg ]\nonumber \\
 & \times  & \exp \Bigg [\frac{i\kappa }{2}\int d^{2}x\sum _{x_{\bot },n}\left\{ \overline{\Psi }_{i}^{\alpha }\ob{x,x_{\bot }+n}\ob{r-i\gamma ^{n}}^{\alpha \beta }U_{ij}\ob{x;x_{\bot },n}\Psi _{j}^{\beta }\ob{x,x_{\bot }}\right.\nonumber \\
 &  & \qquad \qquad \qquad \qquad \left.+\overline{\Psi }_{i}^{\alpha }\ob{x,x_{\bot }}\ob{r+i\gamma ^{n}}^{\alpha \beta }U_{ij}^{\dagger }\ob{x;x_{\bot },n}\Psi _{j}^{\beta }\ob{x,x_{\bot }+n}\right\} \Bigg ],\nonumber 
\end{eqnarray}
 where $h(x-y)$ satisfies $\partial _{-}^{2}h(x-y)=\delta ^{(2)}(x-y)$.
The next step is to carry out the functional integral over the $U$
variables. In the large-$N_{c}$ limit \cite{brezin} \cite{kluberg-stern-etal},
\begin{eqnarray}
Z\left[J\right] & = & \int \left[D\Psi \cdot D\overline{\Psi }\right]\nonumber \\
 & \times  & \exp \Bigg [i\int d^{2}x\sum _{x_{\bot }}\overline{\Psi }_{i}^{\alpha }\left(x,x_{\bot }\right)\left(i\gamma ^{\mu }\partial _{\mu }-m\right)_{ij}^{\alpha \beta }\Psi _{j}^{\beta }\left(x,x_{\bot }\right)\nonumber \\
 &  & \qquad +i\int d^{2}xd^{2}y\sum _{x_{\bot },y_{\bot }}\overline{\Psi }_{i}^{\alpha }\left(x,x_{\bot }\right)J_{ij}^{\alpha \beta }\left(x,x_{\bot };y,y_{\bot }\right)\Psi _{j}^{\beta }\left(y,y_{\bot }\right)\Bigg ]\nonumber \\
 & \times  & \exp \Bigg [-\frac{ig^{2}}{2N_{c}}\int d^{2}xd^{2}y\sum _{x_{\bot },y_{\bot }}\ob{\gamma ^{+}}^{\alpha \beta }\ob{\gamma ^{+}}^{\gamma \delta }\delta _{x_{\bot }y_{\bot }}h\ob{x-y}\label{eq:25.10}\\
 &  & \qquad \cdot \; \overline{\Psi }_{i}^{\alpha }\left(x,x_{\bot }\right)\Psi _{i}^{\delta }\left(y,y_{\bot }\right)\overline{\Psi }_{j}^{\gamma }\left(y,y_{\bot }\right)\Psi _{j}^{\beta }\left(x,x_{\bot }\right)\Bigg ]\nonumber \\
 & \times  & \exp \Bigg [-N_{c}\int d^{2}x\sum _{x_{\bot },n}\mathrm{tr}_{\mathrm{D}}\left\{ \sqrt{1+\frac{\kappa ^{2}}{N_{c}^{2}}B\ob{x;x_{\bot },n}}-1\right.\nonumber \\
 &  & \qquad -\left.\ln \left(\frac{1}{2}+\frac{1}{2}\sqrt{1+\frac{\kappa ^{2}}{N_{c}^{2}}B\ob{x;x_{\bot },n}}\right)\right\} \Bigg ],\nonumber 
\end{eqnarray}
 where $B\ob{x;x_{\bot },n}$ is a matrix in Dirac space with matrix
elements \begin{eqnarray}
B^{\alpha \beta }\ob{x;x_{\bot },n} & = & -\overline{\Psi }_{i}^{\alpha }\ob{x,x_{\bot }+n}\Psi _{i}^{\delta }\ob{x,x_{\bot }+n}\ob{r+i\gamma _{n}^{T}}^{\delta \gamma }\nonumber \\
 &  & \cdot \; \overline{\Psi }_{j}^{\gamma }\ob{x,x_{\bot }}\Psi _{j}^{\tau }\ob{x,x_{\bot }}\ob{r-i\gamma _{n}^{T}}^{\tau \beta }\label{eq:25.11}
\end{eqnarray}
 To perform the functional integral over the fermion variables, we
use a simple trick (see Appendix~B, eq.(\ref{eq:B8})). We first
introduce auxiliary non-local bosonic variables to rewrite the four-fermion
interaction term in terms of fermion bilinears, and then integrate
out the fermion variables. \begin{equation}
Z\rb{J}=\int \rb{D\sigma \cdot D\lambda }\exp \rb{iV\ob{\sigma ,\lambda ;J}},\label{eq:25.12}\end{equation}
 where \begin{equation}
\sigma ^{\alpha \beta }(x,x_{\bot };y,y_{\bot })=\overline{\Psi }_{i}^{\alpha }(x,x_{\bot })\Psi _{i}^{\beta }(y,y_{\bot })\; .\label{eq:25.13}\end{equation}
 Then \begin{eqnarray}
 &  & V\rb{\sigma ,\lambda ;J}=-f_{1}\rb{\sigma }+if_{2}\rb{\sigma }-i\int d^{2}x\; \mathrm{Tr}\Big (\ln \rb{-i\lambda (x,x_{\bot };x,x_{\bot })}\Big )\nonumber \\
 &  & +\int d^{2}xd^{2}y\Bigg [\lambda _{ij}^{\alpha \beta }\left(x,x_{\bot };y,y_{\bot }\right)\sigma _{ij}^{\alpha \beta }\left(x,x_{\bot };y,y_{\bot }\right)\label{eq:25.14}\\
 &  & +\sigma _{ij}^{\alpha \beta }\left(x,x_{\bot };y,y_{\bot }\right)\left(\delta _{x_{\bot }y_{\bot }}\left(i\gamma ^{\mu }\partial _{\mu }-m\right)_{ij}^{\alpha \beta }\delta ^{2}\left(x-y\right)+J_{ij}^{\alpha \beta }\left(x,x_{\bot };y,y_{\bot }\right)\right)\Bigg ],\nonumber 
\end{eqnarray}
 with \begin{eqnarray}
f_{1}\rb{\sigma } & = & \frac{g^{2}}{2N_{c}}\int d^{2}xd^{2}y\sum _{x_{\bot },y_{\bot }}\ob{\gamma ^{+}}^{\alpha \beta }\ob{\gamma ^{+}}^{\gamma \delta }\nonumber \\
 &  & \qquad \cdot \; \delta _{x_{\bot }y_{\bot }}h\ob{x-y}\sigma _{ii}^{\alpha \delta }\ob{x,x_{\bot };y,y_{\bot }}\sigma _{jj}^{\gamma \beta }\ob{y,y_{\bot };x,x_{\bot }}\; ,\label{eq:25.15}\\
f_{2}\rb{\sigma } & = & N_{c}\int d^{2}x\sum _{x_{\bot },n}\mathrm{tr}_{\mathrm{D}}\Big (\tilde{f}_{2}\rb{B\ob{x;x_{\bot },n}}\Big ),\label{eq:25.16}\\
\tilde{f}_{2}\rb{B} & = & \sqrt{1+\frac{\kappa ^{2}}{N_{c}^{2}}B}-1-\ln \ob{\frac{1}{2}+\frac{1}{2}\sqrt{1+\frac{\kappa ^{2}}{N_{c}^{2}}B}}.\label{eq:25.17}
\end{eqnarray}
 The matrix elements of $B$ expressed in terms of $\sigma $ are,
\begin{eqnarray}
B^{\alpha \beta }\ob{x;x_{\bot },n} & = & -\sigma _{ii}^{\alpha \delta }\ob{x,x_{\bot }+n;x,x_{\bot }+n}\ob{r+i\gamma _{n}^{T}}^{\delta \gamma }\nonumber \\
 &  & \cdot \; \sigma _{jj}^{\gamma \tau }\ob{x,x_{\bot };x,x_{\bot }}\ob{r-i\gamma _{n}^{T}}^{\tau \beta }.\label{eq:25.18}
\end{eqnarray}
 In the large-$N_{c}$ limit, $V\rb{\sigma ,\lambda ;J}$ is of $O\ob{N_{c}}$.
Therefore, the $N_{c}\rightarrow \infty $ limit amounts to evaluating
the functional integral at its stationary point. The stationary point
$(\overline{\sigma },\overline{\lambda })$ is determined by \begin{eqnarray}
 &  & \left.\frac{\partial V}{\partial \lambda _{ij}^{\alpha \beta }\left(x,x_{\bot };y,y_{\bot }\right)}\right|_{\overline{\sigma },\overline{\lambda }}=-i\left(\overline{\lambda }^{-1}\right)_{ji}^{\beta \alpha }\left(y,y_{\bot };x,x_{\bot }\right)+\overline{\sigma }_{ij}^{\alpha \beta }\left(x,x_{\bot };y,y_{\bot }\right)=0\; ,\label{eq:25.19}\\
 &  & \left.\frac{\partial V}{\partial \sigma _{ij}^{\alpha \beta }\left(x,x_{\bot };y,y_{\bot }\right)}\right|_{\overline{\sigma },\overline{\lambda }}=\overline{\lambda }_{ij}^{\alpha \beta }\left(x,x_{\bot };y,y_{\bot }\right)+\delta _{x_{\bot }y_{\bot }}\left(i\gamma ^{\mu }\partial _{\mu }-m\right)_{ij}^{\alpha \beta }\delta ^{2}\left(x-y\right)\nonumber \\
 &  & \qquad +J_{ij}^{\alpha \beta }\left(x,x_{\bot };y,y_{\bot }\right)-\frac{g^{2}}{N_{c}}\delta _{ij}\left(\gamma ^{+}\right)^{\alpha \delta }\left(\gamma ^{+}\right)^{\gamma \beta }h(x-y)\delta _{x_{\bot }y_{\bot }}\overline{\sigma }_{kk}^{\gamma \delta }\left(y,y_{\bot };x,x_{\bot }\right)\nonumber \\
 &  & \qquad -iN_{c}\delta _{ij}\delta ^{2}\ob{x-y}\delta _{x_{\bot }y_{\bot }}\label{eq:25.20}\\
 &  & \qquad \cdot \sum _{n}\left\{ \ob{r+i\gamma _{n}^{T}}\overline{\sigma }_{kk}\ob{x,x_{\bot }-n;x,x_{\bot }-n}\ob{r-i\gamma _{n}^{T}}\tilde{f}_{2}'\rb{B\ob{x;x_{\bot }-n,n}}\right.\nonumber \\
 &  & \qquad +\left.\ob{r-i\gamma _{n}^{T}}\tilde{f}_{2}'\rb{B\ob{x;x_{\bot },n}}\overline{\sigma }_{kk}\ob{x,x_{\bot }+n;x,x_{\bot }+n}\ob{r+i\gamma _{n}^{T}}\right\} ^{\beta \alpha }=0\; ,\nonumber 
\end{eqnarray}
 where \begin{equation}
\ob{\tilde{f}_{2}'\rb{B}}^{\beta \alpha }\equiv \frac{\partial \mathrm{tr}_{\mathrm{D}}\tilde{f}_{2}\rb{B}}{\partial B^{\alpha \beta }}=\rb{\frac{\kappa ^{2}}{2N_{c}^{2}\ob{1+\sqrt{1+\frac{\kappa ^{2}}{N_{c}^{2}}B}}}}^{\beta \alpha }.\label{eq:25.21}\end{equation}
 The first stationary point equation (\ref{eq:25.19}) is trivially
solved by \begin{equation}
\overline{\lambda }_{ij}^{\alpha \beta }\ob{x,x_{\bot };y,y_{\bot }}=i\ob{\overline{\sigma }^{-1}}_{ji}^{\beta \alpha }\ob{y,y_{\bot };x,x_{\bot }}\; .\label{eq:25.22}\end{equation}
 Substituting this in the second stationary point equation (\ref{eq:25.20}),
we obtain \begin{eqnarray}
 &  & i\ob{\overline{\sigma }^{-1}}_{ji}^{\beta \alpha }\ob{y,y_{\bot };x,x_{\bot }}+\delta _{x_{\bot }y_{\bot }}\left(i\gamma ^{\mu }\partial _{\mu }-m\right)_{ij}^{\alpha \beta }\delta ^{2}\left(x-y\right)\nonumber \\
 &  & +J_{ij}^{\alpha \beta }\left(x,x_{\bot };y,y_{\bot }\right)-\frac{g^{2}}{N_{c}}\delta _{ij}\left(\gamma ^{+}\right)^{\alpha \delta }\left(\gamma ^{+}\right)^{\gamma \beta }h(x-y)\delta _{x_{\bot }y_{\bot }}\overline{\sigma }_{kk}^{\gamma \delta }\left(y,y_{\bot };x,x_{\bot }\right)\nonumber \\
 &  & -iN_{c}\delta _{ij}\delta ^{2}\ob{x-y}\delta _{x_{\bot }y_{\bot }}\label{eq:25.23}\\
 &  & \cdot \; \sum _{n}\left\{ \ob{r+i\gamma _{n}^{T}}\overline{\sigma }_{kk}\ob{x,x_{\bot }-n;x,x_{\bot }-n}\ob{r-i\gamma _{n}^{T}}\tilde{f}_{2}'\rb{B\ob{x;x_{\bot }-n,n}}\right.\nonumber \\
 &  & \qquad +\left.\ob{r-i\gamma _{n}^{T}}\tilde{f}_{2}'\rb{B\ob{x;x_{\bot },n}}\overline{\sigma }_{kk}\ob{x,x_{\bot }+n;x,x_{\bot }+n}\ob{r+i\gamma _{n}^{T}}\right\} ^{\beta \alpha }=0\; .\nonumber 
\end{eqnarray}
 Solution of this equation will yield $\overline{\sigma }$ as a function
of $J$. The large-$N_{c}$ generating functional can then be succinctly
written as \begin{equation}
Z\rb{J}=\exp \rb{iV\ob{\overline{\sigma },\overline{\lambda };J}}\qquad \Rightarrow \qquad W\rb{J}=V\ob{\overline{\sigma },\overline{\lambda };J}\equiv V_{\mathrm{eff}}\ob{\overline{\sigma };J}\; ,\label{eq:25.24}\end{equation}
 where $V_{\mathrm{eff}}\ob{\overline{\sigma };J}$ is obtained by
substituting $\overline{\lambda }$ in terms of $\overline{\sigma }$.
Modulo $J$-independent terms, \begin{eqnarray}
 &  & V_{\mathrm{eff}}\ob{\overline{\sigma };J}=-f_{1}\rb{\overline{\sigma }}+if_{2}\rb{\overline{\sigma }}+i\int d^{2}x\sum _{x_{\bot }}\left[\mathrm{Tr}\left(\ln \overline{\sigma }\right)(x,x_{\bot };x,x_{\bot })\right]\label{eq:25.25}\\
 &  & +\int d^{2}xd^{2}y\sum _{x_{\bot }y_{\bot }}\overline{\sigma }_{ij}^{\alpha \beta }\left(x,x_{\bot };y,y_{\bot }\right)\left[\delta _{x_{\bot }y_{\bot }}\left(i\gamma ^{\mu }\partial _{\mu }-m\right)_{ij}^{\alpha \beta }\delta ^{2}\left(x-y\right)+J_{ij}^{\alpha \beta }\left(x,x_{\bot };y,y_{\bot }\right)\right].\nonumber 
\end{eqnarray}

The generator of 1PI vertex functions $\Gamma \rb{\varphi }$ (also
known as the effective action) is the Legendre transform of $W\rb{J}$.
Let the effective field $\varphi $ be the variable conjugate to the
external source $J$. Then \begin{equation}
\varphi _{ij}^{\alpha \beta }\ob{x,x_{\bot };y,y_{\bot }}=\frac{\delta W\rb{J}}{\delta J_{ij}^{\alpha \beta }\ob{x,x_{\bot };y,y_{\bot }}}\; ,\label{eq:25.26}\end{equation}
\begin{eqnarray}
\Gamma \rb{\varphi } & = & W\rb{J}-\int d^{2}xd^{2}y\sum _{x_{\bot },y_{\bot }}\varphi _{ij}^{\alpha \beta }\ob{x,x_{\bot };y,y_{\bot }}J_{ij}^{\alpha \beta }\ob{x,x_{\bot };y,y_{\bot }}\nonumber \\
 & = & -f_{1}\rb{\varphi }+if_{2}\rb{\varphi }+i\int d^{2}x\sum _{x_{\bot }}\left[\mathrm{Tr}\left(\ln \varphi \right)(x,x_{\bot };x,x_{\bot })\right]\label{eq:25.27}\\
 & + & \int d^{2}xd^{2}y\sum _{x_{\bot }y_{\bot }}\varphi _{ij}^{\alpha \beta }\left(x,x_{\bot };y,y_{\bot }\right)\rb{\delta _{x_{\bot }y_{\bot }}\left(i\gamma ^{\mu }\partial _{\mu }-m\right)_{ij}^{\alpha \beta }\delta ^{2}\left(x-y\right)}.\nonumber 
\end{eqnarray}
 In the large-$N_{c}$ limit, $\varphi =\overline{\sigma }+\mathrm{O}\ob{1/N_{c}}$.
It is also easy to show that \begin{equation}
\frac{\delta \Gamma \rb{\varphi }}{\delta \varphi _{ij}^{\alpha \beta }(x,x_{\bot };y,y_{\bot })}=-J_{ij}^{\alpha \beta }(x,x_{\bot };y,y_{\bot })\; ,\label{eq:25.28}\end{equation}
\begin{equation}
\rb{W^{(2)}}_{ij;kl}^{\alpha \beta ;\gamma \delta }(xx_{\bot },yy_{\bot };zz_{\bot },ww_{\bot })=-\rb{\ob{\Gamma ^{(2)}}^{-1}}_{ij;kl}^{\alpha \beta ;\gamma \delta }(xx_{\bot },yy_{\bot };zz_{\bot },ww_{\bot })\; ,\label{eq:25.29}\end{equation}
 where we have used the abbreviations, \begin{equation}
\frac{\delta ^{2}W\rb{J}}{\delta J_{ij}^{\alpha \beta }(x,x_{\bot };y,y_{\bot })\delta J_{kl}^{\gamma \delta }(z,z_{\bot };w,w_{\bot })}\equiv \rb{W^{(2)}}_{ij;kl}^{\alpha \beta ;\gamma \delta }(xx_{\bot },yy_{\bot };zz_{\bot },ww_{\bot })\; ,\label{eq:25.30}\end{equation}
\begin{equation}
\frac{\delta ^{2}\Gamma \rb{\varphi }}{\delta \varphi _{kl}^{\gamma \delta }(z,z_{\bot };w,w_{\bot })\delta \varphi _{mn}^{\xi \eta }(u,u_{\bot };v,v_{\bot })}\equiv \rb{\Gamma ^{(2)}}_{kl;mn}^{\gamma \delta ;\xi \eta }(zz_{\bot },ww_{\bot };uu_{\bot },vv_{\bot })\; .\label{eq:25.31}\end{equation}

\chapter{The Chiral Condensate}

\section{Chiral symmetry\label{sec:3.1}}

Consider the transverse lattice action for free naive fermions \begin{eqnarray}
S_{NF}^{\mathrm{free}} & = & a_{\bot }^{2}\int d^{2}x\sum _{x_{\bot }}\overline{\Psi }\ob{x,x_{\bot }}\rb{i\gamma ^{\mu }D_{\mu }-m}\Psi \ob{x,x_{\bot }}\nonumber \\
 & + & \frac{\kappa a_{\bot }}{2}\int d^{2}x\sum _{x_{\bot },n}\left[-i\overline{\Psi }\ob{x,x_{\bot }+a_{\bot }n}\gamma ^{n}\Psi \ob{x,x_{\bot }}\right.\label{eq:31.1}\\
 &  & \qquad \qquad \qquad \left.+i\overline{\Psi }\ob{x,x_{\bot }}\gamma ^{n}\Psi \ob{x,x_{\bot }+a_{\bot }n}\right].\nonumber 
\end{eqnarray}
 On-shell fermion modes of this action satisfy the dispersion relation
\begin{equation}
E\ob{p^{1},p_{\bot }}=\pm \rb{\ob{p^{1}}^{2}+\sum _{n}\frac{\kappa ^{2}}{a_{\bot }^{2}}\sin ^{2}\ob{p_{\bot }\cdot na_{\bot }}+m^{2}}^{\frac{1}{2}},\label{eq:31.2}\end{equation}
 where $E$ and $\ob{p^{1},p_{\bot }}$ refer to the energy and three-momentum
respectively. Since the transverse momenta $p^{n}$ range over the
Brillouin zone, $-\frac{\pi }{a_{\bot }}<p^{n}\leq \frac{\pi }{a_{\bot }}$,
it follows that \begin{equation}
E\ob{p^{1},p_{\bot }}=E\ob{p^{1},p_{\bot }+q_{\bot }^{\pi }}\; ,\label{eq:31.3}\end{equation}
 where $q_{\bot }^{\pi }$ is one of the four transverse momentum
vectors \begin{equation}
q_{\bot }^{\pi }=\cb{\ob{0,0},\ob{\frac{\pi }{a_{\bot }},0},\ob{0,\frac{\pi }{a_{\bot }}},\ob{\frac{\pi }{a_{\bot }},\frac{\pi }{a_{\bot }}}}.\label{eq:31.4}\end{equation}
 The consequence of (\ref{eq:31.3}) is that in the naive transverse
lattice fermions represent four fermion states per field component.
This is the generic {}``fermion doubling'' phenomenon on the lattice.
In order to get a single fermion state per field component, one can
add the irrelevant Wilson term to the naive fermion action (cf.(\ref{eq:24.23})),
\begin{eqnarray}
S_{WF}^{\mathrm{free}} & = & a_{\bot }^{2}\int d^{2}x\sum _{x_{\bot }}\overline{\Psi }\ob{x,x_{\bot }}\rb{i\gamma ^{\mu }D_{\mu }-m}\Psi \ob{x,x_{\bot }}\nonumber \\
 & + & \frac{\kappa a_{\bot }}{2}\int d^{2}x\sum _{x_{\bot },n}\left[\overline{\Psi }\ob{x,x_{\bot }+a_{\bot }n}\ob{r-i\gamma ^{n}}\Psi \ob{x,x_{\bot }}\right.\label{eq:31.5}\\
 &  & \qquad \qquad \qquad \left.+\overline{\Psi }\ob{x,x_{\bot }}\ob{r+i\gamma ^{n}}\Psi \ob{x,x_{\bot }+a_{\bot }n}\right].\nonumber 
\end{eqnarray}
 With the $r$-dependent terms, the dispersion relation for on-shell
fermion modes becomes, \begin{equation}
E\ob{p^{1},p_{\bot }}=\pm \rb{\ob{p^{1}}^{2}+\sum _{n}\frac{\kappa ^{2}}{a_{\bot }^{2}}\sin ^{2}\rb{p_{\bot }\cdot na_{\bot }}+\ob{m-\frac{\kappa r}{a_{\bot }}\sum _{n}\cos \rb{p_{\bot }\cdot na_{\bot }}}^{2}}^{\frac{1}{2}}.\label{eq:31.6}\end{equation}
 This result shows that the $p_{\bot }=0$ mode at the centre of the
Brillouin zone is massless at $m=2\kappa ra_{\bot }^{-1}$. Moreover,
compared to this mode, the other $p_{\bot }=q_{\bot }^{\pi }$ modes
get an additive contribution $n_{\pi }2\kappa ra_{\bot }^{-1}$ to
their masses, where $n_{\pi }$ is the number of components of $q_{\bot }^{\pi }$
equal to $\pi /a_{\bot }$. The modes with $n_{\pi }\neq 0$ thus
become infinitely heavy in the $a_{\bot }\rightarrow 0$ limit, leaving
behind only one physical mode with finite energy at $p_{\bot }=0$.
The usefulness of Wilson fermions comes to fore when there are interactions
between fermions. In an interacting theory, the fermion doublers,
if not eliminated, can affect the physical content of the theory in
a non-trivial way.

Returning to the case of naive lattice fermions, one can show that
their group of chiral symmetries is considerably larger than that
of the continuum theory. Let us find the symmetry groups of individual
terms in the naive fermion transverse lattice QCD action. In terms
of variables scaled using the lattice spacing $a_{\bot }$, \begin{eqnarray}
S_{NF} & = & \int d^{2}x\sum _{x_{\bot }}\overline{\Psi }\ob{x,x_{\bot }}\rb{i\gamma ^{\mu }D_{\mu }-m}\Psi \ob{x,x_{\bot }}\nonumber \\
 & + & \frac{\kappa }{2}\int d^{2}x\sum _{x_{\bot },n}\left[-i\overline{\Psi }\ob{x,x_{\bot }+n}\gamma ^{n}U\ob{x,x_{\bot },n}\Psi \ob{x,x_{\bot }}\right.\label{eq:31.7}\\
 &  & \qquad \qquad \qquad \left.+i\overline{\Psi }\ob{x,x_{\bot }}\gamma ^{n}U^{\dagger }\ob{x,x_{\bot },n}\Psi \ob{x,x_{\bot }+n}\right].\nonumber 
\end{eqnarray}
 With $x_{\bot }\equiv \ob{x^{2},x^{3}}$, let us define new spinor
variables $\chi \ob{x,x_{\bot }}$ and $\overline{\chi }\ob{x,x_{\bot }}$
by \begin{equation}
\left.\begin{array}{c}
 \Psi \ob{x,x_{\bot }}=\ob{\gamma ^{2}}^{x^{2}}\ob{\gamma ^{3}}^{x^{3}}\chi \ob{x,x_{\bot }}\; ,\\
 \overline{\Psi }\ob{x,x_{\bot }}=\overline{\chi }\ob{x,x_{\bot }}\ob{\gamma ^{3}}^{x^{3}}\ob{\gamma ^{2}}^{x^{2}}\; .\end{array}\right.\label{eq:31.8}\end{equation}
 In terms of these new variables, the action takes the form, \begin{eqnarray}
S_{NF} & = & \int d^{2}x\sum _{x_{\bot }}\overline{\chi }\ob{x,x_{\bot }}\rb{i\gamma ^{\mu }D_{\mu }-\ob{-1}^{x^{2}+x^{3}}m}\chi \ob{x,x_{\bot }}\nonumber \\
 & + & \frac{\kappa }{2}\int d^{2}x\sum _{x_{\bot },n}\Delta _{n}\ob{x_{\bot }}\left[-i\overline{\chi }\ob{x,x_{\bot }+n}U\ob{x,x_{\bot },n}\chi \ob{x,x_{\bot }}\right.\label{eq:31.9}\\
 &  & \qquad \qquad \qquad \qquad \quad \left.+i\overline{\chi }\ob{x,x_{\bot }}U^{\dagger }\ob{x,x_{\bot },n}\chi \ob{x,x_{\bot }+n}\right],\nonumber 
\end{eqnarray}
 where \begin{equation}
\Delta _{n}\ob{x_{\bot }}\equiv \Delta _{n}\ob{x^{2},x^{3}}=\left\{ \begin{array}{c}
 \ob{-1}^{1+x^{2}+x^{3}}\qquad :n=2\; ,\\
 \ob{-1}^{1+x^{3}}\qquad \quad :n=3\; .\end{array}\right.\label{eq:31.10}\end{equation}
 In this form, the naive fermion transverse lattice QCD action has
been spin-diagonalized in the transverse spin space. Let a lattice
site be defined as odd (even), if $x^{2}+x^{3}$ is odd (even). The
fermion transverse hopping term has an exact global symmetry $U_{\mathrm{o}}(4)\otimes U_{\mathrm{e}}(4)$,
defined by the transformations \begin{equation}
\left.\begin{array}{c}
 \chi \ob{x,x_{\bot }}\quad \rightarrow \quad \chi '\ob{x,x_{\bot }}=U_{\mathrm{o}(\mathrm{e})}\chi \ob{x,x_{\bot }}\\
 \overline{\chi }\ob{x,x_{\bot }}\quad \rightarrow \quad \overline{\chi }'\ob{x,x_{\bot }}=\overline{\chi }\ob{x,x_{\bot }}U_{\mathrm{e}(\mathrm{o})}^{\dagger }\end{array}\right\} \quad \mathrm{for}\: x_{\bot }\: \mathrm{odd}\, (\mathrm{even}).\label{eq:31.11}\end{equation}
 The fermion mass term is invariant only under the diagonal subgroup
$U_{\Delta }(4)$ of $U_{\mathrm{o}}(4)\otimes U_{\mathrm{e}}(4)$,
given by $U_{\mathrm{o}}=U_{\mathrm{e}}$. To see the symmetry group
of the continuum fermion kinetic term, we choose a particular gamma
matrix representation: \begin{equation}
\gamma ^{0}=\ob{\begin{array}{cc}
 \sigma ^{1} & 0\\
 0 & \sigma ^{1}\end{array}},\; \gamma ^{1}=\ob{\begin{array}{cc}
 i\sigma ^{2} & 0\\
 0 & i\sigma ^{2}\end{array}},\; \gamma ^{2}=\ob{\begin{array}{cc}
 0 & i\sigma ^{3}\\
 i\sigma ^{3} & 0\end{array}},\; \gamma ^{3}=\ob{\begin{array}{cc}
 0 & \sigma ^{3}\\
 -\sigma ^{3} & 0\end{array}}.\label{eq:31.12}\end{equation}
 We also decompose the fermion spinors in two component notation:
\begin{equation}
\chi =\left(\begin{array}{c}
 \Psi _{P}\\
 \Psi _{Q}\end{array}\right),\qquad \overline{\chi }=\ob{\begin{array}{cc}
 \overline{\Psi }_{P} & \overline{\Psi }_{Q}\end{array}},\label{eq:31.13}\end{equation}
 In terms of $2$-dim spinors $\Psi _{P}$ and $\Psi _{Q}$, the naive
fermion transverse lattice QCD action is \begin{eqnarray}
S_{NF} & = & \int d^{2}x\sum _{\lambda ,x_{\bot }}\overline{\Psi }_{\lambda }\ob{x,x_{\bot }}\rb{i\ob{\sigma ^{1}D_{0}+i\sigma ^{2}D_{1}}-\ob{-1}^{x^{2}+x^{3}}m}\Psi _{\lambda }\ob{x,x_{\bot }}\nonumber \\
 & + & \frac{\kappa }{2}\int d^{2}x\sum _{\lambda ,x_{\bot },n}\Delta _{n}\ob{x_{\bot }}\left[-i\overline{\Psi }_{\lambda }\ob{x,x_{\bot }+n}U\ob{x,x_{\bot },n}\Psi _{\lambda }\ob{x,x_{\bot }}\right.\label{eq:31.14}\\
 &  & \qquad \qquad \qquad \qquad \qquad \left.+i\overline{\Psi }_{\lambda }\ob{x,x_{\bot }}U^{\dagger }\ob{x,x_{\bot },n}\Psi _{\lambda }\ob{x,x_{\bot }+n}\right],\nonumber 
\end{eqnarray}
 where the index $\lambda $ takes two values, $P$ and $Q$. It is
now easy to see that the continuum fermion kinetic term is invariant
under the $U_{L}(2)\otimes U_{R}(2)$ transformations: \begin{eqnarray}
\Psi _{\lambda } & \rightarrow  & \rb{\exp \ob{i\theta _{V}^{k}\cdot \tau ^{k}\otimes 1+i\theta _{A}^{k}\cdot \tau ^{k}\otimes \sigma ^{3}(-1)^{x_{2}+x_{3}}}}_{\lambda \lambda '}\Psi _{\lambda '},\nonumber \\
\overline{\Psi }_{\lambda } & \rightarrow  & \overline{\Psi }_{\lambda '}\rb{\exp \ob{-i\theta _{V}^{k}\cdot \tau ^{k}\otimes 1+i\theta _{A}^{k}\cdot \tau ^{k}\otimes \sigma ^{3}(-1)^{x_{2}+x_{3}}}}_{\lambda '\lambda },\label{eq:31.15}
\end{eqnarray}
 where $\tau ^{k}$ ($k=0,1,2,,3$) generate the Lie algebra of the
group $U(2)$ acting on index $\lambda $.

In summary, the symmetries of individual terms in the naive fermion
transverse lattice QCD action are:

\begin{itemize}
\item Fermion transverse hopping term (with coefficient $\kappa $): $U_{\mathrm{o}}\ob{4}\otimes U_{\mathrm{e}}\ob{4}$. 
\item Fermion mass term (with coefficient $m$): $U_{\Delta }(4)$. 
\item Continuum fermion kinetic term: $U_{L}\ob{2}\otimes U_{R}\ob{2}$. 
\end{itemize}
When $m=0$, the combined chiral symmetry of the action is $U_{L}\ob{2}\otimes U_{R}\ob{2}$.
When $m\neq 0$, only the diagonal part of $U_{L}\ob{2}\otimes U_{R}\ob{2}$,
i.e. $U_{V}\ob{2}$, remains an exact symmetry. The non-diagonal part
of the symmetry is broken explicitly by the mass term, and spontaneously
by the strong interactions (as we will show in the following sections).

Eq.(\ref{eq:31.15}) also demonstrates that the doubling problem for
naive fermions can be made less severe by dropping the sum over $\lambda $.
This gives the staggered fermion transverse lattice QCD action, which
has two component spinors, and which gives rise to two degenerate
fermion flavours in the continuum limit.

\section{Quark propagator\label{sec:3.2}}

The quark propagator is obtained from the generating functional of
connected quark-antiquark correlation functions using \begin{eqnarray}
\left\langle \mathrm{T}\ob{\overline{\Psi }_{i}^{\alpha }\left(x,x_{\bot }\right)\Psi _{j}^{\beta }\left(y,y_{\bot }\right)}\right\rangle  & = & \left.\frac{\partial W\left[J\right]}{\partial J_{ij}^{\alpha \beta }\left(x,x_{\bot };y,y_{\bot }\right)}\right|_{J=0}\label{eq:32.1}
\end{eqnarray}
 where $W\rb{J}=V_{eff}\ob{\overline{\sigma };J}$ is given by eq.(\ref{eq:25.25}).
In terms of the solution to the stationary point equation (\ref{eq:25.23}),
\begin{equation}
\left\langle \mathrm{T}\ob{\overline{\Psi }_{i}^{\alpha }\left(x,x_{\bot }\right)\Psi _{j}^{\beta }\left(y,y_{\bot }\right)}\right\rangle =\ob{\overline{\sigma }_{0}}_{ij}^{\alpha \beta }\ob{x,x_{\bot };y,y_{\bot }}\; ,\label{eq:32.2}\end{equation}
 with $\overline{\sigma }_{0}\equiv \overline{\sigma }(J=0)$. Setting
$J=0$ in the stationary point equation (\ref{eq:25.23}), we get
\begin{eqnarray}
 &  & i\ob{\overline{\sigma }_{0}^{-1}}_{ji}^{\beta \alpha }\ob{y,y_{\bot };x,x_{\bot }}+\delta _{x_{\bot }y_{\bot }}\left(i\gamma ^{\mu }\partial _{\mu }-m\right)_{ij}^{\alpha \beta }\delta ^{(2)}\left(x-y\right)\nonumber \\
 &  & -\frac{g^{2}}{N_{c}}\delta _{ij}\left(\gamma ^{+}\right)^{\alpha \delta }\left(\gamma ^{+}\right)^{\gamma \beta }h(x-y)\delta _{x_{\bot }y_{\bot }}\left(\overline{\sigma }_{0}\right)_{kk}^{\gamma \delta }\left(y,y_{\bot };x,x_{\bot }\right)\nonumber \\
 &  & -iN_{c}\delta _{ij}\delta ^{(2)}\ob{x-y}\delta _{x_{\bot }y_{\bot }}\label{eq:32.3}\\
 &  & \cdot \sum _{n}\left\{ \ob{r+i\gamma _{n}^{T}}\left(\overline{\sigma }_{0}\right)_{kk}\ob{x,x_{\bot }-n;x,x_{\bot }-n}\ob{r-i\gamma _{n}^{T}}\tilde{f}_{2}'\rb{\bar{B}_{0}\ob{x;x_{\bot }-n,n}}\right.\nonumber \\
 &  & \quad \; +\left.\ob{r-i\gamma _{n}^{T}}\tilde{f}_{2}'\rb{\bar{B}_{0}\ob{x;x_{\bot },n}}\left(\overline{\sigma }_{0}\right)_{kk}\ob{x,x_{\bot }+n;x,x_{\bot }+n}\ob{r+i\gamma _{n}^{T}}\right\} ^{\beta \alpha }=0\; ,\nonumber 
\end{eqnarray}
 where $\bar{B}_{0}\equiv B(\overline{\sigma }_{0})$. We expect $\overline{\sigma }_{0}$
to obey $(1+1)$-dim Poincar\'{e} invariance and lattice translational
invariance, and so we parametrize it as \begin{eqnarray}
\left(\overline{\sigma }_{0}\right)_{ij}^{\alpha \beta }\left(x,x_{\bot };y,y_{\bot }\right) & = & -i\int \frac{d^{2}pd^{2}p_{\bot }}{\left(2\pi \right)^{4}}\; e^{-ip\cdot \left(y-x\right)+ip_{\bot }\cdot \ob{y_{\bot }-x_{\bot }}}\nonumber \\
 &  & \cdot \; \delta _{ji}\left[\frac{1}{\not p-m-\Sigma \left(p,p_{\bot }\right)+i\epsilon }\right]^{\beta \alpha }.\label{eq:32.4}
\end{eqnarray}
 In the coincidence limit $x\rightarrow y$, the quark propagator
is then proportional to the identity matrix, \begin{equation}
\left(\overline{\sigma }_{0}\right)_{ii}^{\alpha \beta }\left(x,x_{\bot };x,x_{\bot }\right)=N_{c}\tilde{\sigma }_{0}\delta ^{\alpha \beta }.\label{eq:32.5}\end{equation}
 This coincidence limit is singular, and must be regulated. The regulator
must be gauge invariant, and preferably also mass independent. A reasonable
choice is the split-point regularization used in studying the chiral
condensate in the 't~Hooft model \cite{burkardt1}.

Assuming that $\tilde{\sigma }_{0}$ has been properly regulated,
we substitute (\ref{eq:32.5}) in (\ref{eq:25.18}) and (\ref{eq:25.21}),
and obtain \begin{equation}
\bar{B}_{0}\ob{x;x_{\bot },n}=N_{c}^{2}\tilde{\sigma }_{0}^{2}\ob{1-r^{2}}1_{\mathrm{D}}\; ,\label{eq:32.6}\end{equation}
\begin{equation}
\tilde{f}_{2}'\rb{\bar{B}_{0}}=\frac{\kappa ^{2}}{2N_{c}^{2}\ob{1+\sqrt{1+\ob{1-r^{2}}\kappa ^{2}\tilde{\sigma }_{0}^{2}}}}1_{D}\equiv \frac{G_{1}\ob{\tilde{\sigma }_{0}}}{N_{c}^{2}}1_{\mathrm{D}}\; ,\label{eq:32.7}\end{equation}
 where $1_{\mathrm{D}}$ is the identity matrix in Dirac space. Using
these results, the stationary point equation (\ref{eq:32.3}) becomes
\begin{eqnarray}
 &  & i\ob{\overline{\sigma }_{0}^{-1}}_{ji}^{\beta \alpha }\ob{y,y_{\bot };x,x_{\bot }}+\delta _{x_{\bot }y_{\bot }}\left(i\gamma ^{\mu }\partial _{\mu }-m\right)_{ij}^{\alpha \beta }\delta ^{(2)}\left(x-y\right)\nonumber \\
 &  & -\frac{g^{2}}{N_{c}}\delta _{ij}\left(\gamma ^{+}\right)^{\alpha \delta }\left(\gamma ^{+}\right)^{\gamma \beta }h(x-y)\delta _{x_{\bot }y_{\bot }}\ob{\overline{\sigma }_{0}}_{kk}^{\gamma \delta }\left(y,y_{\bot };x,x_{\bot }\right)\label{eq:32.8}\\
 &  & +i\delta _{ij}\delta ^{(2)}\ob{x-y}\delta _{x_{\bot }y_{\bot }}\delta ^{\alpha \beta }4\ob{1-r^{2}}\tilde{\sigma }_{0}G_{1}\ob{\tilde{\sigma }_{0}}=0\; .\nonumber 
\end{eqnarray}
 With the parametrization (\ref{eq:32.4}), this stationary point
equation can be converted to an integral equation for $\Sigma \ob{p,p_{\bot }}$,
\begin{eqnarray}
\Sigma \left(p,p_{\bot }\right) & = & g^{2}\int \frac{d^{2}q}{\left(2\pi \right)^{2}}\; \mathrm{P}\left[\frac{1}{\left(p^{+}-q^{+}\right)^{2}}\right]\gamma ^{+}\left(\frac{1}{\not q-m-\Sigma \left(q\right)+i\epsilon }\right)\gamma ^{+}\label{eq:32.9}\\
 & - & 4i\ob{1-r^{2}}\tilde{\sigma }_{0}G_{1}\ob{\tilde{\sigma }_{0}}\; .\nonumber 
\end{eqnarray}
 This is the same equation as in the 't~Hooft model (see (\ref{eq:B22})),
except that the quark mass has been additively renormalized by a constant
that depends on the chiral condensate. It can be solved exactly as
\begin{equation}
\Sigma \ob{p,p_{\bot }}=\Sigma \ob{p^{+}}=-\frac{g^{2}\gamma ^{+}}{2\pi p^{+}}-4\ob{1-r^{2}}i\tilde{\sigma }_{0}G_{1}\ob{\tilde{\sigma }_{0}}\; .\label{eq:32.10}\end{equation}
 Putting everything together, the quark propagator takes the form
\begin{eqnarray}
 &  & \left\langle \mathrm{T}\ob{\overline{\Psi }_{i}^{\alpha }\left(x,x_{\bot }\right)\Psi _{j}^{\beta }\left(y,y_{\bot }\right)}\right\rangle =\ob{\overline{\sigma }_{0}}_{ij}^{\alpha \beta }\ob{x,x_{\bot };y,y_{\bot }}\label{eq:32.11}\\
 &  & =-i\int \frac{d^{2}p}{\ob{2\pi }^{2}}\delta _{ji}\left[\frac{1}{\not p-\ob{m-4\ob{1-r^{2}}i\tilde{\sigma }_{0}G_{1}\ob{\tilde{\sigma }_{0}}}+\frac{g^{2}\gamma ^{+}}{2\pi p^{+}}+i\epsilon }\right]^{\beta \alpha }e^{-ip\cdot \left(y-x\right)}\delta _{x_{\bot }y_{\bot }}.\nonumber 
\end{eqnarray}
 In particular, for Wilson fermions ($r=1$), the additive correction
to the quark mass vanishes. The quark propagator then reduces to that
for the 't~Hooft model (see (\ref{eq:B25})). This feature is a consequence
of the fact that the Dirac space structure for the Wilson fermion
hopping term has projection operator form.

\section{Chiral condensate\label{sec:3.3}}

Spontaneous chiral symmetry breaking is an important phenomenological
property of the strong interactions (see for instance \cite{weinberg}).
Approximate chiral symmetry, $SU_{L}(2)\otimes SU_{R}(2)$, arises
in QCD, because the two light quarks, $u$ and $d$, have masses much
smaller than $\Lambda _{\mathrm{QCD}}$. (Inclusion of the somewhat
heavier $s$ quark, extends the approximate chiral symmetry to $SU_{L}(3)\otimes SU_{R}(3)$.)
Unbroken chiral symmetry is undesirable, because it predicts nearly
degenerate parity doublets for hadrons---a property that is in contradiction
with experimental facts. The conflict is avoided by realizing that
the QCD vacuum spontaneously breaks the $SU_{L}(2)\otimes SU_{R}(2)$
symmetry to its diagonal subgroup $SU_{V}(2)$, i.e. in the same direction
as the explicit breaking produced by the quark mass term. This pattern
of spontaneous symmetry breaking predicts the existence of nearly
massless pseudo-Goldstone boson with the same quantum numbers as the
broken axial symmetry generators. In fact, the lightest hadrons are
pions with precisely these quantum numbers, and one identifies the
pions as the pseudo-Goldstone bosons. A non-zero vacuum expectation
value of $\overline{\Psi }\Psi $, as $m\rightarrow 0$, is an indicator
of spontaneous chiral symmetry breaking. Demonstration of this behaviour
in QCD involves all the complications of strong interaction dynamics,
and we now analyse this feature in large-$N_{c}$ strong transverse
coupling QCD.

\subsection{Naive fermions}

We have described the chiral symmetry properties of transverse lattice
QCD in section (\ref{sec:3.1}). Only for fermions with $r=0$, chiral
symmetry can be spontaneously broken. We thus have to demonstrate
that that the chiral condensate $\tb{\overline{\Psi }\Psi }$ is non-zero
in the chiral limit for naive fermions. From eq. (\ref{eq:32.11}),
it follows that $\tilde{\sigma }_{0}$ (which is the chiral condensate
up to a proportionality factor) satisfies the recursive relation,
\begin{equation}
iC\tilde{\sigma }_{0}=\int \frac{d^{2}p}{\ob{2\pi }^{2}}\mathrm{tr}_{\mathrm{D}}\left[\frac{1}{\not p-\ob{m-4i\tilde{\sigma }_{0}G_{1}\ob{\tilde{\sigma }_{0}}}+\frac{g^{2}\gamma ^{+}}{2\pi p^{+}}+i\epsilon }\right].\label{eq:33.1}\end{equation}
 where, \begin{equation}
G_{1}\ob{\tilde{\sigma }_{0}}=\frac{\kappa ^{2}}{2\rb{1+\sqrt{1-\kappa ^{2}\ob{i\tilde{\sigma }_{0}}^{2}}}}\; .\label{eq:33.2}\end{equation}
 The integral over $p^{-}$ can be easily evaluated, with the principal
value prescription \cite{coleman}, \begin{eqnarray}
i\tilde{\sigma }_{0} & = & \int \frac{d^{2}p}{\ob{2\pi }^{2}}\left[\frac{m-4i\tilde{\sigma }_{0}G_{1}\ob{\tilde{\sigma }_{0}}}{2p^{+}p^{-}-\ob{m-4i\tilde{\sigma }_{0}G_{1}\ob{\tilde{\sigma }_{0}}}^{2}+\frac{g^{2}}{\pi }+i\epsilon }\right]\nonumber \\
 & = & -i\ob{\frac{m-4i\tilde{\sigma }_{0}G_{1}\ob{\tilde{\sigma }_{0}}}{4\pi }}\int _{0}^{\infty }\frac{dp^{+}}{p^{+}}\; .\label{eq:33.3}
\end{eqnarray}
 The remaining integral over $p^{+}$ is logarithmically divergent.
This divergence is along the light-front, and needs careful regularization
respecting Poincar\'{e} and gauge symmetries.

In the limit $m\rightarrow 0$, and with the introduction of a regularization
kernel $K(p^{+},\tilde{\sigma }_{0})$, eq.(\ref{eq:33.3}) takes
the form \begin{equation}
1=\frac{iG_{1}\ob{\tilde{\sigma }_{0}}}{\pi }\int _{0}^{\infty }\frac{dp^{+}}{p^{+}}\; K(p^{+},\tilde{\sigma }_{0})\; .\label{eq:33.4}\end{equation}
 In case of the 't~Hooft model, Burkardt et al. have evaluated the
chiral condensate of this form, using split-point regularization and
operator formulation \cite{burkardt1}, with the result (\ref{eq:B51}).
A similar procedure would give a non-zero chiral condensate in our
case too, but we have not been able to evaluate it in a compact form.

\subsection{Wilson fermions}

For Wilson fermions, chiral symmetry is explicitly broken. The condensate
$\tb{\overline{\Psi }\Psi }$ therefore does not have direct phenomenological
relevance, but we can still calculate it. For $r=1$, the quark propagator
is identical in structure to that for the $(1+1)$-dim large-$N_{c}$
QCD, except for the dimensionality of Dirac space. Therefore, when
the same regularization is used in $(1+1)$-dim and $(3+1)$-dim theories,
\begin{equation}
\tb{\overline{\Psi }\Psi }_{(3+1)-\mathrm{dim}}^{r=1}=2\tb{\overline{\Psi }\Psi }_{(1+1)-\mathrm{dim}}\mathop {\longrightarrow }^{m\rightarrow 0}-\frac{N_{c}}{a_{\bot }^{3}}\sqrt{\frac{g^{2}}{3\pi }}\; .\label{eq:33.5}\end{equation}

\chapter{The Meson Spectrum}

\section{Quark-antiquark scattering Green's function}

The time-ordered four-point quark-antiquark Green's function can be
obtained from the generating functionals as \begin{eqnarray}
 &  & \left\langle \mathrm{T}\ob{\overline{\Psi }_{i}^{\alpha }\left(x,x_{\bot }\right)\Psi _{j}^{\beta }\left(y,y_{\bot }\right)\overline{\Psi }_{k}^{\gamma }\left(z,z_{\bot }\right)\Psi _{l}^{\delta }\left(w.w_{\bot }\right)}\right\rangle \nonumber \\
 &  & -\left\langle \mathrm{T}\ob{\overline{\Psi }_{i}^{\alpha }\left(x,x_{\bot }\right)\Psi _{j}^{\beta }\left(y,y_{\bot }\right)}\right\rangle \left\langle \mathrm{T}\ob{\overline{\Psi }_{k}^{\gamma }\left(z,z_{\bot }\right)\Psi _{l}^{\delta }\left(w,w_{\bot }\right)}\right\rangle \nonumber \\
 &  & =\left.-i\frac{\partial ^{2}W\left[J\right]}{\partial J_{ij}^{\alpha \beta }\left(x,x_{\bot };y,y_{\bot }\right)\partial J_{kl}^{\gamma \delta }\left(z,z_{\bot };w,w_{\bot }\right)}\right|_{J=0}\nonumber \\
 &  & =\left.i\rb{\ob{\Gamma ^{(2)}}^{-1}}\ob{xx_{\bot },yy_{\bot };zz_{\bot },ww_{\bot }}\right|_{\varphi =\overline{\sigma }_{0}}\; .\label{41.1}
\end{eqnarray}
 Using the effective action (\ref{eq:25.27}), we find \begin{equation}
\left.\Gamma ^{(2)}\right|_{\varphi =\overline{\sigma }_{0}}=S+K\; ,\label{eq:41.2}\end{equation}
\begin{equation}
\ob{S}_{ij,kl}^{\alpha \beta ,\gamma \delta }\ob{xx_{\bot },yy_{\bot };zz_{\bot },ww_{\bot }}=-i\ob{\overline{\sigma }_{0}^{-1}}_{jk}^{\beta \gamma }\ob{y,y_{\bot };z,z_{\bot }}\ob{\overline{\sigma }_{0}^{-1}}_{li}^{\delta \alpha }\ob{w,w_{\bot };x,x_{\bot }}\; ,\label{eq:41.3}\end{equation}
\begin{eqnarray}
 &  & \ob{K}_{ij,kl}^{\alpha \beta ,\gamma \delta }\ob{xx_{\bot },yy_{\bot };zz_{\bot },ww_{\bot }}=\left.\frac{\partial ^{2}\ob{-f_{1}\rb{\varphi }+if_{2}\rb{\varphi }}}{\partial \varphi _{ij}^{\alpha \beta }\ob{x,x_{\bot };y,y_{\bot }}\partial \varphi _{kl}^{\gamma \delta }\ob{z,z_{\bot };w,w_{\bot }}}\right|_{\varphi =\overline{\sigma }_{0}}\nonumber \\
 & = & -\frac{g^{2}}{N_{c}}\delta _{ij}\delta _{kl}h\ob{x-y}\delta _{x_{\bot }y_{\bot }}\delta ^{2}\ob{x-w}\delta _{x_{\bot }w_{\bot }}\delta ^{2}\ob{y-z}\delta _{y_{\bot }z_{\bot }}\ob{\gamma ^{+}}^{\alpha \delta }\ob{\gamma ^{+}}^{\gamma \beta }\nonumber \\
 &  & -i\frac{1}{N_{c}}\delta _{ij}\delta _{kl}\delta ^{2}\ob{x-y}\delta ^{2}\ob{x-z}\delta ^{2}\ob{x-w}\delta _{x_{\bot }y_{\bot }}\delta _{z_{\bot }w_{\bot }}\label{eq:41.4}\\
 &  & \cdot \; \Big [\ob{G_{1}+\ob{1-r^{2}}\tilde{\sigma }_{0}^{2}G_{2}}\Big \{\ob{r-i\gamma ^{n}}^{\alpha \delta }\ob{r+i\gamma ^{n}}^{\gamma \beta }\delta _{x_{\bot }-n,z_{\bot }}\nonumber \\
 &  & \qquad \qquad \qquad +\ob{r+i\gamma ^{n}}^{\alpha \delta }\ob{r-i\gamma ^{n}}^{\gamma \beta }\delta _{x_{\bot }+n,z_{\bot }}\Big \}-4\ob{1-r^{2}}^{2}\tilde{\sigma }_{0}^{2}G_{2}\delta ^{\alpha \delta }\delta ^{\gamma \beta }\delta _{x_{\bot },z_{\bot }}\Big ].\nonumber 
\end{eqnarray}
 Here $G_{1}$ and $G_{2}$ are functions of $\tilde{\sigma }_{0}$
given by \begin{eqnarray}
G_{1}\ob{\tilde{\sigma }_{0}} & = & \frac{\kappa ^{2}}{2\ob{1+\sqrt{1+\ob{1-r^{2}}\kappa ^{2}\tilde{\sigma }_{0}^{2}}}}\; ,\label{eq:41.5}\\
G_{2}\ob{\tilde{\sigma }_{0}} & = & \frac{-\kappa ^{4}}{4\sqrt{1+\ob{1-r^{2}}\kappa ^{2}\tilde{\sigma }_{0}^{2}}\; \ob{1+\sqrt{1+\ob{1-r^{2}}\kappa ^{2}\tilde{\sigma }_{0}^{2}}}^{2}}\; .\label{eq:41.6}
\end{eqnarray}
 The full quark-antiquark scattering Green's function is \begin{eqnarray}
 &  & \left\langle \mathrm{T}\left\{ \overline{\Psi }_{i}^{\alpha }\left(x,x_{\bot }\right)\Psi _{j}^{\beta }\left(y,y_{\bot }\right)\overline{\Psi }_{k}^{\gamma }\left(z,z_{\bot }\right)\Psi _{l}^{\delta }\left(w.w_{\bot }\right)\right\} \right\rangle \label{eq:41.7}\\
 & = & \ob{\overline{\sigma }_{0}}_{ij}^{\alpha \beta }\ob{x,x_{\bot };y,y_{\bot }}\ob{\overline{\sigma }_{0}}_{kl}^{\gamma \delta }\ob{z,z_{\bot };w,w_{\bot }}+i\left.\rb{S+K}^{-1}\right._{ij;kl}^{\alpha \beta ;\gamma \delta }\ob{xx_{\bot },yy_{\bot };zz_{\bot },ww_{\bot }}\; .\nonumber 
\end{eqnarray}
 The meson spectrum is determined by the poles of this Green's function,
while the corresponding residues are related to the meson wavefunctions.
Clearly, the singularities of the quark-antiquark scattering Green's
function are contained in the zeroes of $[S+K]$. The homogenous equation
which determines these zeroes and the corresponding eigenfunctions
is the Bethe-Salpeter equation.

\section{Bethe-Salpeter equation}

In a symbolic form, the Bethe-Salpeter equation is \begin{equation}
\rb{S+K}\chi =0\; .\label{eq:42.1}\end{equation}
 In a more explicit notation, \begin{eqnarray}
\int d^{2}zd^{2}w\sum _{z_{\bot }w_{\bot }} &  & \left[\ob{S}_{ij;kl}^{\alpha \beta ;\gamma \delta }\ob{xx_{\bot },yy_{\bot };zz_{\bot },ww_{\bot }}\right.\nonumber \\
 &  & \left.+\ob{K}_{ij,kl}^{\alpha \beta ,\gamma \delta }\ob{xx_{\bot },yy_{\bot };zz_{\bot },ww_{\bot }}\right]\chi _{kl}^{\gamma \delta }\ob{zz_{\bot },ww_{\bot }}=0\; .\label{eq:42.2}
\end{eqnarray}
 Let us transform this equation to momentum space. It follows from
eq.(\ref{eq:32.11}) that \begin{eqnarray}
\ob{\overline{\sigma }_{0}^{-1}}_{ij}^{\alpha \beta }\ob{x,x_{\bot };y,y_{\bot }} & = & i\int \frac{d^{2}p}{\ob{2\pi }^{2}}\; e^{-ip\cdot \left(y-x\right)}\delta _{x_{\bot }y_{\bot }}\nonumber \\
 &  & \cdot \; \delta _{ji}\left[\gamma ^{\mu }p_{\mu }-\ob{m-4i\ob{1-r^{2}}\tilde{\sigma }_{0}G_{1}}+\frac{g^{2}\gamma ^{+}}{2\pi p^{+}}\right]^{\beta \alpha }.\label{eq:42.3}
\end{eqnarray}
 Define the momentum transform of $\chi _{ij}^{\alpha \beta }\ob{x,x_{\bot };y,y_{\bot }}$
by \begin{equation}
\chi _{ij}^{\alpha \beta }\ob{x,x_{\bot };y,y_{\bot }}=\int \frac{d^{4}k_{1}}{\ob{2\pi }^{4}}\frac{d^{4}k_{2}}{\ob{2\pi }^{4}}\; \tilde{\chi }_{ij}^{\alpha \beta }\ob{k_{1},k_{1\bot };k_{2},k_{2\bot }}e^{-ik_{1}\cdot x+ik_{1\bot }\cdot x_{\bot }-ik_{2}\cdot y+ik_{2\bot }\cdot y_{\bot }}\; .\label{eq:42.4}\end{equation}
 With (\ref{eq:41.3}), (\ref{eq:41.4}), (\ref{eq:42.3}) and (\ref{eq:42.4}),
the momentum space Bethe-Salpeter equation for meson states becomes,
\begin{eqnarray}
 &  & i\left[\not k_{1}-\ob{m_{1}-4\ob{1-r^{2}}i\tilde{\sigma }_{0}G_{1}}+\frac{g^{2}\gamma ^{+}}{2\pi k_{1}^{+}}\right]^{\alpha \delta }\nonumber \\
 &  & \cdot \left[-\not k_{2}-\ob{m_{2}-4\ob{1-r^{2}}i\tilde{\sigma }_{0}G_{1}}-\frac{g^{2}\gamma ^{+}}{2\pi k_{2}^{+}}\right]^{\gamma \beta }\tilde{\chi }_{ji}^{\gamma \delta }\ob{k_{2},k_{2\bot };k_{1},k_{1\bot }}\nonumber \\
 & + & \int \frac{d^{4}l_{1}}{\ob{2\pi }^{4}}\frac{d^{4}l_{2}}{\ob{2\pi }^{4}}\delta ^{(4)}\ob{k_{1}+k_{2}-l_{1}-l_{2}}\frac{1}{N_{c}}\delta _{ij}\Bigg [g^{2}\ob{\gamma ^{+}}^{\alpha \delta }\ob{\gamma ^{+}}^{\gamma \beta }\mathrm{P}\frac{1}{\ob{k_{1}^{+}-l_{1}^{+}}^{2}}\nonumber \\
 &  & -i\Big \{2\ob{G_{1}+\ob{1-r^{2}}\tilde{\sigma }_{0}^{2}G_{2}}\Big (\sum _{n}\rb{r^{2}\delta ^{\alpha \delta }\delta ^{\gamma \beta }+\ob{\gamma ^{n}}^{\alpha \delta }\ob{\gamma ^{n}}^{\gamma \beta }}\cos \left(\ob{k_{1\bot }+k_{2\bot }}\cdot n\right)\nonumber \\
 &  & \qquad \qquad \qquad \qquad \qquad \qquad +r\sum _{n}\rb{\delta ^{\alpha \delta }\ob{\gamma ^{n}}^{\gamma \beta }-\ob{\gamma ^{n}}^{\alpha \delta }\delta ^{\gamma \beta }}\sin \left(\ob{k_{1\bot }+k_{2\bot }}\cdot n\right)\Big )\nonumber \\
 &  & -4\ob{1-r^{2}}^{2}\tilde{\sigma }_{0}^{2}G_{2}\delta ^{\alpha \delta }\delta ^{\gamma \beta }\Big \}\Bigg ]\tilde{\chi }_{ll}^{\gamma \delta }\ob{l_{2},l_{2\bot };l_{1},l_{1\bot }}\; =\; 0\; .\label{eq:42.5}
\end{eqnarray}
 Rewriting the above equation in an obvious matrix notation, we get
\begin{eqnarray}
 &  & \tilde{\chi }_{ii}^{T}\ob{p_{2},p_{2\bot };p_{1},p_{1\bot }}=\frac{\rb{\not p_{1}+\left(m_{1}-4\ob{1-r^{2}}i\tilde{\sigma }_{0}G_{1}\right)+\frac{g^{2}}{2\pi p_{1}^{+}}\gamma ^{+}}}{p_{1}^{2}-\ob{m_{1}-4\ob{1-r^{2}}i\tilde{\sigma }_{0}G_{1}}^{2}+\frac{g^{2}}{\pi }+i\epsilon }\nonumber \\
 &  & \times \int \frac{d^{4}k_{1}d^{4}k_{2}}{\ob{2\pi }^{4}}\; \delta ^{(4)}\ob{p_{1}+p_{2}-k_{1}-k_{2}}\Bigg [ig^{2}\mathrm{P}\frac{1}{\ob{p_{1}^{+}-k_{1}^{+}}^{2}}\; \gamma ^{+}\tilde{\chi }_{jj}^{T}\ob{k_{2},k_{2\bot };k_{1},k_{1\bot }}\gamma ^{+}\nonumber \\
 &  & \quad +2\ob{G_{1}+\ob{1-r^{2}}\tilde{\sigma }_{0}^{2}G_{2}}\nonumber \\
 &  & \qquad \quad \cdot \Big (\sum _{n}\Big [r^{2}\tilde{\chi }_{jj}^{T}\ob{k_{2},k_{2\bot };k_{1},k_{1\bot }}+\gamma ^{n}\tilde{\chi }_{jj}^{T}\ob{k_{2},k_{2\bot };k_{1},k_{1\bot }}\gamma ^{n}\Big ]\cos \ob{\ob{k_{1\bot }+k_{2\bot }}\cdot n}\nonumber \\
 &  & \qquad \quad \; +r\sum _{n}\rb{\tilde{\chi }_{jj}^{T}\ob{k_{2},k_{2\bot };k_{1},k_{1\bot }}\gamma ^{n}-\gamma ^{n}\tilde{\chi }_{jj}^{T}\ob{k_{2},k_{2\bot };k_{1},k_{1\bot }}}\sin \left(\ob{k_{1\bot }+k_{2\bot }}\cdot n\right)\Big )\nonumber \\
 &  & \quad -4\ob{1-r^{2}}^{2}\tilde{\sigma }_{0}^{2}G_{2}\tilde{\chi }_{jj}^{T}\ob{k_{2},k_{2\bot };k_{1},k_{1\bot }}\Bigg ]\label{eq:42.6}\\
 &  & \times \frac{\rb{-\not p_{2}+\left(m_{2}-4\ob{1-r^{2}}i\tilde{\sigma }_{0}G_{1}\right)-\frac{g^{2}}{2\pi p_{2}^{+}}\gamma ^{+}}}{p_{2}^{2}-\ob{m_{2}-4\ob{1-r^{2}}i\tilde{\sigma }_{0}G_{1}}^{2}+\frac{g^{2}}{\pi }+i\epsilon }\; .\nonumber 
\end{eqnarray}
 Without loss of generality, we scale the longitudinal momenta as
\begin{equation}
p_{1}^{+}+p_{2}^{+}=1,\; p_{1}^{+}=x,\; k_{1}^{+}=y=y_{1},\; k_{2}^{+}=y_{2},\label{eq:42.7}\end{equation}
 and choose to work in a reference frame where the total incoming
and outgoing transverse momenta vanish, \begin{equation}
p_{1\bot }+p_{2\bot }=0\; .\label{eq:42.8}\end{equation}
 In such a case, except for the external quark propagators and the
eigenfunction $\tilde{\chi }$, all other factors in eq.(\ref{eq:42.6})
are independent of the {}``-'' and {}``$\bot $'' momentum components.
This instantaneous nature of the Bethe-Salpeter kernel makes it convenient
to project the Bethe-Salpeter equation on the light-front by \cite{brodsky}\begin{eqnarray}
\Phi \ob{x}=\int dp_{1}^{-}dp_{2}^{-}\frac{d^{2}p_{1\bot }}{(2\pi )^{2}}\frac{d^{2}p_{2\bot }}{(2\pi )^{2}}\; \delta \ob{\frac{M^{2}}{2}-p_{1}^{-}-p_{2}^{-}}\delta ^{(2)}\ob{p_{1\bot }+p_{2\bot }} &  & \nonumber \\
\tilde{\chi }_{ii}^{T}\ob{p_{2}^{-},1-x,p_{2\bot };p_{1}^{-},x,p_{1\bot }}. &  & \label{eq:42.9}
\end{eqnarray}
 Performing a simple contour integral over $p_{1}^{-}$, we obtain
the homogeneous equation \begin{eqnarray}
 &  & \rb{M^{2}-\frac{\ob{m_{1}-4\ob{1-r^{2}}i\tilde{\sigma }_{0}G_{1}}^{2}-\frac{g^{2}}{\pi }}{x}-\frac{\ob{m_{2}-4\ob{1-r^{2}}i\tilde{\sigma }_{0}G_{1}}^{2}-\frac{g^{2}}{\pi }}{1-x}}\Phi (x)\nonumber \\
 & = & \frac{1}{2x(1-x)}\rb{\frac{\ob{m_{1}-4\ob{1-r^{2}}i\tilde{\sigma }_{0}G_{1}}^{2}-\frac{g^{2}}{\pi }}{2x}\gamma ^{+}+x\gamma ^{-}+\left(m_{1}-4\ob{1-r^{2}}i\tilde{\sigma }_{0}G_{1}\right)+\frac{g^{2}\gamma ^{+}}{2\pi x}}\nonumber \\
 & \times  & \int _{0}^{1}\frac{dy}{(2\pi )}\Bigg \{g^{2}\mathrm{P}\frac{1}{(x-y)^{2}}\gamma ^{+}\Phi (y)\gamma ^{+}\nonumber \\
 &  & \qquad \quad -i\ob{2\ob{G_{1}+\ob{1-r^{2}}\tilde{\sigma }_{0}^{2}G_{2}}\Big [2r^{2}\Phi (y)+\sum _{n}\gamma ^{n}\Phi (y)\gamma ^{n}\Big ]-4\ob{1-r^{2}}^{2}\tilde{\sigma }_{0}^{2}G_{2}\Phi (y)}\Bigg \}\nonumber \\
 & \times  & \Bigg [-\left(\frac{M^{2}}{2}-\frac{\ob{m_{1}-4\ob{1-r^{2}}i\tilde{\sigma }_{0}G_{1}}^{2}-\frac{g^{2}}{\pi }}{2x}\right)\gamma ^{+}\nonumber \\
 &  & \quad -(1-x)\gamma ^{-}+\left(m_{2}-4\ob{1-r^{2}}i\tilde{\sigma }_{0}G_{1}\right)-\frac{g^{2}\gamma ^{+}}{2\pi (1-x)}\Bigg ]\; .\label{eq:42.10}
\end{eqnarray}
 Solutions to this equation provide the meson spectrum and light-front
wavefunctions.

\section{Meson properties}

The integral equation (\ref{eq:42.10}) is quite complicated, and
it has to be solved numerically in general. But by comparing it to
the corresponding equation for the 't~Hooft model, (\ref{eq:B37}),
we observe that four changes have taken place in going from the $(1+1)$-dim
theory to our $(3+1)$-dim case:

\begin{itemize}
\item The meson wavefunction has become a matrix in Dirac space. Physical
meson states labeled by spin-parity quantum numbers have to be obtained
by diagonalization of the matrix integral equation. 
\item The quark mass has been additively renormalized due to hopping in
the transverse directions (tadpole diagrams), $m\rightarrow \tilde{m}=m-4(1-r^{2})i\tilde{\sigma }_{0}G_{1}$. 
\item The transverse lattice dynamics has shifted the interaction kernel
by a term, independent of the longitudinal momentum fraction $x$
but dependent on the spin-parity of the meson. This wavefunction at
the origin effect is a result of the strong transverse gauge coupling
which does not allow the quark and the antiquark to separate. 
\item The $\delta $-function constraint in the transverse directions makes
the meson orbital angular momentum $L_{z}$ vanish. Then $J_{z}=S_{z}$,
and with spin-half quarks, the meson helicities are restricted to
$0,\pm 1$. Allowed spin-parity quantum numbers for the mesons are
therefore $J^{P}=0^{\pm },1^{\pm }$. 
\end{itemize}
These changes are simple enough, so that even without explicitly solving
(\ref{eq:42.10}), we can infer some of the meson properties, based
on the known results of the 't~Hooft model and noting that extrapolation
in $\kappa $ from zero to non-zero value is smooth. We find it convenient
to describe the meson wavefunctions in the basis that is a direct
product of the usual Clifford bases in the continuum and the lattice
directions: \begin{equation}
\Phi =\sum _{C,L}\Phi _{C;L}\Gamma ^{C;L}\; ,\label{eq:43.1}\end{equation}
\begin{equation}
\Gamma _{M}=\Gamma _{C}\otimes \Gamma _{L}=\cb{1,\gamma ^{+},\gamma ^{-},\frac{1}{2}[\gamma ^{+},\gamma ^{-}]}\otimes \cb{1,\gamma ^{1},\gamma ^{2},\frac{1}{2}[\gamma ^{1},\gamma ^{2}]}\; .\label{eq:43.2}\end{equation}
 In this basis, $\sum _{n}\gamma ^{n}\Phi _{C,L}\gamma ^{n}\propto \Phi _{C,L}$
for each value of $L$. The sixteen components of (\ref{eq:42.10}),
therefore, separate in to four blocks (labeled by the value of $L$)
of four components each. In addition, parity remains an exact symmetry
in each block.

It is then easy to deduce the following features:

\begin{itemize}
\item The singular part of the interaction kernel is the same as in the
`t~Hooft model. It is this part which determines the behaviour of
finite norm solutions at the boundaries at $x=0$ and $x=1$ \cite{thooft2},
and only the components $\Phi _{-;L}$ contribute to it. We thus expect
$\Phi _{-;L}$ to vanish at the boundaries as $x^{\beta _{1}}$ and
$(1-x)^{\beta _{2}}$, where \begin{equation}
\pi \beta _{1}\cot \ob{\pi \beta _{1}}=1-\frac{\tilde{m}_{1}^{2}\pi }{g^{2}}\; ,\quad \pi \beta _{2}\cot \ob{\pi \beta _{2}}=1-\frac{\tilde{m}_{2}^{2}\pi }{g^{2}}\; .\label{eq:43.3}\end{equation}

\item The light-front wavefunctions are gauge invariant by construction,
and they are restricted to the finite box $x\in [0,1]$. The finite
box size guarantees that the spectrum of $M^{2}$ is purely discrete
\cite{coleman}. In each spin-parity sector, meson states can thus
be labeled by a radial excitation quantum number $n=1,2,3,\ldots $. 
\item For large $n$, the behaviour of eigenvalues and eigenfunctions of
the integral equation is governed only by the singular part of the
interaction kernel. Then as in the 't~Hooft model \cite{thooft2},
\begin{equation}
\left.\begin{array}{c}
 M_{n}^{2}\simeq \pi g^{2}n\\
 \Phi _{-;L}^{n}(x)\simeq \sqrt{2}\sin \ob{\pi nx}\end{array}\right\} n\gg 1\; .\label{eq:43.4}\end{equation}
 The non-singular spin-parity dependent part of the interaction kernel,
arising from the transverse lattice dynamics, only provides an $n$-independent
zero-point energy shift. We thus expect parallel infinite towers of
meson states in the $M^{2}-n$ plane, labeled by spin-parity quantum
numbers. 
\item Under the exchange of the quark and the antiquark, the meson wavefunction
transforms as $E_{q\overline{q}}\Phi (x;m_{1},m_{2})=C\Phi ^{T}(1-x;m_{2},m_{1})C^{-1}$,
where $C$ is the charge conjugation operator. Eq.(\ref{eq:42.10})
is invariant under this exchange operation (it is easier to see this
before performing the contour integral over $p_{1}^{-}$). Although
parity is an exact symmetry of our formalism, it is not manifest,
because the light-front is not invariant under parity transformation.
The $(x\leftrightarrow 1-x)$ part of $E_{q\overline{q}}$ can be
associated with parity, however, which implies that meson states in
each tower alternate in parity as in the 't~Hooft model. Since $\Phi _{-;L}^{1}$
is symmetric, and the quark and the antiquark have opposite intrinsic
parities, the lowest meson state in each tower has negative parity
and $P=(-1)^{n}$. 
\item The parity symmetry is exact in every block labeled by $L$, so the
eight possible spin-parity quantum numbers are distributed in to the
four blocks as two states of opposite parity in each block. Explicitly,
in the conventional notation, $\{\pi ,a_{1}(0)\}\in \Phi _{C;n_{1}n_{2}}$,
$\{\rho (n),a_{1}(n)\}\in \Phi _{C;n}$ and $\{\rho (0),\sigma \}\in \Phi _{C;1}$. 
\item Another consequence of the parity symmetry is that the equations for
the components $\Phi _{C;n_{1}}$ and $\Phi _{C;n_{2}}$ are degenerate.
This feature implies identical masses for meson states with helicity$=\pm $,
which correspond to the transverse lattice combinations $(n_{1}\pm in_{2})$
(e.g. in case of vector and axial mesons). There are thus three independent
towers of meson states. Note that helicity can be defined only modulo-$4$
on a transverse square lattice, and explicit breaking of rotational
symmetry makes the equations for the helicity$=0$ state and the helicity$=\pm $
states non-degenerate. 
\item Fitting the meson spectrum to experimental results requires determination
of the parameters $g^{2}$ and $\kappa $, in addition to the quark
masses. It would be convenient to fix $g^{2}$ using the slope of
the infinite towers of states, and $\kappa $ by demanding (to the
extent possible) degeneracy of helicity$=0,\pm $ states. 
\end{itemize}
Now we consider the cases of naive and Wilson fermions explicitly.

\subsection{Naive fermions}

For $r=0$, the meson integral equation becomes \begin{eqnarray}
 &  & \rb{M^{2}-\frac{\ob{m-4i\tilde{\sigma }_{0}G_{1}}^{2}-\frac{g^{2}}{\pi }}{x}-\frac{\ob{m-4i\tilde{\sigma }_{0}G_{1}}^{2}-\frac{g^{2}}{\pi }}{1-x}}\Phi (x)\nonumber \\
 & = & \frac{1}{2x(1-x)}\rb{\frac{\ob{m-4i\tilde{\sigma }_{0}G_{1}}^{2}-\frac{g^{2}}{\pi }}{2x}\gamma ^{+}+x\gamma ^{-}+\left(m-4i\tilde{\sigma }_{0}G_{1}\right)+\frac{g^{2}\gamma ^{+}}{2\pi x}}\label{eq:43.5}\\
 & \times  & \int _{0}^{1}\frac{dy}{(2\pi )}\Bigg \{g^{2}\mathrm{P}\frac{1}{(x-y)^{2}}\gamma ^{+}\Phi (y)\gamma ^{+}-i\ob{2\ob{G_{1}+\tilde{\sigma }_{0}^{2}G_{2}}\sum _{n}\gamma ^{n}\Phi (y)\gamma ^{n}-4\tilde{\sigma }_{0}^{2}G_{2}\Phi (y)}\Bigg \}\nonumber \\
 & \times  & \Bigg [-\left(\frac{M^{2}}{2}-\frac{\ob{m-4i\tilde{\sigma }_{0}G_{1}}^{2}-\frac{g^{2}}{\pi }}{2x}\right)\gamma ^{+}-(1-x)\gamma ^{-}+\left(m-4i\tilde{\sigma }_{0}G_{1}\right)-\frac{g^{2}\gamma ^{+}}{2\pi (1-x)}\Bigg ]\nonumber 
\end{eqnarray}
 In this case, unbroken chiral symmetry of the action implies that
the chiral limit of the theory is at $m=0$.

\subsection{Wilson fermions}

For $r=1$, the meson integral equation becomes \begin{eqnarray}
 &  & \rb{M^{2}-\frac{m^{2}-\frac{g^{2}}{\pi }}{x}-\frac{m^{2}-\frac{g^{2}}{\pi }}{1-x}}\Phi (x)\; =\; \frac{1}{2x(1-x)}\rb{\frac{m^{2}-\frac{g^{2}}{\pi }}{2x}\gamma ^{+}+x\gamma ^{-}+m+\frac{g^{2}\gamma ^{+}}{2\pi x}}\nonumber \\
 &  & \qquad \qquad \qquad \times \int _{0}^{1}\frac{dy}{(2\pi )}\Bigg \{g^{2}\mathrm{P}\frac{1}{(x-y)^{2}}\gamma ^{+}\Phi (y)\gamma ^{+}-i\kappa ^{2}\Big [2\Phi (y)+\sum _{n}\gamma ^{n}\Phi (y)\gamma ^{n}\Big ]\Bigg \}\nonumber \\
 &  & \qquad \qquad \qquad \times \Bigg [-\left(\frac{M^{2}}{2}-\frac{m^{2}-\frac{g^{2}}{\pi }}{2x}\right)\gamma ^{+}-(1-x)\gamma ^{-}+m-\frac{g^{2}\gamma ^{+}}{2\pi (1-x)}\Bigg ]\label{eq:43.6}
\end{eqnarray}
 This equation is quite close in structure to the 't~Hooft equation,
because the tadpole renormalization of quark mass vanishes.

It is easy to see that the transverse part of the interaction kernel
vanishes for the component $\Phi _{-;n_{1}n_{2}}$, for any value
of $\kappa $. It follows that $\Phi _{-;n_{1}n_{2}}$is an eigenstate
of the equation with eigenvalue zero, when $m=0$. This is the pseudoscalar
Goldstone boson of the theory, with the corresponding wavefunction
$\Phi _{-;n_{1}n_{2}}^{1}=1$, just as in the case of the 't~Hooft
model. The fact that the chiral limit of the theory is at $m=0$ is
a remarkable result, quite likely connected to the fact that the Wilson
fermion transverse hopping term is constructed using projection operators
in Dirac space. We feel that with a simple structure and an exact
chiral limit at $m=0$, numerical investigations of (\ref{eq:43.6})
will not be too complicated.

A result of particular interest is the Goldstone boson decay constant,
since that allows us to estimate the overall scale of the theory.
It is the same as that for the 't~Hooft model, except for the change
in dimensionality of Dirac space, \begin{equation}
\ob{f_{\pi }}_{(3+1)-\mathrm{dim}}^{r=1}=\sqrt{2}\ob{f_{\pi }}_{(1+1)-\mathrm{dim}}\mathop {\longrightarrow }^{m\rightarrow 0}\frac{1}{a_{\bot }}\sqrt{\frac{2N_{c}}{\pi }}\; .\label{eq:43.7}\end{equation}
 Using $N_{c}=3$ and the experimental value $f_{\pi }^{\mathrm{exp}}\simeq 132$
MeV, we obtain an estimate of the lattice cut-off, $\pi /a_{\bot }\simeq 300$
MeV.

\section{Outlook}

It is important to study bound states in QCD from first principles,
because detailed quantitative understanding of hadronic phenomena
requires that. We have approached this problem in the transverse lattice
QCD formulation, exploiting properties of large-$N_{c}$ and strong
transverse gauge coupling limits. In these limits, we have explicitly
obtained the generating functional of $(3+1)$-dim QCD, starting from
only quark and gluon degrees of freedom and no ad hoc model assumptions.
Our study has been limited to mesons, and we have derived an integral
equation satisfied by the meson light-front wavefunctions. Extraction
of detailed properties (e.g. masses and form factors) from this equation
requires numerical solutions to this integral equation. To study baryons,
we have to find soliton solutions to the effective action calculated
here in different baryon number sectors, and to study glueballs, we
have to go to subleading orders in strong transverse gauge coupling
expansion.

Other groups have also studied bound states using the transverse lattice
approach. We refer the reader to the detailed review of Burkardt and
Dalley for a survey \cite{burkardt-dalley}. They study transverse
lattice QCD in the Hamiltonian formulation, using the colour-dielectric
expansion. In the dielectric formulation, $SU(N_{c})$ group elements
representing physical gluon fields, are replaced by general matrix
variables corresponding to gluon fields smeared over short distance
scales. These dielectric variables carry colour degrees of freedom,
and form an effective gauge theory with a simple vacuum structure.
The number of possible interaction terms in this effective gauge theory
is much larger than in QCD. The requirements of symmetry restoration
help in organization of these interaction terms, in a hierarchy suitable
for practical solutions. To date, all such solutions have been obtained
in the large-$N_{c}$ limit, which has restricted them to mesons and
glueballs. Moreover, they have not been able to sum the transverse
hopping expansion to all orders, and so they can not extract detailed
chiral properties from their results.

\appendix

\chapter{Notations and Conventions}

\section{Space-time}

A point in $(1+3)$-dim Minkowski space time is specified by the coordinates
$x^{\mu }$, $\mu =0,1,2,3$. $x^{0}$ is the time coordinate, while
$x^{1}$, $x^{2}$, $x^{3}$ are spatial coordinates. The metric tensor
of $(1+3)$-dim Minkowski space is \begin{eqnarray}
g_{\mu \nu }=g^{\mu \nu } & \equiv  & \left(\begin{array}{cccc}
 +1 & 0 & 0 & 0\\
 0 & -1 & 0 & 0\\
 0 & 0 & -1 & 0\\
 0 & 0 & 0 & -1\end{array}\right),\label{eq:A1}
\end{eqnarray}
 and the invariant distance is \begin{equation}
ds^{2}\equiv g_{\mu \nu }dx^{\mu }dx^{\nu }=\ob{dx^{0}}^{2}-\ob{dx^{1}}^{2}-\ob{dx^{2}}^{2}-\ob{dx^{3}}^{2}.\label{eq:A2}\end{equation}
 It is implicitly understood that repeated indices are to be summed
over their appropriate range, unless explicitly stated otherwise.
The space-time indices $\mu ,\nu ,\ldots $ are raised or lowered
by contraction with the metric $g^{\mu \nu }$ or $g_{\mu \nu }$.
\begin{equation}
\begin{array}{cc}
 V^{\mu }=g^{\mu \nu }V_{\nu } & V_{\mu }=g_{\mu \nu }V^{\nu }\\
 V^{\mu }\equiv \ob{V^{0},V^{1},V^{2},V^{3}}\qquad  & \qquad V_{\mu }\equiv \ob{V^{0},-V^{1},-V^{2},-V^{3}}\end{array}\label{eq:A3}\end{equation}
 Derivatives with respect to contravariant $\ob{x^{\mu }}$ or covariant
$\ob{x_{\mu }}$ coordinates are abbreviated as \begin{equation}
\partial _{\mu }\equiv \frac{\partial }{\partial x^{\mu }}\qquad ,\qquad \partial ^{\mu }\equiv \frac{\partial }{\partial x_{\mu }}\; .\label{eq:A4}\end{equation}

The light-front coordinates are defined as \begin{equation}
x^{\pm }=\frac{\ob{x^{0}\pm x^{1}}}{\sqrt{2}}\; ,\quad x_{\pm }=\frac{\ob{x_{0}\pm x_{1}}}{\sqrt{2}}\; ,\quad x^{\bot }=\ob{x^{2},x^{3}},\quad x_{\bot }=\ob{x_{2},x_{3}},\label{eq:A5}\end{equation}
 where $x^{+}$, $x^{-}$ and $x^{\bot }$ are the light-front time,
longitudinal and transverse coordinates respectively. The invariant
distance in terms of light-front coordinates is \begin{equation}
ds^{2}=2dx^{+}dx^{-}-\ob{dx^{2}}^{2}-\ob{dx^{3}}^{2}.\label{eq:A6}\end{equation}
 For any four-vector $V^{\mu }$, $V^{\pm }=V_{\mp }$ while $V^{\bot }=-V_{\bot }$.
In terms of the light-front components, the scalar product of two
arbitrary four-vectors $V^{\mu }$ and $W^{\mu }$ takes the form
\begin{equation}
V^{\mu }W_{\mu }=V^{+}W^{-}+V^{-}W^{+}-V^{2}W^{2}-V^{3}W^{3}\; .\label{eq:A7}\end{equation}

If $p^{\mu }$ is the four-momentum, then $p^{-}$, $p^{+}$ and $p^{\bot }$
are the light-front energy, longitudinal momentum and transverse momentum
respectively. $p^{2}=p_{\mu }p^{\mu }=m^{2}$ for a real particle
of mass $m$. The light cone energy $p^{-}$ is then \begin{equation}
p^{-}=\frac{\ob{p^{\bot }}^{2}+m^{2}}{2p^{+}}.\label{eq:A8}\end{equation}

The $(1+3)$-dim transverse lattice space-time is the direct product
of $(1+1)$-dim Minkowski continuum and planar square lattice with
lattice spacing $a^{\bot }$. We denote the coordinates of a general
point in this space-time by $\ob{x,x^{\bot }}$, where $x\equiv x^{\mu }=\ob{x^{0},x^{1}}$
and $x^{\bot }=\ob{x^{2},x^{3}}$ label the coordinates in $(1+1)$-dim
Minkowski continuum and planar square lattice respectively.

\section{Pauli matrices}

The Pauli sigma-matrices $\sigma ^{1}$, $\sigma ^{2}$ and $\sigma ^{3}$
are the generators of the fundamental representation of $SU(2)$,
i.e. the group of $2\times 2$ unitary matrices, with unit determinant.
\begin{equation}
\begin{array}{ccc}
 \sigma ^{1}=\ob{\begin{array}{cc}
 0 & 1\\
 1 & 0\end{array}}, & \sigma ^{2}=\ob{\begin{array}{cc}
 0 & -i\\
 i & 0\end{array}}, & \sigma ^{3}=\ob{\begin{array}{cc}
 1 & 0\\
 0 & -1\end{array}}.\end{array}\label{eq:A9}\end{equation}
 They satisfy \begin{equation}
\sigma ^{i}\sigma ^{j}=\delta ^{jk}+i\epsilon ^{ijk}\sigma ^{k}.\label{eq:A10}\end{equation}
 The raising and lowering operators are $\sigma ^{\pm }=\frac{1}{2}\ob{\sigma ^{1}\pm i\sigma ^{2}}$,
\begin{equation}
\begin{array}{cc}
 \sigma ^{+}=\ob{\begin{array}{cc}
 0 & 1\\
 0 & 0\end{array}}, & \sigma ^{-}=\ob{\begin{array}{cc}
 0 & 0\\
 1 & 0\end{array}}.\end{array}\label{eq:A11}\end{equation}

\section{Dirac matrices}

The Dirac gamma-matrices $\{\gamma ^{\mu }\}$, with $\mu \in \{0,1,2,3\}$,
form a $4$-dimensional representation of the Clifford algebra defined
by the anticommumtation relations \begin{equation}
\rb{\gamma ^{\mu },\gamma ^{\nu }}_{+}=2g^{\mu \nu }.\label{eq:A12}\end{equation}
 Further define \begin{equation}
\gamma ^{5}=\gamma _{5}=i\gamma ^{0}\gamma ^{1}\gamma ^{2}\gamma ^{3}.\label{eq:A13}\end{equation}
 From the anticommutation relations, it follows that \begin{equation}
\rb{\gamma ^{5},\gamma ^{\mu }}_{+}=0\; ,\quad \ob{\gamma ^{5}}^{2}=1\; .\label{eq:A14}\end{equation}
 We use the particular matrix representation provided by \[
\begin{array}{ccc}
 \gamma ^{0}=\ob{\begin{array}{cc}
 \sigma ^{1} & 0\\
 0 & \sigma ^{1}\end{array}}, & \gamma ^{1}=\ob{\begin{array}{cc}
 i\sigma ^{2} & 0\\
 0 & i\sigma ^{2}\end{array}}, & \gamma ^{2}=\ob{\begin{array}{cc}
 0 & i\sigma ^{3}\\
 i\sigma ^{3} & 0\end{array}},\end{array}\]
\begin{equation}
\begin{array}{cc}
 \gamma ^{3}=\ob{\begin{array}{cc}
 0 & \sigma ^{3}\\
 -\sigma ^{3} & 0\end{array}}, & \gamma ^{5}=\ob{\begin{array}{cc}
 -\sigma ^{3} & 0\\
 0 & \sigma ^{3}\end{array}}.\end{array}\label{eq:A15}\end{equation}
 In this case, $\gamma ^{0},\gamma ^{5}$ are hermitian, while $\gamma ^{1},\gamma ^{2},\gamma ^{3}$
are anti-hermitian. Also \begin{equation}
\gamma ^{0}\ob{\gamma ^{\mu }}^{\dagger }\gamma ^{0}=\gamma ^{\mu }.\label{eq:A16}\end{equation}
 On the light-front, \begin{equation}
\gamma ^{\pm }=\frac{1}{\sqrt{2}}\ob{\gamma ^{0}\pm \gamma ^{1}}.\label{eq:A17}\end{equation}
 It follows that $\ob{\gamma ^{+}}^{2}=\ob{\gamma ^{-}}^{2}=0$, while
$\gamma ^{+}\gamma ^{-}+\gamma ^{-}\gamma ^{+}=2$. Explicitly, \begin{equation}
\begin{array}{cc}
 \gamma ^{+}=\ob{\begin{array}{cc}
 \sqrt{2}\sigma ^{+} & 0\\
 0 & \sqrt{2}\sigma ^{+}\end{array}}, & \gamma ^{-}=\ob{\begin{array}{cc}
 \sqrt{2}\sigma ^{-} & 0\\
 0 & \sqrt{2}\sigma ^{-}\end{array}}\end{array}.\label{eq:A18}\end{equation}

\section{Lie algebra of $SU\ob{N_{c}}$}

The Lie group $SU\ob{N_{c}}$ is the group of $N_{c}\times N_{c}$
matrices of unit determinant. Every element $V$ of $SU\ob{N_{c}}$
can be parametrized as in terms of $N_{c}^{2}-1$ real parameters
labeled as $\omega _{a}$\begin{equation}
V=\exp \ob{i\sum _{a=1}^{N_{c}^{2}-1}t_{a}\omega _{a}},\label{eq:A19}\end{equation}
 where $t_{a}$ are the generators of the Lie algebra of $SU\ob{N_{c}}$
and $\omega _{a}$ are real parameters. The generators are traceless,
Hermitean $N_{c}\times N_{c}$ matrices, i.e. \begin{equation}
\mathrm{tr}\ob{t_{a}}=0\; ,\quad t_{a}^{\dagger }=t_{a}.\label{eq:A20}\end{equation}
 They satisfy the commutation relations, \begin{equation}
\rb{t_{a},t_{b}}_{-}=if_{abc}t_{c}\; .\label{eq:A21}\end{equation}
 where the structure constants $f_{abc}$ are completely antisymmetric
and real. They also obey a completeness relation \begin{equation}
\sum _{a=1}^{N_{c}^{2}-1}\ob{t_{a}}_{ij}\ob{t_{a}}_{kl}=\delta _{il}\delta _{jk}-\frac{1}{N_{c}}\delta _{ij}\delta _{kl}.\label{eq:A22}\end{equation}
 We use the normalization convention \begin{equation}
\mathrm{tr}\ob{t_{a}t_{b}}=\delta _{ab}\; .\label{eq:A23}\end{equation}

For a general matrix $M$ in both colour and Dirac space, we denote
its elements as $M_{ij}^{\alpha \beta }$ (Latin indices for colour
and Greek indices for spinor). For its trace, we use the notation

\begin{itemize}
\item trace in colour space: $\mathrm{tr}(M)$, 
\item trace in Dirac space: $\mathrm{tr}_{\mathrm{D}}(M)$, 
\item trace in colour and Dirac space: $\mathrm{Tr}(M)$. 
\end{itemize}

\section{Parallel transporters}

In a gauge theory, parallel transporters are unitary matrices, $U\ob{C_{y;x}}$,
forming a map from directed curves $C_{y;x}$ to the gauge group.
They satisfy the group properties:

\begin{itemize}
\item $U\ob{\emptyset }=1$ where $\emptyset $ is a curve of zero length. 
\item $U\ob{C_{2}\circ C_{1}}=U\ob{C_{2}}U\ob{C_{1}}$, where $C_{2}\circ C_{1}$
is the path composed of $C_{1}$ followed by $C_{2}$. 
\item $U\ob{-C}=U\ob{C}^{-1}=U\ob{C}^{\dagger }$, where $-C$ denotes the
path $C$ traversed in the opposite direction. 
\end{itemize}
Under a gauge transformation specified by $V(z)$, \begin{equation}
U\ob{C_{y;x}}\rightarrow U'\ob{C_{y;x}}=V\ob{y}U\ob{C_{y;x}}V^{-1}\ob{x}.\label{eq:A24}\end{equation}
 Given the gauge field $A_{\mu }\ob{x}$, for an infinitesimal curve,
\begin{equation}
U\ob{C_{x+dx;x}}=1-A_{\mu }\ob{x}dx^{\mu }+\cdots .\label{eq:A25}\end{equation}
 If $c^{\mu }(t)$ parametrizes the curve $C$ from $c^{\mu }(0)=x^{\mu }$
to $c^{\mu }(s)=y^{\mu }$, then \begin{equation}
\frac{d}{ds}U\ob{C}=-A_{\mu }\ob{c(s)}\left.\frac{dc^{\mu }}{dt}\right|_{t=s}U\ob{C}\label{eq:A26}\end{equation}
 With the initial condition $U\ob{\emptyset }=1$, the solution is
\begin{equation}
U\ob{C}=\mathcal{P}\exp \cb{-\int _{0}^{s}A_{\mu }\ob{c(t)}\frac{dc^{\mu }}{dt}dt}\equiv \mathcal{P}\exp \cb{-\int _{C}A^{\mu }\ob{x,x^{\bot }}dx_{\mu }},\label{eq:A27}\end{equation}
 where the symbol $\mathcal{P}$ denotes path ordering with respect
to the variable $t$.

\chapter{The 't~Hooft Model}

\section{Preliminaries}

The 't~Hooft model is the $(1+1)$-dim $SU\ob{N_{c}}$ gauge theory,
with quarks in the fundamental representation, in the large-$N_{c}$
limit. We describe various aspects of its solution in this Appendix.
We use functional integral methods, since that allows the large-$N_{c}$
limit to be imposed in a straightforward fashion.

The action for this theory in terms of the quark and the gluon fields
is \begin{equation}
S=\frac{N_{c}}{4g^{2}}\int d^{2}x\; \mathrm{tr}\left(F^{\mu \nu }\left(x\right)F_{\mu \nu }\left(x\right)\right)+\int d^{2}x\sum _{a=1}^{N_{f}}\overline{\Psi }_{a}(x)[i\gamma ^{\mu }D_{\mu }-m_{a}]\Psi _{a}(x)\; .\label{eq:B1}\end{equation}
 When the bare quark masses $m_{a}$ are zero, then the action is
invariant under:

\begin{itemize}
\item Single flavour case: $U_{V}(1)\otimes U_{A}(1)$. 
\item Multi-flavour case: $SU_{L}(N_{f})\otimes SU_{R}(N_{f})\otimes U_{V}(1)\otimes U_{A}(1)$. 
\end{itemize}

\section{Generating functional for quark-antiquark \protect \protect \\
 correlation functions}

In presence of an external quark bilinear source term, the generating
functional of the theory is (for sake of simplicity, we work with
a single flavour), \begin{equation}
Z\left[J\right]=\int \left[D\Psi \cdot D\overline{\Psi }\right]\left[DA\right]\; \exp \left[iI\left[\Psi ,\overline{\Psi },A\right]+\ob{\mathrm{gauge}\; \mathrm{terms}}\right],\label{eq:B2}\end{equation}
\begin{equation}
I=S+\int d^{2}xd^{2}y\; \overline{\Psi }_{i}^{\alpha }(x)J_{ij}^{\alpha \beta }(x,y)\Psi _{j}^{\beta }(y)\; .\label{eq:B3}\end{equation}
 Here {}``gauge terms'' stands for both the chosen gauge as well
as the corresponding ghost determinant. Arbitrary connected time-ordered
quark-antiquark correlation functions can be obtained by taking functional
derivatives of $W\left[J\right]\equiv -i\ln Z\left[J\right]$ with
respect to $J$, e.g. \begin{equation}
\left\langle \mathrm{T}\ob{\overline{\Psi }_{i}^{\alpha }\left(x\right)\Psi _{j}^{\beta }\left(y\right)}\right\rangle =\left.\frac{\delta W\left[J\right]}{\delta J_{ij}^{\alpha \beta }\left(x,y\right)}\right|_{J=0},\label{eq:B4}\end{equation}
\begin{eqnarray}
\left\langle \mathrm{T}\ob{\overline{\Psi }_{i}^{\alpha }\left(x\right)\Psi _{j}^{\beta }\left(y\right)\overline{\Psi }_{k}^{\gamma }\left(z\right)\Psi _{l}^{\delta }\left(w\right)}\right\rangle  & - & \left\langle \mathrm{T}\ob{\overline{\Psi }_{i}^{\alpha }\left(x\right)\Psi _{j}^{\beta }\left(y\right)}\right\rangle \left\langle \mathrm{T}\ob{\overline{\Psi }_{k}^{\gamma }\left(z\right)\Psi _{l}^{\delta }\left(w\right)}\right\rangle \nonumber \\
 & = & \left.-i\frac{\delta ^{2}W\left[J\right]}{\delta J_{ij}^{\alpha \beta }\left(x,y\right)\delta J_{kl}^{\gamma \delta }\left(z,w\right)}\right|_{J=0}.\label{eq:B5}
\end{eqnarray}
 In order to compute the quark propagator and the quark-antiquark
scattering amplitude, $W[J]$ should be known to $O(J^{2})$.

\section{Computing $Z\rb{J}$ in large-$N_{c}$ limit }

We work in the ghost-free light-front gauge $A^{+}=A^{0}+A^{1}=0$.
In this gauge, the functional integral over the $A^{-}$ variables
is purely Gaussian; so $A^{-}$ can be integrated out exactly. \begin{eqnarray}
Z\left[J\right] & = & \int \left[D\Psi \cdot D\overline{\Psi }\right]\left[DA^{-}\right]\exp \rb{i\int d^{2}xd^{2}y\; \overline{\Psi }_{i}^{\alpha }(x)J_{ij}^{\alpha \beta }(x,y)\Psi _{j}^{\beta }(y)}\nonumber \\
 &  & \times \exp \left[i\int d^{2}x\; \overline{\Psi }_{i}^{\alpha }(x)\delta _{ij}\ob{i\left(\gamma ^{\mu }\right)^{\alpha \beta }\partial _{\mu }-m\delta ^{\alpha \beta }}\Psi _{j}^{\beta }(x)\right]\label{eq:B6}\\
 &  & \times \exp \left[-\frac{1}{2}\int d^{2}x\; A_{ij}^{-}\left(x\right)\left(-\frac{iN_{c}}{g^{2}}\delta _{il}\delta _{jk}\partial _{-}^{2}A_{kl}^{-}\left(x\right)\right)\right.\nonumber \\
 &  & \qquad \qquad \qquad \left.-\int d^{2}x\; \left(\overline{\Psi }_{i}^{\alpha }\left(x\right)\left(\gamma ^{+}\right)^{\alpha \beta }\Psi _{j}^{\beta }\left(x\right)\right)A_{ij}^{-}\left(x\right)\right].\nonumber 
\end{eqnarray}
 The integration over gauge fields produces a non-local interaction
term which is quadratic in fermion bilinears. Suppressing the overall
normalization constant, \begin{eqnarray}
Z\rb{J} & = & \int \left[D\Psi \cdot D\overline{\Psi }\right]\exp \rb{i\int d^{2}xd^{2}y\; \overline{\Psi }_{i}^{\alpha }(x)J_{ij}^{\alpha \beta }(x,y)\Psi _{j}^{\beta }(y)}\nonumber \\
 & \times  & \exp \left[i\int d^{2}x\; \overline{\Psi }_{i}^{\alpha }(x)\delta _{ij}\ob{i\left(\gamma ^{\mu }\right)^{\alpha \beta }\partial _{\mu }-m\delta ^{\alpha \beta }}\Psi _{j}^{\beta }(x)\right]\label{eq:B7}\\
 & \times  & \exp \rb{-\frac{ig^{2}}{2N_{c}}\int d^{2}xd^{2}y\; h\ob{x-y}\ob{\gamma ^{+}}^{\alpha \delta }\ob{\gamma ^{+}}^{\gamma \beta }\overline{\Psi }_{i}^{\alpha }(x)\Psi _{i}^{\beta }(y)\overline{\Psi }_{j}^{\gamma }(y)\Psi _{j}^{\delta }(x)},\nonumber 
\end{eqnarray}
 where $h(x-y)$ is the solution of $\partial _{-}^{2}h(x-y)=\delta ^{(2)}(x-y)$,
with the principal value prescription as the boundary condition. To
carry out the fermionic integration, we use the identity \cite{kluberg-stern-etal}:

\begin{eqnarray}
 & \int \left[\prod _{a}d\xi _{a}d\overline{\xi }_{a}\right]\exp \mathrm{Tr}f\left(\left\{ \overline{\xi }_{c}\xi _{d}\right\} \right) & \nonumber \\
 & =\int \left[\prod _{a}d\xi _{a}d\overline{\xi }_{a}\right]\rb{\prod _{ab}d\sigma _{ab}\delta \ob{\sigma _{ab}-\overline{\xi }_{a}\xi _{b}}}\exp \mathrm{Tr}f\ob{\cb{\sigma _{cd}}} & \nonumber \\
 & =\int \left[\prod _{a}d\xi _{a}d\overline{\xi }_{a}\right]\rb{\prod _{ab}d\sigma _{ab}d\lambda _{ab}}\exp \rb{i\lambda _{ab}\ob{\sigma _{ab}-\overline{\xi }_{a}\xi _{b}}}\exp \mathrm{Tr}f\ob{\cb{\sigma _{cd}}} & \label{eq:B8}\\
 & =\int \rb{\prod _{ab}d\sigma _{ab}d\lambda _{ab}}\det \ob{-i\lambda }\exp \rb{i\lambda _{ab}\sigma _{ab}+\mathrm{Tr}f\ob{\cb{\sigma _{cd}}}} & \nonumber \\
 & =\int \left[\prod _{ab}d\sigma _{ab}d\lambda _{ab}\right]\exp \ob{\mathrm{Tr}\left[\ln \left(-i\lambda \right)+i\sigma ^{T}\lambda +f\left(\left\{ \sigma _{cd}\right\} \right)\right]} & \nonumber 
\end{eqnarray}
 This identity allows the functional integral to be written only in
terms of bilocal bosonic variables: \begin{equation}
Z\left[J\right]=\int \left[D\sigma \cdot D\lambda \right]\; \exp \left[iV\left(\sigma ,\lambda ;J\right)\right],\label{eq:B9}\end{equation}
 where \begin{eqnarray}
V\ob{\sigma ,\lambda ;J} & = & -i\int d^{2}x\; \mathrm{Tr}\rb{\ln (-i\lambda (x,x))}+\int d^{2}xd^{2}y\Big [\lambda _{ij}^{\alpha \beta }\left(x,y\right)\sigma _{ij}^{\alpha \beta }\left(x,y\right)\nonumber \\
 & + & \sigma _{ij}^{\alpha \beta }\left(x,y\right)\; \Big \{\delta _{ij}\ob{i\ob{\gamma ^{\mu }}^{\alpha \beta }\partial _{\mu }-m\delta ^{\alpha \beta }}\delta ^{(2)}\left(x-y\right)+J_{ij}^{\alpha \beta }\left(x,y\right)\Big \}\label{eq:B10}\\
 & - & \frac{g^{2}}{2N_{c}}h(x-y)\left(\gamma ^{+}\right)^{\alpha \delta }\left(\gamma ^{+}\right)^{\gamma \beta }\sigma _{ii}^{\alpha \beta }\left(x,y\right)\sigma _{jj}^{\gamma \delta }\left(y,x\right)\Big ].\nonumber 
\end{eqnarray}
 $V\left(\sigma ,\lambda ;J\right)$ is of $O\left(N_{c}\right)$,
so the large-$N_{c}$ limit of the functional integral amounts to
evaluating the functional integral at its stationary point. (This
illustrates why the $N_{c}\rightarrow \infty $ limit can be thought
of as a classical limit of a quantum theory; albeit different than
the classical limit corresponding to $\hbar \rightarrow 0$.) The
equations determining the stationary point are: \begin{eqnarray}
\left.\frac{\partial V}{\partial \lambda _{ij}^{\alpha \beta }\left(x,y\right)}\right|_{\overline{\sigma },\overline{\lambda }} & = & -i\left(\overline{\lambda }^{-1}\right)_{ji}^{\beta \alpha }\left(y,x\right)+\overline{\sigma }_{ij}^{\alpha \beta }\left(x,y\right)=0\label{eq:B11}\\
\left.\frac{\partial V}{\partial \sigma _{ij}^{\alpha \beta }\left(x,y\right)}\right|_{\overline{\sigma },\overline{\lambda }} & = & \overline{\lambda }_{ij}^{\alpha \beta }\left(x,y\right)+\delta _{ij}\ob{i\ob{\gamma ^{\mu }}^{\alpha \beta }\partial _{\mu }-m\delta ^{\alpha \beta }}\delta ^{(2)}\left(x-y\right)+J_{ij}^{\alpha \beta }\left(x,y\right)\nonumber \\
 &  & -\frac{g^{2}}{N_{c}}\delta _{ij}\left(\gamma ^{+}\right)^{\alpha \delta }\left(\gamma ^{+}\right)^{\gamma \beta }h(x-y)\; \overline{\sigma }_{kk}^{\gamma \delta }\left(y,x\right)=0\label{eq:B12}
\end{eqnarray}
 The first stationary point equation implies \begin{equation}
\overline{\lambda }_{ij}^{\alpha \beta }(x,y)=i\ob{\overline{\sigma }^{-1}}_{ji}^{\beta \alpha }(y,x)\; .\label{eq:B13}\end{equation}
 Substitution of this in to the second stationary point equation yields
\begin{eqnarray}
 &  & i\ob{\overline{\sigma }^{-1}}_{ji}^{\beta \alpha }(y,x)+\delta _{ij}\ob{i\ob{\gamma ^{\mu }}^{\alpha \beta }\partial _{\mu }-m\delta ^{\alpha \beta }}\delta ^{(2)}\left(x-y\right)+J_{ij}^{\alpha \beta }\left(x,y\right)\nonumber \\
 &  & -\frac{g^{2}}{N_{c}}\delta _{ij}\left(\gamma ^{+}\right)^{\alpha \delta }\left(\gamma ^{+}\right)^{\gamma \beta }h(x-y)\; \overline{\sigma }_{kk}^{\gamma \delta }\left(y,x\right)=0\; .\label{eq:B14}
\end{eqnarray}
 Solution of (\ref{eq:B14}) will yield $\overline{\sigma }$ as a
function of $J$. Thus we have \begin{equation}
Z\left[J\right]=\exp \left[iV\left(\overline{\sigma },\overline{\lambda };J\right)\right]\quad \Rightarrow \quad W\left[J\right]=V\left(\overline{\sigma },\overline{\lambda };J\right)\equiv V_{\mathrm{eff}}\ob{\overline{\sigma };J}\; ,\label{eq:B15}\end{equation}
 where $V_{\mathrm{eff}}\ob{\overline{\sigma };J}$, obtained by substituting
for $\overline{\lambda }$ in terms of $\overline{\sigma }$, is given
by \begin{eqnarray}
V_{\mathrm{eff}}\ob{\overline{\sigma };J} & = & i\int d^{2}x\; \mathrm{Tr}\rb{\ln \overline{\sigma }(x,x)}+\int d^{2}xd^{2}y\nonumber \\
 &  & \Big [\overline{\sigma }_{ij}^{\alpha \beta }\left(x,y\right)\Big \{J_{ij}^{\alpha \beta }\left(x,y\right)+\delta _{ij}\ob{i\ob{\gamma ^{\mu }}^{\alpha \beta }\partial _{\mu }-m\delta ^{\alpha \beta }}\delta ^{(2)}\left(x-y\right)\Big \}\label{eq:B16}\\
 &  & -\frac{g^{2}}{2N_{c}}h(x-y)\left(\gamma ^{+}\right)^{\alpha \delta }\left(\gamma ^{+}\right)^{\gamma \beta }\overline{\sigma }_{ii}^{\alpha \beta }\left(x,y\right)\overline{\sigma }_{jj}^{\gamma \delta }\left(y,x\right)\Big ]\nonumber 
\end{eqnarray}
 Note that $V_{\mathrm{eff}}\left(\overline{\sigma };J\right)$ depends
on $J$ explicitly, as well as implicitly through $\overline{\sigma }$.

The generator of 1PI vertex functions $\Gamma \rb{\varphi }$ is the
Legendre transform of $W\rb{J}$. In terms of the effective field
$\varphi $ conjugate to the external source $J$, \begin{equation}
\varphi _{ij}^{\alpha \beta }(x,y)=\frac{\delta W\rb{J}}{\delta J_{ij}^{\alpha \beta }(x,y)}\; ,\label{eq:B17}\end{equation}
\begin{eqnarray}
\Gamma \rb{\varphi } & = & W\rb{J}-\int d^{2}xd^{2}y\varphi _{ij}^{\alpha \beta }(x,y)J_{ij}^{\alpha \beta }(x,y)\nonumber \\
 & = & i\int d^{2}x\; \mathrm{Tr}\left[\ln \varphi (x,x)\right]\nonumber \\
 &  & +\int d^{2}xd^{2}y\Big [\varphi _{ij}^{\alpha \beta }\left(x,y\right)\delta _{ij}\ob{i\ob{\gamma ^{\mu }}^{\alpha \beta }\partial _{\mu }-m\delta ^{\alpha \beta }}\delta ^{(2)}\left(x-y\right)\label{eq:B18}\\
 &  & -\frac{g^{2}}{2N_{c}}h(x-y)\left(\gamma ^{+}\right)^{\alpha \delta }\left(\gamma ^{+}\right)^{\gamma \beta }\varphi _{ii}^{\alpha \beta }\left(x,y\right)\varphi _{jj}^{\gamma \delta }\left(y,x\right)\Big ].\nonumber 
\end{eqnarray}
 It can be easily seen that in the large-$N_{c}$ limit, $\varphi =\overline{\sigma }+\mathrm{O}\ob{1/N_{c}}$.
Also \begin{equation}
\rb{W^{(2)}}_{ij;kl}^{\alpha \beta ;\gamma \delta }(x,y;z,w)=-\rb{\ob{\Gamma ^{(2)}}^{-1}}_{ij;kl}^{\alpha \beta ;\gamma \delta }(x,y;z,w)\; ,\label{eq:B19}\end{equation}
 where $W^{(2)}$ and $\Gamma ^{(2)}$ are functional second derivatives
of $W[J]$ and $\Gamma [\varphi ]$ respectively.

\section{Quark propagator}

Let us first solve the stationary point equation (\ref{eq:B14}) when
$J=0$. \begin{eqnarray}
 &  & -\ob{\overline{\sigma }_{0}^{-1}}_{ji}^{\beta \alpha }(y,x)+i\delta _{ij}\ob{i\ob{\gamma ^{\mu }}^{\alpha \beta }\partial _{\mu }-m\delta ^{\alpha \beta }}\delta ^{(2)}\left(x-y\right)\nonumber \\
 &  & -i\frac{g^{2}}{N_{c}}\delta _{ij}\left(\gamma ^{+}\right)^{\alpha \delta }\left(\gamma ^{+}\right)^{\gamma \beta }h(x-y)\; \ob{\overline{\sigma }_{0}}_{kk}^{\gamma \delta }\left(y,x\right)=0\; .\label{eq:B20}
\end{eqnarray}
 $\overline{\sigma }_{0}$ is expected to be invariant under space-time
translations, and so can be parametrized as \begin{equation}
\left(\overline{\sigma }_{0}\right)_{ij}^{\alpha \beta }\left(x,y\right)=-i\int \frac{d^{2}p}{\left(2\pi \right)^{2}}\delta _{ji}\left[\frac{1}{\not p-m-\Sigma \left(p\right)+i\epsilon }\right]^{\beta \alpha }e^{-ip\cdot \left(y-x\right)}\label{eq:B21}\end{equation}
 With this parametrization, the $J=0$ stationary point equation becomes
an integral equation for $\Sigma \left(p\right)$, \begin{equation}
\Sigma \left(p\right)=ig^{2}\int \frac{d^{2}q}{\left(2\pi \right)^{2}}\; \mathrm{P}\left[\frac{1}{\left(p^{+}-q^{+}\right)^{2}}\right]\gamma ^{+}\left(\frac{1}{\not q-m-\Sigma \left(q\right)+i\epsilon }\right)\gamma ^{+},\label{eq:B22}\end{equation}
 where the principal value prescription regularizing the singularity
is defined as \begin{equation}
\mathrm{P}\left[\frac{1}{z^{2}}\right]=\frac{1}{2}\left[\frac{1}{(z+i\epsilon )^{2}}+\frac{1}{(z-i\epsilon )^{2}}\right].\label{eq:B23}\end{equation}
 (\ref{eq:B22}) is just the 't~Hooft integral equation for the quark
self-energy. The Dirac structure of the equation requires $\Sigma \ob{p}$
to be proportional to $\gamma ^{+}$. The instantaneous nature of
the kernel demands that $\Sigma \ob{p}$ depend only on $p^{+}$.
The equation can then be solved for $\Sigma \left(p\right)$, \begin{equation}
\Sigma \left(p\right)=-\frac{g^{2}\gamma ^{+}}{2\pi p^{+}}.\label{eq:B24}\end{equation}
 From (\ref{eq:B4}), (\ref{eq:B15}), (\ref{eq:B16}) and (\ref{eq:B20}),
it follows that the $J=0$ stationary point, $\overline{\sigma }_{0}$,
is the quark propagator. Explicitly, \begin{eqnarray}
\left\langle \mathrm{T}\ob{\overline{\Psi }_{i}^{\alpha }\left(x\right)\Psi _{j}^{\beta }\left(y\right)}\right\rangle  & = & \ob{\overline{\sigma }_{0}}_{ij}^{\alpha \beta }\left(x,y\right)\nonumber \\
 & = & -i\int \frac{d^{2}p}{\left(2\pi \right)^{2}}\delta _{ji}\left[\frac{1}{\not p-m+\frac{g^{2}\gamma ^{+}}{2\pi p^{+}}+i\epsilon }\right]^{\beta \alpha }e^{-ip\cdot \left(y-x\right)}\label{eq:B25}\\
 & = & -i\int \frac{d^{2}p}{\ob{2\pi }^{2}}\delta _{ji}\rb{\frac{\not p+m+\frac{g^{2}\gamma ^{+}}{2\pi p^{+}}}{p^{2}-m^{2}+\frac{g^{2}}{\pi }+i\epsilon }}^{\beta \alpha }e^{-ip\cdot (y-x)}\; .\nonumber 
\end{eqnarray}

\section{Meson spectrum}

The time-ordered four-point quark-antiquark Green's function can be
obtained from either $W\left[J\right]$ or $\Gamma \rb{\varphi }$.
Combining (\ref{eq:B5}) and (\ref{eq:B19}), \begin{eqnarray}
\left\langle \mathrm{T}\ob{\overline{\Psi }_{i}^{\alpha }\left(x\right)\Psi _{j}^{\beta }\left(y\right)\overline{\Psi }_{k}^{\gamma }\left(z\right)\Psi _{l}^{\delta }\left(w\right)}\right\rangle  & - & \left\langle \mathrm{T}\ob{\overline{\Psi }_{i}^{\alpha }\left(x\right)\Psi _{j}^{\beta }\left(y\right)}\right\rangle \left\langle \mathrm{T}\ob{\overline{\Psi }_{k}^{\gamma }\left(z\right)\Psi _{l}^{\delta }\left(w\right)}\right\rangle \nonumber \\
 & = & \left.-i\frac{\delta ^{2}W\left[J\right]}{\delta J_{ij}^{\alpha \beta }\left(x,y\right)\delta J_{kl}^{\gamma \delta }\left(z,w\right)}\right|_{J=0}\nonumber \\
 & = & \left.i\rb{\ob{\Gamma ^{(2)}}^{-1}}_{ij;kl}^{\alpha \beta ;\gamma \delta }(x,y;z,w)\right|_{\varphi =\overline{\sigma }_{0}}.\label{eq:B26}
\end{eqnarray}
 Using $\Gamma \rb{\varphi }$ in (\ref{eq:B18}), one can show that
\begin{equation}
\left.\Gamma ^{(2)}\right|_{\varphi =\overline{\sigma }_{0}}=S+K\; ,\label{eq:B27}\end{equation}
 where \begin{equation}
S_{ij,kl}^{\alpha \beta ,\gamma \delta }\left(x,y;z,w\right)=-i\left(\overline{\sigma }_{0}^{-1}\right)_{jk}^{\beta \gamma }\left(y,z\right)\left(\overline{\sigma }_{0}^{-1}\right)_{li}^{\delta \alpha }\left(w,x\right)\; ,\label{eq:B28}\end{equation}
\begin{equation}
K_{ij,kl}^{\alpha \beta ,\gamma \delta }\left(x,y;z,w\right)=-\frac{g^{2}}{N_{c}}\delta _{ij}\delta _{kl}\left(\gamma ^{+}\right)^{\alpha \delta }\left(\gamma ^{+}\right)^{\gamma \beta }h(x-y)\delta ^{(2)}\left(z-y\right)\delta ^{(2)}\left(w-x\right)\; .\label{eq:B29}\end{equation}
 If there are mesons in the spectrum, they will show up as poles in
the four-point quark-antiquark Green's function. To find the positions
of these poles, all we need to do is find the zeros of $\left[S+K\right]$.
The meson spectrum is thus determined by \begin{equation}
\int d^{2}zd^{2}w\left[S+K\right]_{ij,kl}^{\alpha \beta ,\gamma \delta }\left(x,y;z,w\right)\chi _{kl}^{\gamma \delta }\left(z,w\right)=0\; .\label{eq:B30}\end{equation}
 Transforming this equation to momentum space, one gets the integral
equation \begin{eqnarray}
 &  & i\left(\not p_{1}-m+\frac{g^{2}\gamma ^{+}}{2\pi p_{1}^{+}}\right)^{\alpha \delta }\left(-\not p_{2}-m-\frac{g^{2}\gamma ^{+}}{2\pi p_{2}^{+}}\right)^{\gamma \beta }\chi _{ji}^{\gamma \delta }\left(p_{2},p_{1}\right)\nonumber \\
 & + & \delta _{ij}\frac{g^{2}}{N_{c}}\left(\gamma ^{+}\right)^{\alpha \delta }\left(\gamma ^{+}\right)^{\gamma \beta }\int \frac{d^{2}k_{1}d^{2}k_{2}}{\left(2\pi \right)^{2}}\; \delta ^{(2)}\left(p_{1}+p_{2}-k_{1}-k_{2}\right)\nonumber \\
 &  & \cdot \; \mathrm{P}\left[\frac{1}{\left(p_{1}^{+}-k_{1}^{+}\right)^{2}}\right]\chi _{ll}^{\gamma \delta }\left(k_{2},k_{1}\right)=0\; .\label{eq:B31}
\end{eqnarray}
 Without loss of generality, we scale the momenta and choose \begin{eqnarray}
p_{1}=\left(p_{1}^{+},p_{1}^{-}\right),\; p_{2}=\left(p_{2}^{+},p_{2}^{-}\right),\; p_{1}^{+}+p_{2}^{+}=1,\; p_{1}^{+}=x,\; k_{1}^{+}=y=y_{1},\; k_{2}^{+}=y_{2}. &  & \label{eq:B32}
\end{eqnarray}
 Then the momentum space integral equation (\ref{eq:B31}) becomes
\begin{eqnarray}
\chi _{ii}^{\alpha \beta }\left(1-x,p_{2}^{-};x,p_{1}^{-}\right) & = & \frac{\left[p_{1}^{-}\gamma ^{+}+x\gamma ^{-}+m+\frac{g^{2}\gamma ^{+}}{2\pi x}\right]^{\beta \gamma }}{2p_{1}^{-}x-m^{2}+\frac{g^{2}}{\pi }+i\epsilon }\nonumber \\
 & \times  & \frac{\left[-p_{2}^{-}\gamma ^{+}-\left(1-x\right)\gamma ^{-}+m-\frac{g^{2}\gamma ^{+}}{2\pi \left(1-x\right)}\right]^{\delta \alpha }}{2p_{2}^{-}\left(1-x\right)-m^{2}+\frac{g^{2}}{\pi }+i\epsilon }\label{eq:B33}\\
 & \times  & ig^{2}\left(\gamma ^{+}\right)^{\gamma \sigma }\left(\gamma ^{+}\right)^{\tau \delta }\int \frac{dk_{1}^{-}dk_{2}^{-}dy_{1}dy_{2}}{\left(2\pi \right)^{2}}\; \delta \left(p_{1}^{-}+p_{2}^{-}-k_{1}^{-}-k_{2}^{-}\right)\nonumber \\
 & \times  & \delta \left(1-y_{1}-y_{2}\right)\mathrm{P}\left[\frac{1}{\left(x-y_{1}\right)^{2}}\right]\chi _{ll}^{\tau \sigma }\left(y_{1},k_{2}^{-};y_{1},k_{1}^{-}\right)\nonumber 
\end{eqnarray}
 The light-front wavefunction is defined through the projection \begin{equation}
\Phi ^{\beta \alpha }\left(x\right)=\int dp_{1}^{-}dp_{2}^{-}\; \delta \left(\frac{M^{2}}{2}-p_{1}^{-}-p_{2}^{-}\right)\chi _{ii}^{\alpha \beta }\left(1-x,p_{2}^{-};x,p_{1}^{-}\right).\label{eq:B34}\end{equation}
 The content of (\ref{eq:B33}) can then be written solely in terms
of $\Phi ^{\alpha \beta }\left(x\right)$ as, \begin{eqnarray}
\left[M^{2}-\frac{\left(m^{2}-\frac{g^{2}}{\pi }\right)}{x}-\frac{\left(m^{2}-\frac{g^{2}}{\pi }\right)}{1-x}\right]\Phi ^{\beta \alpha }\left(x\right) & = & \frac{\left(x\gamma ^{-}+m\right)^{\beta \gamma }\left(-\left(1-x\right)\gamma ^{-}+m\right)^{\delta \alpha }}{2x\left(1-x\right)}\label{eq:B35}\\
 & \times  & \frac{g^{2}}{2\pi }\left(\gamma ^{+}\right)^{\gamma \sigma }\left(\gamma ^{+}\right)^{\tau \delta }\int _{0}^{1}dy\; \mathrm{P}\left[\frac{1}{\left(x-y\right)^{2}}\right]\Phi ^{\sigma \tau }\left(y\right).\nonumber 
\end{eqnarray}
 Using standard matrix notation, (\ref{eq:B35}) can be written in
a compact form, \begin{eqnarray}
M^{2}\Phi (x) & = & \left[\frac{\left(m^{2}-\frac{g^{2}}{\pi }\right)}{x}+\frac{\left(m^{2}-\frac{g^{2}}{\pi }\right)}{1-x}\right]\Phi \left(x\right)\label{eq:B36}\\
 & + & \frac{g^{2}}{4\pi x(1-x)}\int _{0}^{1}dy\; \mathrm{P}\rb{\frac{1}{(x-y)^{2}}}\ob{x\gamma ^{-}+m}\gamma ^{+}\Phi (y)\gamma ^{+}\ob{-(1-x)\gamma ^{-}+m}\; .\nonumber 
\end{eqnarray}
 Until now we were working with a single flavour. The calculation
can easily be extended to the multi-flavour case, since quark-antiquark
pair production/annihilation vanishes in the large-$N_{c}$ limit.
If $m_{a}$ and $m_{b}$ are the masses of flavour $a$ and $b$ respectively,
then the multi-flavour analogue of (\ref{eq:B36}) is \begin{eqnarray}
M^{2}\Phi (x) & = & \left[\frac{\left(m_{a}^{2}-\frac{g^{2}}{\pi }\right)}{x}+\frac{\left(m_{b}^{2}-\frac{g^{2}}{\pi }\right)}{1-x}\right]\Phi \left(x\right)\label{eq:B37}\\
 & + & \frac{g^{2}}{4\pi x(1-x)}\int _{0}^{1}dy\; \mathrm{P}\rb{\frac{1}{(x-y)^{2}}}\ob{x\gamma ^{-}+m_{a}}\gamma ^{+}\Phi (y)\gamma ^{+}\ob{-(1-x)\gamma ^{-}+m_{b}}\; .\nonumber 
\end{eqnarray}
 From the structure of above equation, one can see that only one component
of $\Phi ^{\alpha \beta }\left(x\right)$ is independent. We expand
$\Phi $ as \begin{equation}
\Phi \left(x\right)=\Phi _{1}\left(x\right)1+\Phi _{+}\left(x\right)\gamma ^{+}+\Phi _{-}\left(x\right)\gamma ^{-}+\Phi _{+-}\left(x\right)\frac{1}{2}\rb{\gamma ^{+},\gamma ^{-}}_{-}.\label{eq:B38}\end{equation}
 Substituting this expansion in (\ref{eq:B37}), one can easily see
that $\Phi _{-}\left(x\right)$ satisfies the 't~Hooft equation,
\begin{eqnarray}
H\Phi _{-}(x)\equiv M^{2}\Phi _{-}\left(x\right) & = & \left[\frac{\left(m_{a}^{2}-\frac{g^{2}}{\pi }\right)}{x}+\frac{\left(m_{b}^{2}-\frac{g^{2}}{\pi }\right)}{1-x}\right]\Phi _{-}\left(x\right)\nonumber \\
 & - & \frac{g^{2}}{\pi }\int _{0}^{1}dy\; \mathrm{P}\left[\frac{1}{\left(x-y\right)^{2}}\right]\Phi _{-}\left(y\right)\; .\label{eq:B39}
\end{eqnarray}
 The remaining components $\Phi _{1}\left(x\right)$, $\Phi _{+}\left(x\right)$
and $\Phi _{+-}\left(x\right)$ are algebraically determined in terms
of $\Phi _{-}\left(x\right)$ as \begin{eqnarray}
\Phi _{1}\left(x\right) & = & \frac{1}{2}\ob{\frac{m_{a}}{x}-\frac{m_{b}}{1-x}}\Phi _{-}\left(x\right)\; ,\label{eq:B40}\\
\Phi _{+}\left(x\right) & = & -\frac{m_{a}m_{b}}{2x(1-x)}\Phi _{-}\left(x\right)\; ,\label{eq:B41}\\
\Phi _{+-}\left(x\right) & = & \frac{1}{2}\ob{\frac{m_{a}}{x}+\frac{m_{b}}{1-x}}\Phi _{-}\left(x\right)\; .\label{eq:B42}
\end{eqnarray}

\section{'t~Hooft solution}

Over the years, the properties and solutions of the 't~Hooft equation
(\ref{eq:B39}) have been investigated in great detail. Prominent
results are \cite{thooft2}\cite{ccg}\cite{einhorn1}:

\begin{itemize}
\item For finite norm solutions, $\Phi _{-}(x)$ must vanish at the boundaries
$x=0$ and $x=1$ as $x^{\beta _{a}}$ and $(1-x)^{\beta _{b}}$ respectively,
where \begin{equation}
\pi \beta _{a}\cot \ob{\pi \beta _{a}}=1-\frac{m_{a}^{2}\pi }{g^{2}}\; ,\quad \pi \beta _{b}\cot \ob{\pi \beta _{b}}=1-\frac{m_{b}^{2}\pi }{g^{2}}\; .\label{eq:B43}\end{equation}

\item The spectrum of $M^{2}$ is purely discrete. The light-front wavefunctions
$\Phi _{-}^{n}(x)$ are gauge invariant, and they form a complete
and orthogonal set. \begin{equation}
\sum _{n}\Phi _{-}^{n}(x)\Phi _{-}^{n}(x')=\delta (x-x')\label{eq:B44}\end{equation}
\begin{equation}
\int _{0}^{1}dx\Phi _{-}^{n*}(x)\Phi _{-}^{n'}(x)=\delta ^{nn'}\label{eq:B45}\end{equation}

\item When $m_{a}=m_{b}=0$, the lowest meson state has zero energy. The
corresponding wavefunction is $\Phi _{-}^{1}(x)=1$. This is the Goldstone
boson corresponding to chiral symmetry breaking. 
\item For large $n$ (large energy), the eigenvalues and eigenfunctions
can be approximated as \begin{equation}
\left.\begin{array}{c}
 M_{n}^{2}\simeq \pi g^{2}n\\
 \Phi _{-}^{n}(x)\simeq \sqrt{2}\sin \ob{\pi nx}\end{array}\right\} n\gg 1\; .\label{eq:B46}\end{equation}
 Thus the eigenvalues form an almost linear trajectory in the $M^{2}-n$
plane. 
\item Using the operator $K$ defined by \begin{equation}
K\Phi _{-}(x)=\mathrm{P}\int _{0}^{1}dy\frac{\Phi _{-}(y)}{(y-x)}\; ,\label{eq:B47}\end{equation}
 and the fact that \begin{equation}
\int _{0}^{1}dx\; \Phi _{-}^{n}(x)\rb{H,K}\Phi _{-}^{n}(x)=0\; ,\label{eq:B48}\end{equation}
 one can prove that \begin{equation}
m_{b}\int _{0}^{1}dy\frac{\Phi _{-}^{n}(y)}{1-y}=-\mathrm{P}_{n}m_{a}\int _{0}^{1}dy\frac{\Phi _{-}^{n}(y)}{y}\; .\label{eq:B49}\end{equation}
 Here $\mathrm{P}_{n}=(-1)^{n}$ is the parity of the eigenstate $n$.
The lightest meson is a pseudoscalar, $\mathrm{P}_{1}=-1$. 
\item Continuous symmetries cannot be spontaneously broken in two dimensions.
In the 't~Hooft model, chiral symmetry is realized in the Berezinski\u{\i}-Kosterlitz-Thouless
mode \cite{zhitnitsky}, \begin{equation}
\tb{\overline{\Psi }\Psi (x)\; \overline{\Psi }\Psi (0)}\; \sim \; x^{-1/N_{c}}\; .\label{eq:B50}\end{equation}
 To obtain results for the broken chiral symmetry phase (e.g. non-zero
chiral condensate), one must take the $N_{c}\rightarrow \infty $
limit before taking the $m\rightarrow 0$ limit. 
\item The chiral condensate, regularized by a split-point definition and
by subtracting the free field theory contribution, is \cite{burkardt1}\begin{eqnarray}
\tb{\overline{\Psi }\Psi } & = & -N_{c}\frac{m}{2\pi }\Big [1+\gamma _{E}+\ln \ob{\frac{m^{2}}{g^{2}}}\nonumber \\
 &  & +\ob{1-\frac{m^{2}\pi }{g^{2}}}\Big \{\frac{1}{\beta _{1}}+\sum _{n=2}^{\infty }\ob{\frac{1}{\beta _{n}}-\frac{1}{n}}\Big \}\Big ],\label{eq:B51}
\end{eqnarray}
 where $\beta _{n}\in [n-1,n]$ are the solutions to the 't~Hooft
boundary condition $\pi \beta \cot \ob{\pi \beta }=1-(m^{2}\pi /g^{2})$.
In the chiral limit, \begin{equation}
\tb{\overline{\Psi }\Psi }=-\frac{N_{c}}{\sqrt{12}}\ob{\frac{g^{2}}{\pi }}^{\frac{1}{2}}.\label{eq:B52}\end{equation}

\item Mesonic matrix elements of quark-antiquark bilinear operators are
easily obtained from (\ref{eq:B40}-\ref{eq:B42}) \cite{ccg}. In
particular, the axial current coupling is saturated by the Goldstone
boson in the chiral limit, and the pion decay constant is given by
\begin{equation}
\tb{0|\overline{\Psi }_{i}\gamma ^{\mu }\gamma ^{5}\Psi _{i}|\pi (p)}=if_{\pi }p^{\mu }\quad \Longrightarrow \quad f_{\pi }=\sqrt{\frac{N_{c}}{\pi }}\int _{0}^{1}dx\; \Phi _{-}^{1}(x)\; .\label{eq:B53}\end{equation}

\item When one of the quark flavours is much heavier than the other, $M_{Q}\gg m$,
one can take the static limit of the 't~Hooft equation \cite{burkardt2}.
In terms of the non-relativistic variables, $E_{\mathrm{stat}}=M-m_{Q}$,
$t=(1-x)m_{Q}$ and $\chi _{-}(t)=\frac{1}{\sqrt{m_{Q}}}\Phi _{-}(1-\frac{1}{m_{Q}})$,
\begin{equation}
E_{\mathrm{stat}}\chi _{-}(t)=\left[\frac{\left(m^{2}-\frac{g^{2}}{\pi }\right)}{2t}+\frac{t}{2}\right]\chi _{-}(t)-\frac{g^{2}}{\pi }\int _{0}^{\infty }dy\; \mathrm{P}\left[\frac{1}{\left(t-s\right)^{2}}\right]\chi _{-}(s)\; .\label{eq:B54}\end{equation}

\item Detailed results for physical properties have to be obtained numerically. 
\end{itemize}

\chapter{Lattice QCD at strong coupling and large-$N_{c}$}

\section{Preliminaries}

The $d$-dim Euclidean lattice QCD action in the strong coupling limit
is \begin{eqnarray}
S_{E} & = & -\sum _{x,\mu }\frac{i}{2}\Big [\overline{\Psi }(x+\mu )\ob{r-i\gamma ^{\mu }}U(x,\mu )\Psi (x)+\overline{\Psi }(x)\ob{r+i\gamma ^{\mu }}U^{\dagger }(x,\mu )\Psi (x+\mu )\Big ]\nonumber \\
 &  & +im\sum _{x}\overline{\Psi }(x)\Psi (x)\; .\label{eq:C1}
\end{eqnarray}
 Here $x\equiv \ob{x^{1},x^{2},\ldots ,x^{d}}$ are points on a $d$-dimensional
hypercubic lattice, and the lattice directions are labeled by $\mu =1,2,\ldots d$.
We choose to work with even $d$, and use units such that the lattice
spacing $a=1$. The Dirac matrices have dimensions $C=2^{d/2}$, and
satisfy \begin{eqnarray}
\rb{\gamma ^{\mu },\gamma ^{\nu }}_{+}=-2\delta ^{\mu \nu } & , & \ob{\gamma ^{\mu }}^{\dagger }=-\gamma ^{\mu }.\label{eq:C2}
\end{eqnarray}
 The gauge links $U(x,\mu )$ connect the nearest neighbour points
on the lattice. They are independent random variables in the strong
coupling limit (there is no pure gauge term in the action). The fermion
nearest neighbour hopping term is written using the Wilson parameter
$r$. With naive fermions ($r=0$), the theory gives rise to $2^{d}$
fermion zero modes, one for each corner of the Brillouin zone. Choosing
$r\neq 0$ removes all fermion doublers, leaving behind only the fermion
mode corresponding to the continuum theory.

For naive fermions, (\ref{eq:C1}) reduces to \begin{eqnarray}
S_{EN} & = & -\sum _{x,\mu }\frac{1}{2}\Big [\overline{\Psi }(x+\mu )\gamma ^{\mu }U(x,\mu )\Psi (x)-\overline{\Psi }(x)\gamma ^{\mu }U^{\dagger }(x,\mu )\Psi (x+\mu )\Big ]\nonumber \\
 &  & +im\sum _{x}\overline{\Psi }(x)\Psi (x)\label{eq:C3}
\end{eqnarray}
 The global symmetry group of this naive fermion action, $S_{EN}$,
can be easily deduced by rewriting it in terms of spin-diagonalized
fermion variables $\chi $ and $\overline{\chi }$. Let \begin{equation}
\Psi (x)=A(x)\chi (x)\; ,\quad \overline{\Psi }(x)=\overline{\chi }(x)A^{\dagger }(x)\; .\label{eq:C4}\end{equation}
 The choice \begin{equation}
A(x)=\ob{\gamma ^{1}}^{x^{1}}\ob{\gamma ^{2}}^{x^{2}}\cdots \ob{\gamma ^{d}}^{x^{d}},\quad A^{\dagger }(x)A(x)=1\; ,\label{eq:C5}\end{equation}
 reduces the $\gamma $-matrices in the action to position dependent
phase factors, \begin{equation}
\begin{array}{c}
 A^{\dagger }(x+\mu )\gamma ^{\mu }A(x)=\Delta ^{\mu }(x)\cdot 1\\
 A^{\dagger }(x)\gamma ^{\mu }A(x+\mu )=-\Delta ^{\mu }(x)\cdot 1\end{array},\label{eq:C6}\end{equation}
 where \begin{equation}
\Delta ^{\mu }(x)=(-1)^{\sum _{\nu <\mu }x^{\nu }}.\label{eq:C7}\end{equation}
 Thus the action reduces to \begin{eqnarray}
S_{EN} & = & -\sum _{x,\mu }\frac{\Delta ^{\mu }(x)}{2}\Big [\overline{\chi }(x+\mu )U(x,\mu )\chi (x)+\overline{\chi }(x)U^{\dagger }(x,\mu )\chi (x+\mu )\Big ]\nonumber \\
 &  & +im\sum _{x}\overline{\chi }(x)\chi (x)\; .\label{eq:C8}
\end{eqnarray}
 It is easy to see from this form of the action that the fermion hopping
term is invariant under $U_{\mathrm{o}}(C)\otimes U_{\mathrm{e}}(C)$
({}``$\mathrm{o}$'' and {}``$\mathrm{e}$'' respectively stand
for odd and even points): \begin{equation}
\left.\begin{array}{c}
 \chi (x)\quad \rightarrow \quad \chi '(x)=U_{\mathrm{o}(\mathrm{e})}\; \chi (x)\\
 \overline{\chi }(x)\quad \rightarrow \quad \overline{\chi }'(x)=\overline{\chi }(x)\; U_{\mathrm{e}(\mathrm{o})}^{\dagger }\end{array}\right\} \quad \mathrm{for}\; \ob{\sum _{\mu =1}^{d}x^{\mu }}\; \mathrm{odd}\, (\mathrm{even})\label{eq:C9}\end{equation}
 This invariance group describes the chiral symmetry of naive lattice
fermions, and it is considerably larger than the corresponding chiral
invariance group in the $d$-dim continuum theory. The fermion mass
term is invariant only the diagonal subgroup $U_{\Delta }(C)$ of
$U_{\mathrm{o}}(C)\otimes U_{\mathrm{e}}(C)$, defined by $U_{\mathrm{o}}=U_{\mathrm{e}}$.

For fermions with $r\neq 0$, the action $S_{E}$ is invariant only
under $U_{\Delta }(C)$, even if $m=0$. The $r$-dependent terms
break the non-diagonal part of $U_{\mathrm{o}}(C)\otimes U_{\mathrm{e}}(C)$
explicitly.

\section{Generating functional for quark-antiquark \protect \protect \\
 correlation functions}

Since we are interested in meson correlation functions, we add an
appropriate meson source term to the starting action. The partition
function of the theory is \begin{equation}
Z\rb{J}=\int \rb{D\Psi \cdot D\overline{\Psi }}\rb{DU}\exp \rb{-S_{E}+\sum _{x}J^{\alpha \beta }(x)\overline{\Psi }_{i}^{\alpha }(x)\Psi _{i}^{\beta }(x)}.\label{eq:C10}\end{equation}
 The integral over the gauge links $U(x,\mu )$ completely factorizes
to a separate integral for each individual link, and so can be evaluated
explicitly in the large-$N_{c}$ limit. The result is \begin{eqnarray}
Z\rb{J} & = & \int \rb{D\Psi \cdot D\overline{\Psi }}\exp \rb{-im\sum _{x}\overline{\Psi }(x)\Psi (x)+\sum _{x,y}J^{\alpha \beta }(x)\overline{\Psi }_{i}^{\alpha }(x)\Psi _{i}^{\beta }(x)}\nonumber \\
 &  & \qquad \qquad \quad \times \exp \rb{-N_{c}\sum _{x,\mu }\mathrm{tr}_{\mathrm{D}}\tilde{f}\rb{B(x,\mu )}}\label{eq:C11}
\end{eqnarray}
 where \begin{equation}
\tilde{f}\rb{B}=\sqrt{1+\frac{1}{N_{c}^{2}}B}-1-\ln \rb{\frac{1}{2}+\frac{1}{2}\sqrt{1+\frac{1}{N_{c}^{2}}B}}.\label{eq:C12}\end{equation}
 Matrix elements of $B(x,\mu )$ in Dirac space are \begin{equation}
B^{\alpha \beta }(x,\mu )=-\overline{\Psi }_{i}^{\alpha }(x+\mu )\Psi _{i}^{\delta }(x+\mu )\ob{r+i\gamma _{\mu }^{T}}^{\delta \gamma }\overline{\Psi }_{j}^{\gamma }(x)\Psi _{j}^{\sigma }(x)\ob{r-i\gamma _{\mu }^{T}}^{\sigma \beta }.\label{eq:C13}\end{equation}
 The integration over gauge link has produced non-local interactions
between fermion bilinears. To perform the fermionic functional integral
in the large~$N_{c}$ limit, we first rewrite the fermionic integral
in terms of bosonic variables using the identity (\ref{eq:B8}): \begin{equation}
Z\rb{J}=\int \rb{D\sigma \cdot D\lambda }\; \exp \rb{-V(\sigma ,\lambda )}\; ,\label{eq:C14}\end{equation}
\begin{eqnarray}
V(\sigma ,\lambda ) & = & -\sum _{x}\mathrm{Tr}\rb{\ln (-i\lambda \otimes 1_{c})(x)}-\sum _{x}i\sigma ^{\alpha \beta }(x)\lambda ^{\alpha \beta }(x)\nonumber \\
 &  & +\sum _{x}\ob{im\delta ^{\alpha \beta }-J^{\alpha \beta }(x)}\sigma ^{\alpha \beta }(x)+N_{c}f\rb{\sigma }\; .\label{eq:C15}
\end{eqnarray}
 Here \begin{equation}
f[\sigma ]=\sum _{x,\mu }\mathrm{Tr}\tilde{f}\rb{B(x,\mu )},\label{eq:C16}\end{equation}
 with matrix elements of $B^{\alpha \beta }(x,\mu )$ in terms of
$\sigma $ given by \begin{equation}
B^{\alpha \beta }(x,\mu )=-\sigma _{ii}^{\alpha \delta }(x+\mu )\ob{r+i\gamma _{\mu }^{T}}^{\delta \gamma }\sigma _{jj}^{\gamma \tau }(x)\ob{r-i\gamma _{\mu }^{T}}^{\tau \beta }\; .\label{eq:C17}\end{equation}
 In the large-$N_{c}$ limit, the bosonized functional integral reduces
to its stationary point value. The stationary point equations are:
\begin{equation}
\left.\frac{\partial V}{\partial \lambda ^{\alpha \beta }(x)}\right|_{\overline{\sigma },\overline{\lambda }}=-i\overline{\sigma }^{\alpha \beta }(x)-N_{c}\ob{\overline{\lambda }^{-{1}}}^{\beta \alpha }(x)=0\; ,\label{eq:C18}\end{equation}
\begin{eqnarray}
\left.\frac{\partial V}{\partial \sigma ^{\alpha \beta }(x)}\right|_{\overline{\sigma },\overline{\lambda }} & = & -i\overline{\lambda }^{\alpha \beta }(x)+\ob{im\delta ^{\alpha \beta }-J^{\alpha \beta }(x)}\nonumber \\
 &  & -N_{c}\sum _{\mu }\Big [\ob{r+i\gamma _{\mu }^{T}}\sigma (x-\mu )\ob{r-i\gamma _{\mu }^{T}}\tilde{f}'\rb{B(x-\mu ,\mu )}\nonumber \\
 &  & +\ob{r-i\gamma _{\mu }^{T}}\tilde{f}'\rb{B(x,\mu )}\sigma (x+\mu )\ob{r+i\gamma _{\mu }^{T}}\Big ]^{\beta \alpha }=0\; .\label{eq:C19}
\end{eqnarray}
 Here \begin{equation}
\tilde{f}'[B]\equiv \frac{\partial \tilde{f}[B]}{\partial B}=\frac{1}{2N_{c}^{2}\ob{1+\sqrt{1+B/N_{c}^{2}}}}.\label{eq:C20}\end{equation}
 Solution of the first stationary point equation (\ref{eq:C17}) gives
\begin{equation}
\overline{\lambda }^{\alpha \beta }(x)=iN_{c}\ob{\overline{\sigma }^{-1}}^{\beta \alpha }(x)\; .\label{eq:C21}\end{equation}
 Substituting this in the second stationary point equation (\ref{eq:C18})
yields a self-consistent equation for $\overline{\sigma }$, \begin{eqnarray}
 &  & N_{c}\ob{\overline{\sigma }^{-1}}^{\beta \alpha }(x)+\ob{im\delta ^{\alpha \beta }-J^{\alpha \beta }(x)}\nonumber \\
 &  & -N_{c}\sum _{\mu }\Big [\ob{r+i\gamma _{\mu }^{T}}\sigma (x-\mu )\ob{r-i\gamma _{\mu }^{T}}\tilde{f}'\rb{B(x-\mu ,\mu )}\nonumber \\
 &  & +\ob{r-i\gamma _{\mu }^{T}}\tilde{f}'\rb{B(x,\mu )}\sigma (x+\mu )\ob{r+i\gamma _{\mu }^{T}}\Big ]^{\beta \alpha }=0\; .\label{eq:C22}
\end{eqnarray}
 Solution of this equation will yield $\overline{\sigma }$ as a function
of $J$. The large-$N_{c}$ generating functional of connected meson
correlations is thus given by \begin{equation}
W\rb{J}=-\ln Z\rb{J}=V\ob{\overline{\sigma },\overline{\lambda }}\equiv V_{\mathrm{eff}}\ob{\overline{\sigma };J}\; ,\label{eq:C23}\end{equation}
 with $V_{\mathrm{eff}}\ob{\overline{\sigma };J}$ (obtained by substituting
$\overline{\lambda }$ in terms of $\overline{\sigma }$ ) given by,
\begin{eqnarray}
V_{\mathrm{eff}}\ob{\overline{\sigma };J} & = & \sum _{x}\mathrm{Tr}\rb{\ln \ob{\overline{\sigma }\otimes 1_{c}}(x)}+im\sum _{x}\mathrm{tr}_{\mathrm{D}}\overline{\sigma }(x)-\sum _{x}\mathrm{tr}_{\mathrm{D}}J^{T}(x)\overline{\sigma }(x)\label{eq:C24}\\
 & + & N_{c}\sum _{x,\mu }\mathrm{tr}_{\mathrm{D}}\rb{\ob{1+\frac{\overline{B}(x,\mu )}{N_{c}^{2}}}^{\frac{1}{2}}-1-\ln \ob{\frac{1}{2}+\frac{1}{2}\ob{1+\frac{\overline{B}(x,\mu )}{N_{c}^{2}}}^{\frac{1}{2}}}}.\nonumber 
\end{eqnarray}
 The matrix $\overline{B}$ is $B$, eq.(\ref{eq:C17}), evaluated
at the stationary point, \begin{equation}
\overline{B}^{\alpha \beta }(x,\mu )=-\overline{\sigma }^{\alpha \delta }(x+\mu )\ob{r+i\gamma _{\mu }^{T}}^{\delta \gamma }\overline{\sigma }^{\gamma \tau }(x)\ob{r-i\gamma _{\mu }^{T}}^{\tau \beta }\label{eq:C25}\end{equation}

The generator of 1PI vertex functions $\Gamma \rb{\varphi }$ is the
Legendre transform of $W\rb{J}$. In terms of the effective field
$\varphi $ conjugate to the external source $J$, \begin{equation}
\varphi ^{\alpha \beta }(x)=\frac{\partial W\rb{J}}{\partial J^{\alpha \beta }(x)}\; ,\label{eq:C26}\end{equation}
\begin{eqnarray}
\Gamma \rb{\varphi } & = & W\rb{J}+\sum _{x}J^{\alpha \beta }(x)\; \varphi ^{\alpha \beta }(x)\nonumber \\
 & = & \sum _{x}\mathrm{Tr}\rb{\ln \ob{\varphi \otimes 1_{c}}}(x)+im\sum _{x}\mathrm{tr}_{\mathrm{D}}\varphi (x)\label{eq:C27}\\
 &  & +N_{c}\sum _{x,\mu }\mathrm{tr}_{\mathrm{D}}\rb{\ob{1+\frac{\overline{B}(x,\mu )}{N_{c}^{2}}}^{\frac{1}{2}}-1-\ln \ob{\frac{1}{2}+\frac{1}{2}\ob{1+\frac{\overline{B}(x,\mu )}{N_{c}^{2}}}^{\frac{1}{2}}}},\nonumber 
\end{eqnarray}
 with \begin{equation}
\overline{B}^{\alpha \beta }(x,\mu )=-\varphi ^{\alpha \delta }(x+\mu )\ob{r+i\gamma _{\mu }^{T}}^{\delta \gamma }\varphi ^{\gamma \tau }(x)\ob{r-i\gamma _{\mu }^{T}}^{\tau \beta }.\label{eq:C28}\end{equation}
 Note that $\varphi \simeq \overline{\sigma }$ in the large-$N_{c}$
limit. Also \begin{equation}
\rb{W^{(2)}}^{\alpha \beta ;\gamma \delta }(x,y)=\rb{\ob{\Gamma ^{(2)}}^{-1}}^{\alpha \beta ;\gamma \delta }(x,y)\; ,\label{eq:C29}\end{equation}
 where $W^{(2)}$ and $\Gamma ^{(2)}$ are functional second derivatives
of $W[J]$ and $\Gamma [\varphi ]$ respectively.

\section{Chiral condensate}

When $J=0$, we expect the stationary point to be translationally
invariant. Then equation (\ref{eq:C22}) gives, \begin{equation}
\frac{1}{\tilde{\sigma }_{0}}+im+\frac{d\ob{1-r^{2}}\tilde{\sigma }_{0}}{1+\sqrt{1+\ob{1-r^{2}}\tilde{\sigma }_{0}^{2}}}=0\; ,\label{eq:C30}\end{equation}
 where $\overline{\sigma }_{0}^{\alpha \beta }(x)=N_{c}\tilde{\sigma }_{0}\delta ^{\alpha \beta }$.
This equation can be solved to yield \begin{equation}
\tilde{\sigma }_{0}=i\frac{-m(d-1)+d\sqrt{m^{2}+(2d-1)\ob{1-r^{2}}}}{d^{2}\ob{1-r^{2}}+m^{2}}\; .\label{eq:C31}\end{equation}
 The chiral condensate is given by \begin{equation}
\tb{\overline{\Psi }_{i}^{\alpha }(x)\Psi _{i}^{\alpha }(x)}=\left.\frac{\partial W\rb{J}}{\partial J^{\alpha \alpha }(x)}\right|_{J=0}=\ob{\overline{\sigma }_{0}}^{\alpha \alpha }(x)=N_{c}C\tilde{\sigma }_{0}\; ,\label{eq:C32}\end{equation}
 In particular,

\begin{itemize}
\item Naive fermions ($r=0$):\begin{equation}
\tb{\overline{\Psi }\Psi }=iN_{c}C\frac{-m(d-1)+d\sqrt{m^{2}+2d-1}}{d^{2}+m^{2}}\; .\label{eq:C33}\end{equation}
 In the chiral limit $m\rightarrow 0$, a non-zero chiral condensate
signals spontaneous breakdown of $U_{\mathrm{o}}(C)\otimes U_{\mathrm{e}}(C)$
to its diagonal subgroup $U_{\Delta }(C)$, \begin{equation}
\lim _{m\rightarrow 0}\tb{\overline{\Psi }\Psi }=iN_{c}C\frac{\sqrt{2d-1}}{d}\; .\label{eq:C34}\end{equation}

\item Wilson fermions ($r=1$): \begin{equation}
\tb{\overline{\Psi }\Psi }=N_{c}C\frac{i}{m}\; .\label{eq:C35}\end{equation}
 Chiral symmetry is explicitly broken in this case. 
\end{itemize}

\section{Meson propagator}

The meson propagator is given by \begin{eqnarray}
\tb{\overline{\Psi }_{i}^{\alpha }(x)\Psi _{i}^{\beta }(x)\overline{\Psi }_{j}^{\gamma }(y)\Psi _{j}^{\delta }(y)} & = & \left.\frac{\partial W\rb{J}}{\partial J^{\alpha \beta }(x)}\right|_{J=0}\left.\frac{\partial W\rb{J}}{\partial J^{\gamma \delta }(y)}\right|_{J=0}+\left.\frac{\partial ^{2}W\rb{J}}{\partial J^{\alpha \beta }(x)\partial J^{\gamma \delta }(y)}\right|_{J=0}\nonumber \\
 & = & \ob{N_{c}\tilde{\sigma }_{0}}^{2}\delta ^{\alpha \beta }\delta ^{\gamma \delta }+\left.\rb{\ob{\Gamma ^{(2)}}^{-1}}^{\alpha \beta ,\gamma \delta }(x,y)\right|_{\varphi =\overline{\sigma }_{0}},\label{eq:C36}
\end{eqnarray}
 where \begin{eqnarray}
\left.\ob{\Gamma ^{(2)}}^{\alpha \beta ,\gamma \delta }(x,y)\right|_{\varphi =\overline{\sigma }_{0}} & = & -\frac{1}{N_{c}\tilde{\sigma }_{0}^{2}}\delta ^{\alpha \delta }\delta ^{\gamma \beta }\delta _{x,y}\nonumber \\
 &  & -\frac{1}{N_{c}}\Big [\ob{G_{1}+\ob{1-r^{2}}\tilde{\sigma }_{0}^{2}G_{2}}\sum _{\mu }\Big \{\ob{r-i\gamma _{\mu }}^{\alpha \delta }\ob{r+i\gamma _{\mu }}^{\gamma \beta }\delta _{x-\mu ,y}\nonumber \\
 &  & \qquad \qquad \qquad \qquad \qquad \qquad \qquad +\ob{r+i\gamma _{\mu }}^{\alpha \delta }\ob{r-i\gamma _{\mu }}^{\gamma \beta }\delta _{x+\mu ,y}\Big \}\nonumber \\
 &  & \qquad \quad -2d\ob{1-r^{2}}^{2}\tilde{\sigma }_{0}^{2}G_{2}\delta ^{\alpha \delta }\delta ^{\gamma \beta }\delta _{x,y}\Big ],\label{eq:C37}
\end{eqnarray}
 with \begin{equation}
G_{1}=\frac{1}{2\ob{1+\sqrt{1+\ob{1-r^{2}}\tilde{\sigma }_{0}^{2}}}}\; ,\label{eq:C38}\end{equation}
\begin{equation}
G_{2}=\frac{-1}{4\sqrt{1+\ob{1-r^{2}}\tilde{\sigma }_{0}^{2}}\ob{1+\sqrt{1+\ob{1-r^{2}}\tilde{\sigma }_{0}^{2}}}^{2}}\; .\label{eq:C39}\end{equation}
 Eq.(\ref{eq:C37}) is the inverse of the connected meson propagator.
The poles of the meson propagator therefore correspond to the zero
eigenvalues of $\Gamma ^{(2)}(p)$. Transforming $\Gamma ^{(2)}(x,y)$
to momentum space, we get \begin{eqnarray}
\ob{\Gamma ^{(2)}}^{\alpha \beta ,\gamma \delta }(p) & = & -\frac{1}{N_{c}}\Big [\Big (\frac{1}{\tilde{\sigma }_{0}^{2}}-2d\ob{1-r^{2}}^{2}\tilde{\sigma }_{0}^{2}G_{2}\Big )\delta ^{\alpha \delta }\delta ^{\gamma \beta }\label{eq:C40}\\
 &  & \qquad +\ob{G_{1}+\ob{1-r^{2}}\tilde{\sigma }_{0}^{2}G_{2}}\sum _{\mu }\Big \{2\ob{r^{2}\delta ^{\alpha \delta }\delta ^{\gamma \beta }+\gamma _{\mu }^{\alpha \delta }\gamma _{\mu }^{\gamma \beta }}\cos \ob{p_{\mu }}\nonumber \\
 &  & \qquad \qquad \qquad \qquad \qquad \qquad \qquad +2r\ob{\delta ^{\alpha \delta }\gamma _{\mu }^{\gamma \beta }-\gamma _{\mu \textrm{ }}^{\alpha \delta }\delta ^{\gamma \beta }}\sin \ob{p_{\mu }}\Big \}\Big ].\nonumber 
\end{eqnarray}
 $\Gamma ^{(2)}(p)$ is a matrix in the $C^{2}$-dimensional space
of the Clifford algebra generated by the $d$ gamma matrices. This
matrix has to be diagonalized in order to find the spectrum of meson
states. In $4$-dim, the commonly chosen Clifford algebra basis is:
\begin{equation}
\Gamma _{M}=\{S=1,\; V=\gamma _{\mu },\; T=\frac{1}{2}[\gamma _{\mu },\gamma _{\nu }],\; A=\gamma _{\mu }\gamma _{5},\; P=\gamma _{5}\}\; .\label{eq:C41}\end{equation}
 There is at most one meson state for each spin-parity sector, i.e.
no radial excitations survive in the strong coupling limit.

For naive fermions ($r=0$), the inverse connected meson propagator
is \begin{equation}
\ob{\Gamma ^{(2)}}^{\alpha \beta ,\gamma \delta }(p)=-N_{c}\Big [\Big (\frac{1}{\tilde{\sigma }_{0}^{2}}-2d\tilde{\sigma }_{0}^{2}G_{2}\Big )\delta ^{\alpha \delta }\delta ^{\gamma \beta }+2\ob{G_{1}+\tilde{\sigma }_{0}^{2}G_{2}}\sum _{\mu }\gamma _{\mu }^{\alpha \delta }\gamma _{\mu }^{\gamma \beta }\cos \ob{p_{\mu }}\Big ],\label{eq:C42}\end{equation}
 and it is exactly diagonalized by the basis (\ref{eq:C41}). The
resultant dispersion relation is \begin{equation}
4\sum _{\mu }\eta _{\mu }^{A}\sin ^{2}\ob{\frac{p_{\mu }}{2}}+\cb{\frac{im}{\tilde{\sigma }_{0}\ob{G_{1}+\tilde{\sigma }_{0}^{2}G_{2}}}+4n}=0\; ,\label{eq:C43}\end{equation}
 where $n$ is $0$ (pseudoscalar), $1$ (vector), $\ldots $ , $d-1$
(axial), $d$ (scalar), and $\eta _{\mu }^{A}$ is $+1(-1)$ when
$\Gamma _{M}$ anticommutes (commutes) with $\gamma _{\mu }$. This
result shows that, the fermion doubling phenomenon for naive fermions,
produces $C^{2}$ pseudo-Goldstone bosons with mass \begin{equation}
M^{2}=\frac{im}{\tilde{\sigma }_{0}\ob{G_{1}+\tilde{\sigma }_{0}^{2}G_{2}}}=im\frac{4\sqrt{1+\tilde{\sigma }_{0}^{2}}}{\tilde{\sigma }_{0}}\; ,\label{eq:C44}\end{equation}
 one for each corner of the Brillouin zone. The pseudoscalar located
at $p=0$ is identified as the physical pion, while the other modes
at $p=0$ (i.e. $n\neq 0$) are interpreted as the heavier meson states
\cite{kluberg-stern-etal}\cite{martin1}.

For Wilson fermions, we restrict ourselves to $d=4$. For $r\neq 0$,
the inverse connected meson propagator separates into three block
matrices corresponding to $S$, $PA$ and $VT$ channels. As $r$
increases from $0$ to $1$, states within each channel get more and
more strongly coupled. For $r\neq 0$, the {}``chiral limit'' is
non-perturbatively defined as the value $m=m_{c}$ for which the pseudoscalar
mass vanishes: \begin{equation}
r\rightarrow 0\; :\; m_{c}\sim \frac{3\sqrt{7}}{4}r\; ,\qquad r=1\; :\; m_{c}=2\; .\label{eq:C45}\end{equation}
 At $m=m_{c}$, $\tilde{\sigma }_{0}$ decreases smoothly from $\frac{\sqrt{7}}{4}$
to $\frac{1}{2}$ as $r$ increases from $0$ to $1$. For $r\neq 0$,
all physical meson states are located at $p=0$. At $r=1$ in particular,
there are only four meson states, one pseudoscalar and three vectors
\cite{kawamoto}\cite{kawamoto-smit}\cite{martin2}.

\end{document}